\documentclass[journal]{IEEEtran}
\usepackage{times}
\usepackage{arydshln}
\usepackage{tikz}
\usetikzlibrary{calc}
\usepackage[numbers,sort,compress]{natbib}
\usepackage{multicol}
\usepackage[bookmarks=true]{hyperref}

\usepackage{comment}
\usepackage{amsthm}
\usepackage{mathtools}
\usepackage{amsmath}
\usepackage{amsfonts}
\usepackage{soul} %for strike-through text
\usepackage{amsmath,amsfonts,amssymb}
\usepackage[mathscr]{euscript}
\usepackage{hyperref}
\usepackage{graphicx}
\usepackage[dvipsnames]{xcolor} % added for forest green colour.

\usepackage{algorithm}
\usepackage[noend]{algpseudocode}
\algrenewcommand{\algorithmicrequire}{\textbf{Input:}}
\algrenewcommand{\algorithmicensure}{\textbf{Output:}}

\hypersetup{
    colorlinks=true,
    linkcolor=black,
    citecolor=blue,
    urlcolor=black
}

\newtheorem{remark}{Remark}[section]
\newtheorem{theorem}{Theorem}[section]
\newtheorem{problem}{Problem}[section]

\newtheorem{condition}{Condition}[section]
\newtheorem{lemma}{Lemma}[section]
\newtheorem{definition}{Definition}[section]
\newcounter{subproblem}[problem] % Create a counter for subproblems
\usepackage{xcolor}

\renewcommand{\bf}[1]{\textbf{#1}}
\newcommand{\mycomment}[1]{}

\begin{document}

\title{Efficient Probabilistic Planning with Maximum-Coverage Distributionally Robust Backward Reachable Trees}

% You will get a Paper-ID when submitting a pdf file to the conference system
%\author{Author Names Omitted for Anonymous Review. Paper-ID [add your ID here]}
\author{Alex Rose, Naman Aggarwal, Christopher Jewison
 and Jonathan P. How, \IEEEmembership{IEEE Fellow} 
\thanks{This work was supported by the National Science Foundation Graduate Research Fellowship under grant no. 2141064 and by the Draper Scholars program. Alex Rose, Naman Aggarwal and Jonathan P. How are with Laboratory for Information and Decision Systems, Massachusetts
Institute of Technology, Cambridge MA 02139. Alex Rose is also a Draper Scholar and Christopher Jewison is with Draper, Cambridge MA 02139. Email: $\{\texttt{ameredit}, \texttt{namanagg}, \texttt{jhow} \}$@mit.edu, cjewison@draper.com.
}}

%\author{\authorblockN{Michael Shell}
%\authorblockA{School of Electrical and\\Computer Engineering\\
%Georgia Institute of Technology\\
%Atlanta, Georgia 30332--0250\\
%Email: mshell@ece.gatech.edu}
%\and
%\authorblockA{Twentieth Century Fox\\
%\and
%\authorblockN{James Kirk\\ and Montgomery Scott}
%\authorblockA{Starfleet Academy\\
%San Francisco, California 96678-2391\\

% avoiding spaces at the end of the author lines is not a problem with
% conference papers because we don't use \thanks or \IEEEmembership

%Homer Simpson\authorrefmark{2},
%Montgomery Scott\authorrefmark{3} and
%Eldon Tyrell\authorrefmark{4}}
%Telephone: (800) 555--1212, Fax: (888) 555--1212}
%\authorblockA{\authorrefmark{4}Tyrell Inc., 123 Replicant Street, Los Angeles, California 90210--4321}}

\maketitle

\begin{abstract}
    This paper presents a new multi-query motion planning algorithm for linear Gaussian systems with the goal of reaching a Euclidean ball with high probability. We develop a new formulation for ball-shaped ambiguity sets of Gaussian distributions and leverage it to develop a distributionally robust belief roadmap construction algorithm. This algorithm synthesizes robust controllers which are certified to be safe for maximal size ball-shaped ambiguity sets of Gaussian distributions. Our algorithm achieves better coverage than the maximal coverage algorithm for planning over Gaussian distributions \cite{aggarwal2024sdp}, and we identify mild conditions under which our algorithm achieves strictly better coverage. For the special case of no process noise or state constraints, we formally prove that our algorithm achieves maximal coverage. 
    In addition, we present a second multi-query motion planning algorithm for linear Gaussian systems with the goal of reaching a region parameterized by the Minkowski sum of an ellipsoid and a Euclidean ball with high probability. This algorithm plans over ellipsoidal sets of maximal size ball-shaped ambiguity sets of Gaussian distributions, and provably achieves equal or better coverage than the best-known algorithm for planning over ellipsoidal ambiguity sets of Gaussian distributions \cite{aggarwal2025tac}. We demonstrate the efficacy of both methods in a wide range of conditions via extensive simulation experiments.
\end{abstract}

\IEEEpeerreviewmaketitle
%\section{Introduction}
\section{Introduction}
Robots navigating in environments with obstacles often rely on roadmaps of dynamically feasible and collision-free trajectories through the environment. Roadmaps are typically graph-structured and constructed offline. At runtime, a robot connects its initial and goal configurations to the precomputed roadmap and performs a graph search to find a feasible path to the goal configuration. For successful multi-query planning, it is desirable for a roadmap to have high \textit{coverage} (such that a large portion of the state space can be easily connected to the roadmap, enabling reuse) while maintaining \textit{efficiency} (such that the roadmap has relatively few nodes, enabling efficient search and easy storage of the roadmap).

Many prior works address multi-query roadmap construction for deterministic systems \cite{kuffner2000rrt, kavraki1996probabilistic, karaman2011sampling, tedrake2010lqr, majumdar2017funnel}. The most similar to ours  \cite{tedrake2010lqr, majumdar2017funnel} construct invariant ``funnels'' from linear quadratic regular (LQR) feedback controllers, and build roadmaps with funnels as edges. In these roadmaps, nodes represent regions in $\mathbb{R}^n$ that are inside the ``mouths'' or ``tails'' of corresponding edge funnels. If a robot begins inside the mouth of any funnel in the roadmap, its trajectory is guaranteed to remain within a sequence of invariant funnels as it proceeds to the goal. However, these algorithms do not account for noise in the system dynamics.  

Other works have addressed multi-query roadmap construction for stochastic systems, including chance-constrained multi-query planning with open-loop control trajectories \cite{luders2010chance, luders2013robust}. FIRM \cite{agha2011firm} introduces feedback control, constructing stationary belief nodes connected by linear quadratic Gaussian (LQG) edge controllers. CS-BRM \cite{csbrm_tro2024} uses non-stationary Gaussian belief nodes with covariance steering edge controllers, but it does not handle state or control constraints. In the presence of state and control constraints, the mean and covariance steering problems are coupled \cite{okamoto2018optimal} and node feasibility must be explicitly established.

Our prior works \cite{aggarwal2024sdp, rose2024revise, aggarwal2025tac} establish the notion of roadmap coverage and present algorithms that synthesize distributionally robust edge controllers to maximize roadmap coverage when node feasibility must be explicitly established. MAXCOVAR \cite{aggarwal2024sdp} characterizes roadmap coverage in terms of backward reachable sets of distributions, and develops an edge controller that provably maximizes coverage. Our more recent work MAXELLIPSOID \cite{aggarwal2025tac} introduces an algorithm that relies on a new type of ambiguity set of Gaussian distributions. Roadmaps constructed by this algorithm achieve provably maximal volume projections in the state space. %Both MAXCOVAR \cite{aggarwal2024sdp} and MAXELLIPSOID \cite{aggarwal2025tac} leverage node representations that decouple first- and second-order moment uncertainty. 

In this work, we consider a problem where the objective is to reach a goal region in Euclidean space with success probability of at least $1-\epsilon$. This is a common goal set formulation in real-word robotics and aerospace scenarios, and is typically consistent with NASA safety requirements \cite{stamatelatos2011probabilistic}. 
In order to satisfy this stochastic goal constraint, distributions with a mean close to the boundary of the goal region must have a small covariance, while distributions with a mean close to the center of the goal region can safely have a larger covariance. In other words, the geometry of the goal ambiguity set in the joint mean-covariance space includes a coupling between the first- and second-order moments of distributions.
%We must capture this coupling between the first- and second-order moments of distributions to accurately represent the geometry of the goal ambiguity set in the joint mean-covariance space. %Our prior works on planning for maximal coverage \cite{aggarwal2024sdp, aggarwal2025tac} cannot represent the coupling in the goal set, and so do not achieve maximal coverage for this problem (see Section \ref{sec: experiments}).

When the first- and second-order moments of distributions are coupled, prior works on distributionally robust control have relied on Gelbrich uncertainty sets \cite{pan2023distributionally} and Wasserstein uncertainty sets \cite{balci2020covariance, hakobyan2024wasserstein, pilipovsky2024distributionally, yang2020wasserstein} to represent uncertainty. However, we show that Gelbrich and 2-Wasserstein uncertainty sets under-approximate the set of distributions that can reach a Euclidean ball with high probability under a given control policy (see Remark \ref{remark: w2_projections} for details). Further, our prior works on planning for maximal coverage \cite{aggarwal2024sdp, aggarwal2025tac} fully decouple first- and second-order moment uncertainty, and so also under-approximate this set of distributions. As such, we define a new ball-shaped ambiguity set that represents this set of distributions more accurately than Gelbrich or 2-Wasserstein uncertainty sets or uncertainty sets which decouple first- and second-order moment uncertainty (see Section \ref{subsec: ambiguity_sets}). 

We leverage our novel ambiguity set representation to construct two distributionally robust belief roadmap construction algorithms. The first, called MAX-COV-BALL, grows a backward reachable tree of maximal size ball-shaped ambiguity sets. This algorithm guarantees equal or better coverage than MAXCOVAR \cite{aggarwal2024sdp} under any conditions. 
%We also identify mild conditions where MAX-COV-BALL strictly outperforms MAXCOVAR. We present a formal proof that MAX-COV-BALL achieves maximal coverage in the special case of no process noise and no state constraints. 
The second, called MAX-ELL-BALL, plans over ellipsoidal sets of maximal size ball-shaped ambiguity sets, and always achieves equal or better coverage than MAXELLIPSOID \cite{aggarwal2025tac}. Our contributions are: 
\begin{itemize}
\item We provide a novel ambiguity set representation and leverage it to construct a novel belief roadmap construction algorithm (MAX-COV-BALL) that constructs backward reachable trees of maximal ball-shaped ambiguity sets of distributions. We also develop a related method (MAX-ELL-BALL) that constructs backward reachable trees of maximal ellipsoidal sets of ball-shaped ambiguity sets of distributions.
\item We provide a formal proof that MAX-COV-BALL achieves maximal coverage in the absence of process noise and state constraints, and guarantees strictly better coverage than state-of-the-art algorithms under mild conditions.
\item We demonstrate that our methods outperform state-of-the-art methods \cite{aggarwal2024sdp, aggarwal2025tac} in a variety of practical scenarios via extensive simulation experiments.
\end{itemize}

%\section{Related Work}
%\input{related_work.tex}
\section{Problem Statement}\label{sec:problem_statement}
\subsection{Notation and Mathematical Preliminaries}
We describe positive semidefinite and positive definite matrices in $\mathbb{R}^{n \times n}$ as $\mathbb{S}^n_+$ and $\mathbb{S}^n_{++}$, respectively. We denote diagonal matrices in $\mathbb{R}^{n \times n}$ as $\mathbf{D}^n$ and the special orthogonal group of dimension $n$ as $\textrm{SO}(n)$. We use the notation that 
$\mathbb{B}_n(\mu, r) := \{\mathbf{x} \in \mathbb{R}^n: ||\mathbf{x} - \mu||_2 \leq r \}$ represents an $n$-ball with respect to the Euclidean norm. We define $n$-dimensional ellipsoids by $\mathscr{E}(\mu, \mathscr{P}) := \{\mathbf{x} \in \mathbb{R}^n: (\mathbf{x} - \mu)^T\mathscr{P}^{-1}(\mathbf{x} - \mu) \leq 1 \}$. The Minkowski sum of two sets $\mathcal{A}$ and $\mathcal{B}$ is denoted by $\mathcal{A} \oplus \mathcal{B} := \{a + b : a \in \mathcal{A}, b \in \mathcal{B} \}$. We denote n-dimensional multivariate Gaussian distributions by $\mathcal{N}(\mu, \Sigma)$, where $\mu \in \mathbb{R}^n$ is the distribution mean and $\Sigma \in \mathbb{S}^n_+$ is the covariance.

We assume that all random objects are defined on a common
probability space $(\Omega, \mathcal{F}, \mathbb{P})$. An $n$-dimensional real random variable $\mathbf{x}$ is defined as a measurable function from the sample space $\Omega$ to $\mathbb{R}^n$, $\mathbf{x} : \Omega \to \mathbb{R}^n$, such that $\mathbb{P}_\mathbf{x}$ denotes the measure induced by $\mathbf{x}$ on $\mathbb{R}^n$. In this paper, we work in finite-dimensional Euclidean spaces, and so assume that $\mathcal{F} = \mathcal{B}(\mathbb{R}^n)$, where $\mathcal{B}$ is the Borel $\sigma-$algebra on $\mathbb{R}^n$ for some finite $n$. As such, we define $\mathbb{P}_\mathbf{x} := \mathbb{P}(\mathbf{x}^{-1}(\mathcal{B}(\mathbb{R}^n)))$. We define $\mathcal{P}(\mathcal{W})$ as the set of all probability distributions on any measurable Borel space $\mathcal{W}$.
\subsection{Ambiguity Sets of Distributions}\label{subsec: ambiguity_sets}
An ambiguity set is a set of probability distributions defined on the same measurable space. As in \cite{aggarwal2025tac}, we use the notation that for a random variable $\mathbf{x} : \Omega \to \mathbb{R}^n$, $\mathcal{P}^\mathbf{x}$ denotes the ambiguity set of distributions, such that $\mathbb{P}_\mathbf{x} \in \mathcal{P}^\mathbf{x} \subset \mathcal{P}(\mathcal{B}(\mathbb{R}^n))$.

We focus here on Chebyshev ambiguity sets, which are fully parameterized by uncertainty in the first two moments of the probability distributions in the set. These sets take the form
\begin{equation}
\mathcal{P}_\mathbf{x} := \{\mathbb{P} \in \mathcal{P}(\mathcal{B}(\mathbb{R}^n)): (\mu_\mathbb{P}, \Sigma_{\mathbb{P}}) \in \mathscr{S} \},
\end{equation}
where $\mathscr{S} \subset \mathbb{R}^n \times \mathbb{S}^n_+$ is a convex set, $\mu_\mathbb{P} = \mathbb{E}_\mathbb{P}[\mathbf{x}]$, and $\Sigma_\mathbb{P} = \mathbb{E}_\mathbb{P}[(\mathbf{x} - \mu_\mathbb{P})(\mathbf{x} - \mu_\mathbb{P})^T]$. We will define the projection operator $\text{Proj}_{(\mathcal{A}, \mathcal{B})}(\mathscr{S}) := \{\mu: (\mu, \Sigma) \in \mathscr{S} \cap \mathcal{A} \times \mathcal{B}\}$ on Chebyshev ambiguity sets for $\mathcal{A} \subseteq \mathbb{R}^n, \mathcal{B} \subseteq \mathbb{S}^n_+$. For ease of notation, we will define $\text{Proj}_{(\cdot, \mathcal{B})}(\mathscr{S}) = \text{Proj}_{(\mathbb{R}^n, \mathcal{B})}$ for $\mathcal{B} \subseteq \mathbb{S}^n_+$.

We define a novel ball-shaped ambiguity set of distributions to tightly under-approximate the set of distributions that reach a Euclidean ball with probability of at least $1-\epsilon$. Recall that the probability of a random sample from an $n$-dimensional Gaussian distribution having a squared Mahalanobis distance less than some threshold $d$ is distributed according to a $\chi^2$ distribution with $n$ degrees of freedom \cite{gallego2013mahalanobis}. Then, we can define
\begin{align}
&\mathcal{P}^{\text{BALL}, \chi^2}(\hat{\mu}, \hat{r}) \\ \notag 
&:= \{\mathbb{P}: ||\mu_{\mathbb{P}} - \hat{\mu}||_2 + \sqrt{f^{-1}(1-\epsilon, n)}\sqrt{\lambda_{\max}(\Sigma_{\mathbb{P}})} \leq \hat{r} \},
\end{align}
where $f^{-1}(\cdot, n)$ is the inverse cumulative distribution function of the $\chi^2$ distribution with $n$ degrees of freedom (assuming $\hat{\mu} \in \mathbb{R}^n$). By this definition, $\mathcal{P}^{\text{BALL}, \chi^2}(\hat{\mu}, \hat{r})$ represents a set of Gaussian distributions whose $(1-\epsilon)$ covariance contours fall strictly inside the Euclidean ball $\mathbb{B}(\hat{\mu}, \hat{r})$. This is sufficient to establish that for a random sample $\mathbf{x}$ drawn from any distribution belonging to $\mathcal{P}^{\text{BALL}, \chi^2}(\hat{\mu}, \hat{r})$, $\mathbf{x} \in \mathbb{B}(\hat{\mu}, \hat{r})$ with probability of at least $1-\epsilon$.

However, in the presence of polytopic state and control constraints, sets of the form $\mathcal{P}^{\text{BALL}, \chi^2}(\hat{\mu}, \hat{r})$ do not accurately represent the set of Gaussian distributions that can reach the goal region while satisfying all polytopic constraints with probability of at least $1-\epsilon$. Therefore, throughout this work, we will instead represent uncertainty with ambiguity sets 
\begin{align}
&\mathcal{P}^{\text{BALL}}(\hat{\mu}, \hat{r}) \\ \notag 
&:= \{\mathbb{P}: ||\mu_{\mathbb{P}} - \hat{\mu}||_2 + \Phi^{-1}(1-\epsilon)\sqrt{\lambda_{\max}(\Sigma_{\mathbb{P}})} \leq \hat{r} \},
\end{align}
where $\Phi^{-1}(\cdot)$ is the inverse cumulative distribution function of the normal distribution. In the presence of polytopic state or control constraints, the largest set with ball-shaped projections belonging to the backward reachable set of distributions from $\mathcal{P}^{\text{BALL}, \chi^2}(\hat{\mu}_\mathcal{G}, \hat{r}_\mathcal{G})$ approximately (and in the case of no process noise, exactly) takes the form $\mathcal{P}^{\text{BALL}}(\hat{\mu}_\mathcal{I}, \hat{r}_\mathcal{I})$ (see Lemma \ref{lemma: maximal_brs_subset}).

Further, maximal radius sets of the form $\mathcal{P}^{\text{BALL}}$ and $\mathcal{P}^{\text{BALL}, \chi^2}$ have strictly larger projections into the mean space than maximal radius 2-Wasserstein or Gelbrich uncertainty sets (which are equivalent when only considering Gaussian distributions) in most realistic scenarios (see Remarks \ref{remark: w2_projections} and \ref{remark: w2_projections_phi}). 2-Wasserstein ambiguity sets are constructed with some centroid distribution $\mathcal{N}(\hat{\mu}, \hat{\Sigma})$ and radius $\hat{r}$, such that
\begin{align}
&\mathcal{P}^{\mathbb{W}_2}(\hat{\mu}, \hat{\Sigma}, \hat{r}) \notag \\ 
&= \left\{\sqrt{||\mu_\mathbb{P} - \hat{\mu}||_2^2 + \text{tr}(\Sigma_\mathbb{P} + \hat{\Sigma}  - 2(\hat{\Sigma}^{1/2}\Sigma_\mathbb{P}\hat{\Sigma}^{1/2})^{1/2})} \leq \hat{r} \right\}.
\end{align}

For an appropriate choice of $\hat{r}$, $\mathcal{P}^{\text{BALL}}$ also subsumes ambiguity sets of the type used in the MAXCOVAR \cite{aggarwal2024sdp} algorithm. Although MAXCOVAR is described as planning over distributions, not ambiguity sets of distributions, each node in a backward reachable tree constructed using MAXCOVAR represents an ambiguity set of distributions with a fixed mean and a bound on the maximum eigenvalue of the covariance. Such ambiguity sets can be represented as
\begin{equation}
\mathcal{P}^{\text{MC}}(\hat{\mu}, \hat{\Sigma}) := \left\{\mathbb{P}:  \begin{array}{l}
\mu_\mathbb{P} = \hat{\mu} \\
\lambda_{\max}(\Sigma_{\mathbb{P}}) \leq \lambda_{\min}(\hat{\Sigma})
\end{array}\right\}.
\end{equation}
%When $\hat{r} \geq \Phi^{-1}(1-\epsilon)\sqrt{\lambda_{\min}(\hat{\Sigma})}$, $\mathcal{P}^{\text{BALL}}(\hat{\mu}, \hat{r}) \subset \mathcal{P}^{\text{MC}}(\hat{\mu}, \hat{\Sigma})$.

Suppose that the goal region is not a Euclidean ball, but is instead parameterized by the Minkowski sum of an ellipsoid $\mathcal{E}(\hat{\mu}, \hat{\mathscr{P}})$ and a Euclidean ball $\mathbb{B}_n(\hat{\mu}, \hat{r})$, with an additional constraint on the covariance at the goal region, such that $\Phi^{-1}(1-\epsilon)\sqrt{\lambda_{\max}(\Sigma)} \leq \hat{r}$. Then, we can construct an ellipsoidal generalization of our ball-shaped ambiguity set formulation as follows:
\begin{align}
\mathcal{P}^{\text{ELL-BALL}}(\hat{\mu}, \hat{\mathscr{P}}, \hat{r}) := \left\{\mathbb{P}: \begin{array}{l} (s -\hat{\mu})^T\hat{\mathscr{P}}^{-1}(s -\hat{\mu}) \leq 1 \\
(\mu_{\mathbb{P}}, \Sigma_{\mathbb{P}}) \in \mathcal{P}^{\text{BALL}}(s, \hat{r})\end{array} \right\}.
%||\mu_{\mathbb{P}} - s||_2 + \Phi^{-1}(1-\epsilon)\sqrt{\lambda_{\max}(\Sigma_{\mathbb{P}})} \leq \hat{r}  
\end{align}
%Intuitively, these sets allow for $(\mu_{\mathbb{P}} -\hat{\mu})^T\hat{\mathscr{P}}^{-1}(\mu_{\mathbb{P}} -\hat{\mu}) > 1 $ if $\lambda_{\max}(\Sigma_{\mathbb{P}})$ is sufficiently small. In other words, this ambiguity set does not fully decouple first- and second-order moment uncertainty, instead allowing for additional flexibility in certain cases.
%
Consider the ellipsoidal ambiguity sets used in the MAXELLIPSOID algorithm \cite{aggarwal2025tac}, which take the form
\begin{equation}
\mathcal{P}^{\text{ELL}}(\hat{\mu}, \hat{\Sigma}, \hat{\mathscr{P}}) := \left\{\mathbb{P}: \begin{array}{l} (\mu_{\mathbb{P}} -\hat{\mu})^T\hat{\mathscr{P}}^{-1}(\mu_{\mathbb{P}} -\hat{\mu}) \leq 1 \\ \lambda_{\max}(\Sigma_{\mathbb{P}}) \leq \lambda_{\min}(\hat{\Sigma}) \end{array} \right\}.
\end{equation}
For an appropriate choice of $\hat{r}$, our ambiguity set formulation $\mathcal{P}^{\text{ELL-BALL}}$ subsumes $\mathcal{P}^{\text{ELL}}$, as our formulation allows for $\mu_\mathbb{P} \notin \mathcal{E}(\hat{\mu}, \hat{\mathscr{P}})$
if $\lambda_{\max}(\Sigma_{\mathbb{P}})$ is sufficiently small. In other words, $\mathcal{P}^{\text{ELL-BALL}}$ includes distributions with high first-order moment deviation but low second-order moment uncertainty that $\mathcal{P}^{\text{ELL}}$ excludes, allowing for additional flexibility in certain cases.

\subsection{Problem}
Consider a discrete-time linear system with dynamics
\begin{equation}\label{eq: linear_dynamics}
\mathbf{x}_{t+1} = A\mathbf{x}_t + B\mathbf{u}_t + D\mathbf{w}_t
\end{equation}
where $\mathbf{x}_t \in \mathcal{X} \subseteq \mathbb{R}^n$ for all $t$, $\mathbf{u}_t \in \mathcal{U} \subseteq \mathbb{R}^m$ for all $t$, and $\mathbf{w}_t \sim \mathcal{N}(0, I_{n})$, with $\{\mathbf{w}_t\}$ i.i.d. and $A$ non-singular.

We define a generic N-step finite-horizon steering problem in Problem \ref{prob: optimal_ambiguity_set_steering} where an initial Gaussian distribution $\mathcal{I}$ is steered to a goal region $\mathcal{G}$, subject to the dynamics in \eqref{eq: linear_dynamics}. We assume that the goal region $\mathcal{G}$ is given by the Minkowski sum of an $n$-dimensional ellipsoid and an $n$-dimensional Euclidean ball. More formally, $\mathcal{G} = \mathcal{E}(\mu_\mathcal{G}, \mathscr{P}_\mathcal{G}) \oplus \mathbb{B}_n(0, r_\mathcal{G})$, such that any point $x \in \mathcal{G}$ satisfies $||y - x||_2 \leq r_\mathcal{G}$ for at least one $y \in \mathcal{E}(\mu_\mathcal{G}, \mathscr{P}_\mathcal{G})$. We also assume a constraint on the maximum eigenvalue of the system covariance at the goal.

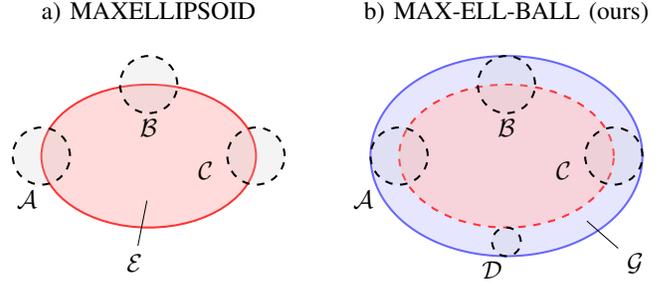
\begin{figure}[tp]
\centering
\resizebox {\columnwidth} {!} {
\begin{tikzpicture}[scale=1]
    % Define ellipse parameters
    \def\a{1.5}    % semi-major axis
    \def\b{1}    % semi-minor axis
    \def\r{0.4}  % radius of circle for Minkowski sum
    
    % Draw the Minkowski sum using proper offset curve calculation
    % For an ellipse, the offset curve at distance r is computed by
    % moving distance r along the outward normal at each point
    \node at (-5, 2) {a) MAXELLIPSOID};
    \draw[thick, red!80, fill=red!20, fill opacity=0.7] 
        (-5,0) ellipse (1.5 and 1);
    \node at (-5, -0.5) (E){};
    \node at (-5.2, -1.5) (Elabel){$\mathcal{E}$};
    \draw (E) -- (Elabel);
    \draw[thick, dashed, black, fill=black!20, fill opacity=0.2] 
        ({-6.5},{0}) circle (\r);
    \node at (-6.7, -0.6) {$\mathcal{A}$};
    \draw[thick, dashed, black, fill=black!20, fill opacity=0.2] 
        ({-3.5},{0}) circle (\r);
    \node at (-4.2, -0.2) {$\mathcal{C}$};
    \draw[thick, dashed, black, fill=black!20, fill opacity=0.2] 
        ({-5},{1}) circle (\r);
    \node at (-5, 0.4) {$\mathcal{B}$};

    \node at (0, 2) {b) MAX-ELL-BALL (ours)};
    \draw[thick, blue!60, fill=blue!10] 
        plot[domain=0:360, samples=360, smooth cycle] 
        ({
            \a*cos(\x) + \r*\b*cos(\x)/sqrt(\a*\a*sin(\x)*sin(\x) + \b*\b*cos(\x)*cos(\x))
        }, 
        {
            \b*sin(\x) + \r*\a*sin(\x)/sqrt(\a*\a*sin(\x)*sin(\x) + \b*\b*cos(\x)*cos(\x))
        });

    % Draw the original ellipse
    \draw[thick, dashed, red!80, fill=red!20, fill opacity=0.7] 
        (0,0) ellipse (1.5 and 1);
        
    \draw[thick, dashed, black, fill=black!20, fill opacity=0.2] 
        ({1.5},{0}) circle (\r);
    \node at (0.8, -0.2) {$\mathcal{C}$};
    %\draw[thick, dashed, black, fill=black!20, fill opacity=0.2] 
        %({0},{0}) circle (\r);
    \draw[thick, dashed, black, fill=black!20, fill opacity=0.2] 
        ({0},{1}) circle (\r);
    \node at (0, 0.4) {$\mathcal{B}$};
    \draw[thick, dashed, black, fill=black!20, fill opacity=0.2] 
        ({-1.5},{0}) circle (\r);
    \node at (-2, -0.6) {$\mathcal{A}$};
    \draw[thick, dashed, black, fill=black!20, fill opacity=0.2] 
        ({0},{-1-\r/2}) circle (\r/2);
    \node at (-0.2, -1.6) {$\mathcal{D}$};
    \node at (1, -0.8) (G){};
    \node at (1.8, -1.5) (Glabel){$\mathcal{G}$};
    \draw (G) -- (Glabel);
        
\end{tikzpicture}
}
\vspace*{-0.25in}
\caption{\textbf{a}) MAXELLIPSOID \cite{aggarwal2025tac} decouples first- and second-order moment uncertainty, requiring the mean of each distribution to fall inside the ellipsoid $\mathcal{E}$. $\mathcal{A}$, $\mathcal{B}$, and $\mathcal{C}$ represent $(1-\epsilon)$ percentile covariance contours for distributions that satisfy the goal constraint. \textbf{b}) When the goal region $\mathcal{G}$ is parameterized by the Minkowski sum of $\mathcal{E}$ and a Euclidean ball, some distributions where the mean falls outside $\mathcal{E}$ can still satisfy the goal constraint. $\mathcal{D}$ represents the $(1-\epsilon)$ percentile covariance contour for one such valid distribution.}
\vspace{-0.2in}
\end{figure}
Note that if $\mathscr{P}_\mathcal{G} = 0$, then $\mathcal{G}$ represents a ball in Euclidean space that must be reached with probability of at least $1 - \epsilon$ (see Section \ref{sec: max_cov_prob_reach} for further details). Note also that $\mathcal{G}$ subsumes an ellipsoidal goal set of the type discussed in \cite{aggarwal2025tac}, as $\mathcal{G}$ allows the final mean of the distribution to fall outside of the ellipsoid if the final covariance is small enough. Intuitively speaking, the tolerance for the mean deviation $||\mathbf{x}_N - \mu_\mathcal{G}||_2$ depends on the final covariance $\Sigma_N$, with a higher tolerance when the final state uncertainty is lower.

\begin{problem}\label{prob: optimal_ambiguity_set_steering}
\textit{Steer from initial distribution $\mathcal{I}$ to reach a goal region $\mathcal{G}$ with probability of at least $(1-\epsilon)$,  subject to controller parameterization $f_k(\cdot)$ such that $\mathbf{u}_k = f_k(\mathbf{x}_k) \ \forall k$ and cost function $c_k(\mathbf{u}_k, \mathbf{x}_{k+1})$.
The goal region $\mathcal{G}$ is assumed to be the Minkowski sum of an $n$-dimensional ellipsoid and $n$-ball, such that $\mathcal{G} := \mathcal{E}(\mu_\mathcal{G}, \mathscr{P}_\mathcal{G}) \oplus \mathbb{B}_n(0, r_\mathcal{G})$. Note that in the special case $\mathscr{P}_\mathcal{G} = 0$, $\mathcal{G}$ represents an $n$-ball in Euclidean space $\mathbb{B}_n(\mu_\mathcal{G}, r_\mathcal{G})$.}
\begin{equation}
\min_{f_k} J = \mathbb{E}[\sum_{k=0}^{N-1} c_k(\mathbf{u}_k, \mathbf{x}_{k+1})]
\end{equation}
such that:
\begin{align}
&\mathbf{x}_{k+1} = A\mathbf{x}_k + B\mathbf{u}_k + D\mathbf{w}_k, \notag \\
&\mathbf{x}_0 \sim \mathcal{N}(\mu_\mathcal{I}, \Sigma_\mathcal{I}), \quad \mathbf{x}_N \sim \mathcal{N}(\mu_N, \Sigma_N), \notag \\
&\mathbb{P}(\mathbf{x}_N \in \mathcal{G}) \geq 1- \epsilon_\mathcal{G}, \quad \lambda_{\max}(\Sigma_N) \leq \lambda_{\mathcal{G}}, \notag \\
%&\exists y \in \mathcal{E}(\mu_\mathcal{G}, \mathscr{P}_\mathcal{G}) \text{ s.t. } \mathbb{P}(||\mathbf{x}_N - y||_2 \leq r_\mathcal{G}) \geq 1 - \epsilon_\mathcal{G}, \notag \\
&\mathbb{P}(\mathbf{x}_k \in \mathcal{X}) \geq 1 - \epsilon_x \ \forall k, \quad \mathbb{P}(\mathbf{u}_k \in \mathcal{U}) \geq 1 - \epsilon_u \ \forall k.
\end{align}
\end{problem}
Under state and control constraints, Problem \ref{prob: optimal_ambiguity_set_steering} may not have a feasible solution. However, even if Problem \ref{prob: optimal_ambiguity_set_steering} is infeasible, it may still be possible to steer from $\mathcal{I}$ to $\mathcal{G}$ by composing multiple N-step steering maneuvers. As such, in this paper we solve the following problem:
\begin{problem}\label{prob: find_all_paths}
Find paths from all initial distributions to (the ambiguity set representing) the goal region $\mathcal{G}$ for which paths exist (subject to state and control constraints).
\end{problem}
We solve this problem by constructing a belief roadmap backwards from the ambiguity set representing the goal region $\mathcal{G}$. Each node in the roadmap represents an ambiguity set of distributions, and edges in the roadmap represent distributionally robust control policies that steer between ambiguity sets of distributions. Our roadmap construction algorithm maximizes the size of each ambiguity set that is added to the roadmap, resulting in maximal roadmap coverage.

This rest of the paper is organized as follows: in Section \ref{sec: max_cov_prob_reach} we develop a novel multi-query motion planning algorithm for distributionally robust planning when the goal region is a Euclidean ball, and in Section \ref{sec: coverage_proofs} we provide maximal coverage proofs for this algorithm. In Section \ref{sec: max_ell_prob_reach}, we discuss an extension of our multi-query motion planning algorithm to distributionally robust planning when the goal region is parameterized by the Minkowski sum of an ellipsoid and a Euclidean ball. Section \ref{sec: experiments} presents the results of our simulation experiments.
%\section{Approach}

\section{MAX-COV-BALL: Constructing a BRT of Ball-shaped Ambiguity Sets of Distributions}\label{sec: max_cov_prob_reach}
To solve Problem \ref{prob: find_all_paths}, we build a roadmap of ambiguity sets of distributions backwards from the goal region $\mathcal{G}$. In this section, we discuss an algorithm for constructing a roadmap of ball-shaped ambiguity sets of distributions. See Section \ref{sec: max_ell_prob_reach} for the extension of this algorithm to construct a roadmap of ellipsoidal sets of ball-shaped ambiguity sets of distributions. We begin by constructing a nonlinear program to maximize the size of the ball-shaped ambiguity set of distributions.
\begin{problem}\label{prob: ball_ambiguity_nlp}
Maximize the size of the ambiguity set of distributions $\mathcal{P}^{\text{BALL}}(\mu_\mathcal{I}, r_\mathcal{I})$ that can reach a goal ambiguity set of distributions $\mathcal{P}^{\text{BALL}}({\mu}_\mathcal{G}, {r}_\mathcal{G})$.
\begin{align}
\max_{f_k} r_\mathcal{I}
\end{align}
such that:
\begin{align}
\mathbf{x}_{k+1} &= A\mathbf{x}_k + B\mathbf{u}_k + D\mathbf{w}_k, \notag \\
\mathbf{x}_0 &\sim \mathbb{P}_{\mathbf{x}_0} \text{ s.t. } \mathbb{P}_{\mathbf{x}_0} \in \mathcal{P}^{\text{BALL}}(\mu_\mathcal{I}, r_\mathcal{I}), \notag \\
\mathbf{u}_k &= f(\mathbf{x}_k), \quad \mathbb{P}_{\mathbf{x}_N} \in \mathcal{P}^{\text{BALL}}(\mu_\mathcal{G}, r_\mathcal{G}), \notag \\
\inf_{\mathcal{P}^{\mathbf{x}_k}} &\mathbb{P}(\mathbf{x}_k \in \mathcal{X}) \geq 1-\epsilon, \quad
\inf_{\mathcal{P}^{\mathbf{u}_k}} \mathbb{P}(\mathbf{u}_k \in \mathcal{U}) \geq 1-\epsilon. 
\end{align}
\end{problem}
We adopt a control law of the form:
\begin{equation}\label{eq: control_law}
    \mathbf{u}_k = K_k(\mathbf{x}_k - \overline{\mathbf{x}}_k) + \overline{\mathbf{u}}_k,
\end{equation}
where $\overline{\mathbf{x}}_k$ refers to the reference state. Under a control law of the type presented in \eqref{eq: control_law} and dynamics of the form presented in \eqref{eq: linear_dynamics}, any Gaussian distribution remains Gaussian for all time.

Consider a discrete-time linear system with dynamics given by \eqref{eq: linear_dynamics} and a control policy ($\overline{\mathbf{x}}_{0:N}, \overline{\mathbf{u}}_{0:N}, K_{0:N})$, such that $\overline{\mathbf{x}}_{t+1} = A\overline{\mathbf{x}}_t + B\overline{\mathbf{u}}_t$ $\forall t$ and with the control law given by \eqref{eq: control_law}. Note that we do \textit{not} require $\overline{\mathbf{x}}_t = \mathbb{E}[\mathbf{x}_t]$; the same reference state trajectory $\overline{\mathbf{x}}_{0:N}$ can be used for stochastic trajectories originating from different initial probability distributions. If a Gaussian distribution (or ambiguity set of distributions) $\mathcal{A}$ is driven to a Gaussian distribution (or ambiguity set of distributions) $\mathcal{B}$ under a control policy $\mathscr{C} := (\overline{\mathbf{x}}_{0:N}, \overline{\mathbf{u}}_{0:N}, K_{0:N})$, we denote this as $\mathcal{A} \xrightarrow[]{\mathscr{C}} \mathcal{B}$. Any control policy $\mathscr{C}$ applied to an ambiguity set of distributions with centroid $\mu_\mathcal{I}$ can be equivalently written such that $\overline{\mathbf{x}}_0 = \mu_\mathcal{I}$. We assume this form and denote the final ambiguity set centroid as $\mu_N =\overline{\mathbf{x}}_N$.

Optimizing directly over a ball-shaped ambiguity set $\mathcal{P}^{\text{BALL}}(\mu_\mathcal{I}, r_\mathcal{I})$ is challenging. Instead, we optimize over the projection of a ball-shaped ambiguity set into mean space $\text{Proj}_{(\cdot, \mathbb{S}^n_+)}(\mathcal{P}^{\text{BALL}}(\mu_\mathcal{I}, r_\mathcal{I})) \subset \mathbb{R}^n$. This allows us to tractably optimize while maintaining recursive feasibility, because $\text{Proj}_{(\cdot, \mathbb{S}^n_+)}(\mathcal{P}^{\text{BALL}}(\mu_\mathcal{I}, r_\mathcal{I}))$ reaching a set $\mathcal{P}^{\text{BALL}}(\mu_N, r_N) \subset \mathcal{G}$ is sufficient for the entire initial ambiguity set $\mathcal{P}^{\text{BALL}}(\mu_\mathcal{I}, r_\mathcal{I})$ to reach the goal set.

More generally, we will show that for any $\lambda \geq 0$, if the projection of the initial ball-shaped ambiguity set $\mathcal{P}^{\text{BALL}}(\mu_\mathcal{I}, r_\mathcal{I})$ with fixed $\Sigma = \lambda I$ reaches $\mathcal{P}^{\text{BALL}}(\mu_N, r_N) \subseteq \mathcal{P}^{\text{BALL}}(\mu_\mathcal{G}, r_\mathcal{G})$, so does any distribution with $\Sigma \succeq \lambda I$  belonging to the initial set $\mathcal{P}^{\text{BALL}}(\mu_\mathcal{I}, r_\mathcal{I})$. This property is formalized in Lemma \ref{lemma: ball_invariance}. It follows from this property that if the projection of the initial ambiguity set with $\Sigma = 0$ reaches the $\mathcal{P}^{\text{BALL}}(\mu_N, r_N) \subseteq \mathcal{P}^{\text{BALL}}(\mu_\mathcal{G}, r_\mathcal{G})$, then the entire initial ambiguity set must reach the goal ambiguity set.
\begin{lemma}\label{lemma: ball_invariance}
Consider an initial ambiguity set $\mathcal{P}^{\text{BALL}}(\mu_\mathcal{I}, r_\mathcal{I})$ and a goal ambiguity set $\mathcal{P}^{\text{BALL}}(\mu_\mathcal{G}, r_\mathcal{G})$. Suppose that all state and control constraints are polytopic.
Then, $\exists \mathscr{C}, \lambda \geq 0$ such that $\text{Proj}_{(\cdot, \{\lambda I\})}(\mathcal{P}^{\text{BALL}}(\mu_\mathcal{I}, r_\mathcal{I})) \xrightarrow[]{\mathscr{C}} \mathcal{P}^{\text{BALL}}(\mu_N, r_N) \subseteq \mathcal{P}^{\text{BALL}}(\mu_\mathcal{G}, r_\mathcal{G})$, with all state and control chance constraints satisfied at each timestep with probability of at least $(1-\epsilon)$ if and only if $\text{Proj}_{(\cdot, \{\Sigma: \Sigma \succeq \lambda I \})}(\mathcal{P}^{\text{BALL}}(\mu_\mathcal{I}, r_\mathcal{I})) \xrightarrow[]{\mathscr{C}} \mathcal{P}^{\text{BALL}}(\mu_N, r_N) \subseteq \mathcal{P}^{\text{BALL}}(\mu_\mathcal{G}, r_\mathcal{G})$, with all state and control chance constraints satisfied at each timestep with probability of at least $(1-\epsilon)$. 
\end{lemma}
\begin{proof}
See Appendix \ref{sec:appendix}.
\end{proof}
Furthermore, in the absence of process noise, any projection of the initial ball-shaped ambiguity set $\mathcal{P}^{\text{BALL}}(\mu_\mathcal{I}, r_\mathcal{I})$ with fixed $\Sigma = \lambda I$ (for $\lambda \geq 0$) reaching $\mathcal{P}^{\text{BALL}}(\mu_N, r_N) \subseteq \mathcal{G}$, with $\mathcal{G}$ of the form $\mathcal{P}^\text{BALL}$ or $\mathcal{P}^{\text{BALL}, \chi^2}$ is both necessary and sufficient to show that the entire initial ambiguity set $\mathcal{P}^{\text{BALL}}(\mu_\mathcal{I}, r_\mathcal{I})$ reaches the goal set. This property is formally stated below in Lemma \ref{lemma: ball_invariance_no_process_noise}. %Also, in the absence of process noise and in the presence of polytopic constraints, the backward reachable set of distributions from a goal set of the form $\mathcal{P}^{\text{BALL}, \chi^2}$ under a fixed control policy $\mathscr{C}$ is exactly represented by an ambiguity set of the form $\mathcal{P}^{\text{BALL}}(\mu_\mathcal{I}, r_\mathcal{I})$. In Section \ref{sec: coverage_proofs} of this paper, we will leverage this property to formally prove that our method achieves maximal coverage in the absence of process noise  and state constraints (see Theorem \ref{thm: maximal_coverage} for details). This property is formally stated below in Lemma \ref{lemma: ball_invariance_no_process_noise}.
\begin{lemma}\label{lemma: ball_invariance_no_process_noise}
Suppose that the process noise is zero ($D = 0$). Consider an initial ambiguity set $\mathcal{P}^{\text{BALL}}(\mu_\mathcal{I}, r_\mathcal{I})$ and a goal ambiguity set $\mathcal{G} = \mathcal{P}^{\text{BALL}}(\mu_\mathcal{G}, r_\mathcal{G})$ or $\mathcal{G} = \mathcal{P}^{\text{BALL}, \chi^2}(\mu_\mathcal{G}, r_\mathcal{G})$. Suppose that all state and control constraints are polytopic. 
Then, $\exists \mathscr{C}, \lambda \geq 0$ such that $\text{Proj}_{(\cdot, \{\lambda I\})}(\mathcal{P}^{\text{BALL}}(\mu_\mathcal{I}, r_\mathcal{I})) \xrightarrow[]{\mathscr{C}} \mathcal{P}^{\text{BALL}}(\mu_N, r_N) \subseteq \mathcal{G}$,  with all state and control chance constraints satisfied at each timestep with probability of at least $(1-\epsilon)$ if and only if $\mathcal{P}^{\text{BALL}}(\mu_\mathcal{I}, r_\mathcal{I}) \xrightarrow[]{\mathscr{C}} \mathcal{G}$ with all state and control chance constraints satisfied at each timestep with probability of at least $(1-\epsilon)$. %and if and only if $\exists \mu_{\mathcal{G}, \chi^2}$ such that $\mathcal{P}^{\text{BALL}}(\mu_\mathcal{I}, r_\mathcal{I}) \xrightarrow[]{\mathscr{C}} \mathcal{P}^{\text{BALL}, \chi^2}\left(\mu_{\mathcal{G}, \chi^2}, c_\mathcal{G}r_\mathcal{G}\right)$ with all state and control chance constraints satisfied at each timestep with probability of at least $(1-\epsilon)$, for $c_\mathcal{G} = \frac{\sqrt{f^{-1}(1-\epsilon, n)}}{\Phi^{-1}(1-\epsilon)}$.
\end{lemma}
\begin{proof}
See Appendix \ref{sec:appendix}.
\end{proof}
From Lemma \ref{lemma: ball_invariance}, establishing that the projection of an initial ambiguity set into mean space $\text{Proj}_{(\cdot, \mathbb{S}^n_+)}(\mathcal{P}_\mathcal{I}^{\text{BALL}})$ reaches $\mathcal{P}_N^{\text{BALL}}\subseteq \mathcal{P}^{\text{BALL}}_\mathcal{G}$ is always necessary and sufficient to show that the entire initial ambiguity set $\mathcal{P}_\mathcal{I}^{\text{BALL}}$ reaches $\mathcal{P}_N^{\text{BALL}}\subseteq \mathcal{P}_\mathcal{G}^{\text{BALL}}$, assuming polytopic state and control constraints. %From Lemma \ref{lemma: ball_invariance_no_process_noise}, in the absence of process noise, $\mathcal{P}_\mathcal{I}^{\text{BALL}}$ reaching $\mathcal{P}_N^{\text{BALL}}\subseteq \mathcal{P}_\mathcal{G}^{\text{BALL}}$ is necessary to show that $\mathcal{P}_\mathcal{I}^{\text{BALL}}$ reaches $\mathcal{P}_\mathcal{G}^{\text{BALL}}$.
So, maximizing the radius $r_\mathcal{I}$ of a Euclidean ball $\mathbb{B}_n(\mu_\mathcal{I}, r_\mathcal{I})$ such that any point in the Euclidean ball reaches $\mathcal{P}_N^{\text{BALL}}\subseteq \mathcal{P}_\mathcal{G}^{\text{BALL}}$ maximizes the radius of the initial ambiguity set $\mathcal{P}^{\text{BALL}}(\mu_\mathcal{I}, r_\mathcal{I})$ that reaches $\mathcal{P}_N^{\text{BALL}}\subseteq \mathcal{P}_\mathcal{G}^{\text{BALL}}$. 

Accordingly, we reformulate Problem \ref{prob: ball_ambiguity_nlp} to maximize the radius $r$ of a Euclidean ball $\mathbb{B}_n(\mu_0, r)$ such that $\mathbb{B}_n(\mu_0, r) \subseteq \mathcal{E}(\mu_0, \mathscr{R}_0) \xrightarrow{\mathscr{C}} \mathcal{P}^{\text{BALL}}(\mu_N, r_N) \subseteq \mathcal{P}^{\text{BALL}}(\mu_\mathcal{G}, r_\mathcal{G})$. At each timestep, we propagate the decoupled first-order moment uncertainty by updating $\mu_k$ and $\mathscr{R}_k$ according to the state dynamics and control policy. Similarly, we propagate the second-order moment uncertainty by updating $\Sigma_k$. The ambiguity set at timestep $k$ is represented by the set of Gaussian distributions such that $\mu \in \mathcal{E}(\mu_k, \mathscr{R}_k)$ and $\Sigma = \Sigma_k$. A full description of the reformulation is given below in Problem \ref{prob: ball_nlp}.
\begin{problem}\label{prob: ball_nlp} (MAX-COV-BALL) Find $\mathscr{C}$ for fixed $\mu_\mathcal{I}$ that maximizes the radius of the initial ambiguity set $\mathcal{P}_\mathcal{I}^{\text{BALL}}(\mu_\mathcal{I}, r)$ such that $\mathcal{P}_\mathcal{I}^{\text{BALL}}(\mu_\mathcal{I}, r) \xrightarrow{\mathscr{C}} \mathcal{P}^\text{BALL}_\mathcal{G}$ by maximizing the radius of the ball $\mathbb{B}_n(\mu_\mathcal{I}, r)$ such that $\mathbb{B}_n(\mu_\mathcal{I}, r) \xrightarrow{\mathscr{C}} \mathcal{P}^\text{BALL}_\mathcal{G}$.
\begin{align}
\max_{\overline{\mathbf{u}}_k, K_k, \Sigma_k, \mathscr{R}_k} r = \lambda_{\min}(\mathscr{R}_0)
\end{align}
such that:
\begin{align}
\mu_{k+1} &= A\mu_k + B\overline{\mathbf{u}}_k, \label{eq: mean_mcb} \\
\mathscr{R}_{k+1} &= (A + BK_k)\mathscr{R}_k(A + BK_k)^T, \label{eq: Rk_mcb} \\
\Sigma_{k+1} &= (A + BK_k)\Sigma_k(A + BK_k)^T + DD^T, \label{eq: sigmak_mcb} \\
\mu_0 &= \mu_\mathcal{I}, \ \Sigma_0 = 0, \label{eq: init_mb} \\
\mu &\in \mathcal{E}(\mu_N, \mathscr{R}_N) \implies (\mu, \Sigma_N) \in \mathcal{P}^{\text{BALL}}(\mu_\mathcal{G}, r_\mathcal{G}),  \\
%||\mu_N - \mu_\mathcal{G}||_2 &+ \Phi^{-1}(1-\epsilon)\sqrt{\lambda_{\max}(\Sigma_N)} + \sqrt{\lambda_{\max}(\mathscr{R}_N)} \leq r_\mathcal{G}, \notag \\
{a_i^u}^T\overline{\mathbf{u}}_k &+ \Phi^{-1}(1-\epsilon)\sqrt{{a_i^u}^T K_k \Sigma_k K_k a_i^u} \leq b_i^u, \label{eq: ctrl_mcb} \\
{a_i^x}^T\mu_k &+ \Phi^{-1}(1-\epsilon)\sqrt{{a_i^x}^T \Sigma_k a_i^x} + \sqrt{{a_i^x}^T \mathscr{R}_k a_i^x} \leq b_i^x. \label{eq: state_mcb}
\end{align}
\end{problem}
This problem is still nonlinear because of the shape matrix and covariance dynamics, chance constraints, and final ambiguity set containment constraint. In the presence of process noise, it is not generally tractable to find a lossless relaxation of the shape matrix and covariance dynamics unless the control feedback gain $K$ is fixed. We take the general approach of first finding a control policy that maximizes an upper bound on the radius $r$ of the initial ambiguity set $\mathcal{P}^{\text{BALL}}(\mu_\mathcal{I}, r)$, and then maximizing the radius $r$ for a fixed control policy. 

We construct a nonlinear optimization problem to find a control policy that maximizes $\lambda_{\min}(\Sigma_0)$ such that a Gaussian distribution $\mathcal{N}(\mu_\mathcal{I}, \Sigma_0)$ reaches $\mathcal{P}_N^{\text{BALL}}\subseteq \mathcal{P}^{\text{BALL}}_\mathcal{G}$. In the absence of process noise, $\mathcal{N}(\mu_\mathcal{I}, \Sigma_0)$ reaching $\mathcal{P}_N^{\text{BALL}}\subseteq \mathcal{P}^{\text{BALL}}_\mathcal{G}$ implies that an initial ambiguity set $\mathcal{P}^{\text{BALL}}(\mu_\mathcal{I}, r)$ with $r = \Phi^{-1}(1-\epsilon)\sqrt{\lambda_{\min}(\Sigma_0)}$ reaches the goal ambiguity set (by Lemma \ref{lemma: ball_invariance_no_process_noise}), so maximizing $\lambda_{\min}(\Sigma_0)$ is exactly equivalent to maximizing the radius of the initial ambiguity set $\mathcal{P}^{\text{BALL}}(\mu_\mathcal{I}, r)$. In the presence of process noise, $r \leq \Phi^{-1}(1-\epsilon)\sqrt{\lambda_{\min}(\Sigma_0)}$ for any initial ambiguity set $\mathcal{P}^{\text{BALL}}(\mu_\mathcal{I}, r)$ that reaches the goal ambiguity set, so maximizing $\lambda_{\min}(\Sigma_0)$ is equivalent to maximizing an upper bound on the radius of the initial ambiguity set $\mathcal{P}^{\text{BALL}}(\mu_\mathcal{I}, r)$.

\begin{problem}\label{prob: ball_nlp_loose}
(MAX-COV-BALL-UB) Find a control policy $\mathscr{C}$ for fixed $\mu_\mathcal{I}$ that maximizes an upper bound on the radius $r$ of the initial ambiguity set $\mathcal{P}^{\text{BALL}}(\mu_\mathcal{I}, r)$ by maximizing $\lambda_{\min}(\Sigma_0)$ such that the Gaussian distribution $\mathcal{N}(\mu_\mathcal{I}, \Sigma_0)$ can reach the goal ambiguity set. The upper bound is given by $r \leq \Phi^{-1}(1-\epsilon)\sqrt{\lambda_{\min}(\Sigma_0)}$ and is always tight when the process noise is zero.
\begin{align}
\max_{\overline{\mathbf{u}}_k, K_k, \Sigma_k} r = \lambda_{\min}(\Sigma_0)
\end{align}
such that:
\begin{align}
%\mu_{k+1} &= A\mu_k + B\overline{\mathbf{u}}_k, \notag \\
%\Sigma_{k+1} &= (A + BK_k)\Sigma_k(A + BK_k)^T + DD^T, \notag \\
&\mu_0 = \mu_\mathcal{I}, \quad {a_i^x}^T\mu_k + \Phi^{-1}(1-\epsilon)\sqrt{{a_i^x}^T \Sigma_k a_i^x} \leq b_i^x ,\notag \\
&||\mu_N - \mu_\mathcal{G}||_2  + \Phi^{-1}(1-\epsilon)\sqrt{\lambda_{\max}(\Sigma_N)} \leq r_\mathcal{G}, \notag \\
%{a_i^u}^T\overline{\mathbf{u}}_k  &+ \Phi^{-1}(1-\epsilon)\sqrt{{a_i^u}^T K_k \Sigma_k K_k a_i^u} \leq b_i^u, \notag \\
%{a_i^x}^T\mu_k &+ \Phi^{-1}(1-\epsilon)\sqrt{{a_i^x}^T \Sigma_k a_i^x}  \leq b_i^x, \\
&\text{subject to (\ref{eq: mean_mcb}), (\ref{eq: sigmak_mcb}), (\ref{eq: ctrl_mcb}).}
\end{align}
\end{problem}
Next, we will construct a semidefinite relaxation of Problem \ref{prob: ball_nlp_loose} that finds a control policy that maximizes an upper bound on $r$. Following \cite{rapakoulias2023discrete, aggarwal2024sdp}, we can losslessly transform the nonlinear dynamics constraint $\Sigma_{k+1} = (A + BK)\Sigma_k(A + BK)^T + DD^T$ into a convex equality constraint and a semidefinite constraint by substituting $U_k = K_k \Sigma_k$ and adding variable $Y_k \succeq U_k\Sigma_k^{-1}U_k^T = K_k\Sigma_kK_K^T$. Again following \cite{rapakoulias2023discrete, aggarwal2024sdp}, we can convexify the state and chance constraints by linearizing the square root around a reference value and then using the tangent line as a global overestimator. For any symmetric positive semidefinite matrix $M_k$, we can define
\begin{equation}\label{eq: M1}
\mathbb{M}_1(a, M_r, M_k) := \frac{1}{2 \sqrt{{a}^T M_r a}}{a}^T M_k a + \frac{\sqrt{{a M_r a^T}}}{2}
\end{equation}
such that $\mathbb{M}_1(a, M_r, M_k) \geq \sqrt{a^TM_ka}$ for all $a \in \mathbb{R}^n$, $M_r \in \mathbb{S}^n_+$. It follows that $\mathbb{M}_1(a_i^x, \Sigma_r, \Sigma_k) \geq \sqrt{{a_i^x}^T\Sigma_k{a_i^x}}$ for any $\Sigma_r$ and that $\mathbb{M}_1(a_i^u, Y_r, Y_k) \geq \sqrt{{a_i^u}^TY_k{a_i^u}}$ for any $Y_r$. We use a similar approach to convexify the terminal constraint, where by the tangent line overestimation property, $\forall \Sigma_r$:
\begin{equation}
\sqrt{\lambda_{\max}(\Sigma_N)} \leq \frac{1}{2 \sqrt{\lambda_{\max}(\Sigma_r)}} \lambda_{\max}(\Sigma_N) + \frac{\sqrt{\lambda_{\max}(\Sigma_r)}}{2}
\end{equation}
For the terminal constraint, we select $\sqrt{\lambda_{\max}(\Sigma_r)} = r/(\Phi^{-1}(1-\epsilon))$. Then, we can construct a semidefinite relaxation of Problem \ref{prob: ball_nlp_loose}, such that any solution to the semidefinite relaxation is also a valid solution to Problem \ref{prob: ball_nlp_loose}.
\begin{problem}\label{prob: ball_sdp_loose}
Semidefinite relaxation of MAX-COV-BALL-UB.% Find a control policy $\mathscr{C}$ which maximizes an upper bound on the radius $r$ of the initial ambiguity set $\mathcal{P}^{\text{BALL}}(\mathbf{x}_0, r)$ by maximizing $\lambda_{\min}(\Sigma_0)$ such that the Gaussian distribution $\mathcal{N}(\mu_0, \Sigma_0)$ can reach the goal ambiguity set. %The upper bound is given by $r \leq \Phi^{-1}(1-\epsilon)\sqrt{\lambda_{\min}(\Sigma_0)}$ and is always tight when the process noise is zero.
\begin{align}
\max_{\overline{\mathbf{u}}_k, K_k, \Sigma_k} r = \lambda_{\min}(\Sigma_0)
\end{align}
such that:
\begin{align}
&\mu_{k+1} = A\mu_k + B\overline{\mathbf{u}}_k, \notag \\
&\Sigma_{k+1} = A\Sigma_kA^T + BU_kA^T + AU_kTB^T + BY_kB^T + DD^T, \notag \\
&\begin{bmatrix}
\Sigma_k & U_k^T \\
U_k & Y_k
\end{bmatrix}  \succeq 0, \ \mu_0 = \mu_\mathcal{I}, \notag \\
&||\mu_N - \mu_\mathcal{G}||_2 + \left(\frac{(\Phi^{-1}(1-\epsilon))^2}{2r} \lambda_{\max}(\Sigma_N) + \frac{r}{2} \right) \leq r_\mathcal{G}, \notag \\
&{a_i^u}^T\overline{\mathbf{u}}_k + \Phi^{-1}(1-\epsilon)\mathbb{M}_1(a_i^u, Y_r, Y_k) \leq b_i^u, \notag \\
%&+ \frac{\Phi^{-1}(1-\epsilon)}{2 \sqrt{{a_i^u}^T Y_r a_i^u}}{a_i^u}^T Y_k a_i^u \notag \\ &+ \frac{\Phi^{-1}(1-\epsilon) \sqrt{{a_i^u}^T Y_r a_i^u}}{2}\leq b_i^u, \notag \\
&{a_i^x}^T\mu_k + \Phi^{-1}(1-\epsilon)\mathbb{M}_1(a_i^x, \Sigma_r, \Sigma_k) \leq b_i^x.
\end{align}
\end{problem}
Recall, however, that Problem \ref{prob: ball_nlp_loose} yields an upper bound on the ambiguity set radius $r$, which may not be exact in the presence of process noise. So, we construct an optimization problem to maximize the exact value of $r$ given a fixed control policy. This problem, given in Problem \ref{prob: ball_sdp}, is a semidefinite relaxation of Problem \ref{prob: ball_nlp} with a fixed control policy.

Because $K$ is no longer a variable but is now a constant, we do not need to substitute $U_k = K_k\Sigma_k$, but still need to convexify the terminal and chance constraints. Because $K$ and $\Sigma_0$ are fixed, $\Sigma_k$ is also a constant. So, the only nonlinearities arise from square roots of $\mathscr{R}_k$. The terminal constraint is convexified through rearrangement:
\begin{align}\label{eq: terminal_constraint}
&\lambda_{\max}(\mathscr{R}_N) \\ \notag
&\leq \left(r_\mathcal{G} - ||\mu_N - \mu_\mathcal{G}||_2 - \Phi^{-1}(1-\epsilon)\sqrt{\lambda_{\max}(\Sigma_N)}\right)^2.
\end{align}
The state chance constraints can be convexified by linearizing the square root around a reference value.
\begin{problem}\label{prob: ball_sdp}
 Semidefinite relaxation of MAX-COV-BALL with fixed control policy $\mathscr{C}$.%, maximize the radius of the initial ambiguity set $\mathcal{P}_0^{\text{BALL}}(\mu_0, r)$ such that $\mathcal{P}_0^{\text{BALL}}(\mu_0, r) \xrightarrow{\mathscr{C}} \mathcal{P}^\text{BALL}_\mathcal{G}$.% by maximizing the radius of the ball $\mathbb{B}_n(\overline{\mathbf{x}}_0, r)$ such that $\mathbb{B}_n(\mathbf{x}_0, r) \xrightarrow{\mathscr{C}} \mathcal{P}^\text{BALL}_\mathcal{G}$.
\begin{align}
\max_{\mathscr{R}_k} r = \lambda_{\min}(\mathscr{R}_0)
\end{align}
such that:
\begin{align}
%\mu_{k+1} &= A\mu_k + B\overline{\mathbf{u}}_k, \notag \\
%\mathscr{R}_{k+1} &= (A + BK)\mathscr{R}_k(A + BK)^T, \notag \\
%\Sigma_{k+1} &= (A + BK)\Sigma_k(A + BK)^T + DD^T, \notag \\
%\mu_0 &= \mathbf{x}_0, \ \Sigma_0 = 0, \notag \\
%\lambda_{\max}(\mathscr{R}_N) &\leq \notag \\
%&\left(r_\mathcal{G} - ||\mu_N - \mu_\mathcal{G}||_2 - \Phi^{-1}(1-\epsilon)\sqrt{\lambda_{\max}(\Sigma_N)}\right)^2, \notag \\
%{a_i^u}^T\overline{\mathbf{u}}_k &+ \Phi^{-1}(1-\epsilon)\sqrt{{a_i^u}^T K_k \Sigma_k K_k a_i^u} \leq b_i^u, \notag \\
{a_i^x}^T\mu_k &+ \Phi^{-1}(1-\epsilon)\sqrt{{a_i^x}^T \Sigma_k a_i^x} + \mathbb{M}_1(a_i^x, \mathscr{R}_r, \mathscr{R}_k) \leq b_i^x, \notag \\
&\text{subject to (\ref{eq: mean_mcb}), (\ref{eq: Rk_mcb}), (\ref{eq: sigmak_mcb}), (\ref{eq: init_mb}), (\ref{eq: ctrl_mcb})}, (\ref{eq: terminal_constraint}). \notag
\end{align}
\end{problem}
As long as a solution exists to Problem \ref{prob: ball_sdp_loose}, a solution exists to Problem \ref{prob: ball_sdp} for an appropriate choice of $\mathscr{R}_r$. However, in the presence of process noise, the upper bound on $r$ maximized in Problem \ref{prob: ball_sdp_loose} may be larger than the actual $r$ found by solving Problem \ref{prob: ball_sdp}. When the upper bound is loose, the certified ambiguity set $\mathcal{P}^{\text{BALL}}(\mu_0, r = \lambda_{\min}(\mathscr{R}_0))$ will be overly conservative, as it always excludes the distribution $\mathcal{N}(\mu_0, \lambda_{\min}(\Sigma_0)I)$, which is also certifiably part of the safe backward reachable set. This represents a tradeoff between including the backward reachable distribution with maximal covariance $\mathcal{N}(\mu_0, \lambda_{\min}(\Sigma_0)I)$ and including other backward reachable distributions with $\mu \neq \mu_0$.

We resolve this tradeoff by taking the best of both representations: that is, we always include maximal covariance backward reachable distributions in our ambiguity sets, yet we also represent other backward reachable distributions with $\mu \neq \mu_0$. We achieve this by constructing a mixed-integer program that leverages multiple ambiguity set representations and has worst-case performance equivalent to MAXCOVAR.  Algorithm \ref{alg:construct_brt_max_prob_reach} is our method for constructing backward reachable trees with MAX-COV-BALL.

Algorithm \ref{alg:construct_brt_max_prob_reach} tracks six elements at each node, which are parameters to the CREATENODE routine. The first three elements, $\mu$, $\Sigma$, and $r$, are parameters representing the size and shape of the ambiguity set at each node. Each node is represented by an ambiguity set $\mathcal{P}^{\text{MC}}(\mu, \Sigma) \cup \mathcal{P}^{\text{BALL}}(\mu, r)$, where $\Sigma$ is the maximal backward reachable covariance that reaches the next node with some control policy $\mathscr{C}$, and $r$ is the maximal radius for a ball-shaped ambiguity set that reaches the next node under the same control policy $\mathscr{C}$. The next three elements, idx, $k$, and $\text{ch}_{\text{idx}}$ are a pointer to the parent node, a pointer to the control policy $\mathscr{C}$, and a pointer to the set of child nodes. At each iteration, the RAND routine is used to select a node for expansion and the RANDMEANAROUND routine is used to select a query node mean $\mu_q$. Then, MAXCOVAR is run in parallel with MAX-COV-BALL-UB in order to get two candidate initial covariances $\Sigma_{0, \text{MC}}$ and $\Sigma_{0, \text{ours}}$. Algorithm \ref{alg:construct_brt_max_prob_reach} then selects the distribution (and corresponding control policy) that maximizes $\lambda_{\min}(\Sigma_0)$, before running MAX-COV-BALL to find the maximal radius $r$ associated with the chosen distribution and control policy.

\setlength{\textfloatsep}{10pt}
\begin{algorithm}[t]
\caption{Constructing a BRT with MAX-COV-BALL}
\label{alg:construct_brt_max_prob_reach}
\begin{algorithmic}[1]
    \Require $\mathcal{G}$, $N$, $n_\mathrm{iter}$ %Input
    \Ensure  $\mathcal{T}$ %Output
    \State $\nu \gets \phi$, $\varepsilon \gets \phi$
    \State $ \nu_{0} \gets \operatorname{CREATENODE}(\mu_{\mathcal{G},c}, \Sigma_{\mathcal{G}}, r_{\mathcal{G}}, \emptyset, \emptyset, \{\})$
    \State $\nu \gets \nu \cup \{\nu_{0}\}$
    \For{$i \gets 1 \textrm{ to } n_\mathrm{iter}$}
        \State %$\mathrm{idx}_\textrm{sampled}, \nu_{\textrm{sampled}} \gets \operatorname{RAND}(\nu)$
        $\nu_{k} \gets \operatorname{RAND}(\nu)$ %{\tt rand}
        \State $ \mu_{q} \gets \operatorname{RANDMEANAROUND}(\nu_{k}, r_{\textrm{sample}}) $
        \State $\textrm{status}_\text{ours}, \Sigma_{0, \text{ours}}, \mathscr{C}_{\text{ours}} \gets$\\
        \hspace*{2em}$\operatorname{MAX-COV-BALL-UB}(\mu_q, \nu_k, N)$
        \State $\textrm{status}_\text{MC}, \Sigma_{0, \text{MC}}, \mathscr{C}_{\text{MC}} \gets \operatorname{MAXCOVAR}(\mu_q, \nu_k, N)$
        \State $\textrm{status} \gets \textrm{status}_\text{ours} \vee \textrm{status}_\text{MC}$
        \State $\textrm{outputs} \gets \{(\Sigma_{0, \text{ours}}, \mathscr{C}_{\text{ours}}), (\Sigma_{0, \text{MC}}, \mathscr{C}_{\text{MC}})\}$
        \If{$\textrm{status} \neq \textrm{infeasible}$}
            \State $\Sigma_0, \mathscr{C} \gets \arg \max_{\text{outputs}} \lambda_{\min}(\Sigma_0)$
            \State $r \gets \operatorname{MAX-COV-BALL}(\Sigma_0, \mathscr{C}, \nu_k)$
            \State $ \textrm{idx} \gets \operatorname{size}(\mathcal{\nu}) + 1 $, $ \mathrm{ch}_{\mathrm{idx}} \gets \{\}$
            \State $ \nu_{\mathrm{new}} \gets \operatorname{CREATENODE}(\mu_{q}, \Sigma_0, r, \textrm{idx}, k, \mathrm{ch}_{\mathrm{idx}})$
            \State $ \nu \gets \nu \bigcup \{ \nu_{\mathrm{new}} \},  \mathrm{ch}_{k} \gets \mathrm{ch}_{k} \bigcup \{ \textrm{idx}\} $
            \State $ \varepsilon_{\textrm{idx},k} \gets \mathscr{C} $, $ \varepsilon \gets \varepsilon \bigcup \{ \varepsilon_{\textrm{idx},k} \} $
        \EndIf
    \EndFor
    \State $  \mathcal{T} \gets \{ \nu, \varepsilon \} $
    \State $ \textrm{\textbf{Return}} \ \  \mathcal{T}$
\end{algorithmic}
%\vspace{-0.5cm}
\end{algorithm}
\section{Proof of Maximal Coverage of MAX-COV-BALL}\label{sec: coverage_proofs}
We can characterize the coverage of an ambiguity set of distributions or of a tree of ambiguity sets of distributions in terms of its $h$-BRS, or ``h-hop'' backward reachable set of distributions. We define the $h$-BRS of an ambiguity set of distributions as in \cite{aggarwal2025tac}:
\begin{definition}\label{def: h_BRS_ambiguity_set}
[$h$-BRS of an ambiguity set of distributions $\mathcal{P}$]
The $h$-hop backward reachable set of distributions ($h$-BRS) of an ambiguity set $\mathcal{P}$ is defined as the set of all initial distributions for which there exists a feasible control sequence that can steer the system to $\mathcal{P}$ in hN timesteps.
\end{definition}
Similarly, we can define the $h$-BRS of a tree of ambiguity sets of distributions as:
\begin{definition}\label{def: h_BRS_tree}
[$h$-BRS of a tree of ambiguity sets of distributions $\mathcal{T}$]
The $h$-hop backward reachable set of distributions ($h$-BRS) of a tree of ambiguity sets of distributions $\mathcal{T}$ is defined as the set of all initial distributions for which there exists a feasible control sequence that can steer the system to the ambiguity set stored at the root node of $\mathcal{T}$ in hN timesteps. More formally, 
%$h$-BRS($\mathcal{T}$) can be described as
\begin{equation}
h-\text{BRS}(\mathcal{T}) := \bigcup_{i \in v(\mathcal{T})} (h - d_i)-\text{BRS}(v_i)
\end{equation}
where $v(\mathcal{T})$ is the set of all vertices in the tree (each vertex represents an ambiguity set of distributions), and $d_i$ represents the number of hops from the $i$th vertex to the root node.
\end{definition}

We define maximal coverage in terms of the size of the $h$-BRS, meaning that if $\mathscr{S}$ has maximal coverage over some class $\mathcal{S}$ of ambiguity sets, then $h$-BRS($\mathscr{S}$) $\supset$ $h$-BRS($\mathscr{S}'$) for all $\mathscr{S}' \in \mathcal{S}$ such that $\mathscr{S}' \neq \mathcal{S}$. We denote h-BRS($\mathcal{T}^{(r)}_{\mathcal{E}}$) as the $h$-backward reachable set of a tree $\mathcal{T}$ constructed under edge controller $\mathcal{E}$, after $r$ iterations.

Suppose we plan backwards from a goal node $\mathcal{P}^{\text{BALL}}(\mu_\mathcal{G}, r_\mathcal{G})$, and our method returns an ambiguity set $\mathcal{P}^{\text{BALL}}(\mu_q, r_q)$. We define the ANY procedure such that it returns a candidate ambiguity set $\mathcal{P}^{ANY}(\mu_q, \cdot)$ with ball-shaped projections that are invariant to transformations under $\textrm{SO}(n)$, such that $\forall \hat{\Sigma} = VDV^T \in \mathbb{S}^n_+$, with $V \in \textrm{SO}(n), D \in \mathbf{D}^n$, $\text{Proj}_{(\cdot, \{\hat{\Sigma}\})}(\mathcal{P}^{ANY}) = \mathbb{B}_n(\mu_q, \cdot) = \text{Proj}_{(\cdot, \{QDQ^T: Q \in \mathrm{SO}(n)\})}(\mathcal{P}^{ANY})$. We further restrict the class ANY such that any procedure in it differs from our set in its projection into the mean space, with $\text{Proj}_{(\cdot, \mathbb{S}^n_+)} (\mathcal{P}^{ANY}) \neq \text{Proj}_{(\cdot, \mathbb{S}^n_+)} (\mathcal{P}^{\text{BALL}}(\mu_q, r_q))$. Note that any procedure returning a MAXCOVAR ambiguity set, a maximal radius 2-Wasserstein ambiguity set (which will have $\hat{\Sigma} = 0$), or a set of the form $\mathcal{P}^{\text{BALL}, \chi^2}$ falls into this category. 

Under these definitions, in the absence of process noise and state constraints, our method achieves maximal coverage, achieving coverage greater than any procedure belonging to the class ANY described above. We state this property formally in Theorem \ref{thm: maximal_coverage}.

\begin{theorem}\label{thm: maximal_coverage}
(Maximum Coverage) In the absence of process noise and state constraints, h-BRS($\mathcal{T}^{(r)}_{\text{OURS}}$) $\supset$ h-BRS($\mathcal{T}^{(r)}_{\text{ANY}}$) for all planning scenes $\forall h \geq 1, r \geq 1$.
\end{theorem}

Theorem \ref{thm: maximal_coverage} is a consequence of two properties. First, in the absence of process noise, the maximal-radius ambiguity set $\mathcal{P}^{\text{BALL}}(\mu, r)$ found by our method has provably maximal projections into the state space for any fixed $\Sigma$. Second, in the absence of process noise and state constraints, any ball-shaped ambiguity set achieves greater coverage than any set $\mathcal{P}^{ANY}$ with smaller projections, as long as $\mathcal{P}^{ANY}$ has a strictly smaller projection into the mean space. These facts are stated formally below in Lemmas \ref{lemma: maximal_brs_subset} and \ref{lemma: no_process_noise}.

\begin{lemma}\label{lemma: maximal_brs_subset}
In the absence of process noise but in the presence of state or control constraints, consider a goal set $\mathcal{G} = \mathcal{P}^{\text{BALL}, \chi^2}(\mu_\mathcal{G}, r_\mathcal{G})$ or $\mathcal{G} = \mathcal{P}^{\text{BALL}}(\mu_\mathcal{G}, r_\mathcal{G})$ and $\mu_\mathcal{I} \in \mathbb{R}^n$. There exists $r_\mathcal{I} \in \mathbb{R}_{\geq 0}$, $\mathscr{C}$ such that $\mathcal{P}^{\text{BALL}}(\mu_\mathcal{I}, r_\mathcal{I}) \xrightarrow{\mathscr{C}} \mathcal{G}$ and such that $\forall \mathcal{P}^{\text{ANY}}(\mu_\mathcal{I}, \cdot), \mathscr{C}'$ such that $\mathcal{P}^{\text{ANY}}(\mu_\mathcal{I}, \cdot) \xrightarrow{\mathscr{C}'} \mathcal{P}^{\text{ANY}}(\mu_N, \cdot) \subseteq\mathcal{G}$, $\text{Proj}_{(\cdot, \hat{\Sigma})}(\mathcal{P}^{\text{ANY}}(\mu_\mathcal{I}, \cdot)) \subseteq \text{Proj}_{(\cdot, \hat{\Sigma})}(\mathcal{P}^{\text{BALL}}(\mu_\mathcal{I}, r_\mathcal{I}))$ for any $\hat{\Sigma} \in \mathbb{S}^n_+$.
%Consider the backward reachable set of distributions $\mathscr{S} := \{(\mu, \Sigma) \in \mathbb{R}^n \times \mathbb{S}^n_+: (\mu, \Sigma) \xrightarrow{\mathscr{C}} \mathcal{G}\}$. For any $\mu_\mathcal{I} \in \text{Proj}_{(\cdot, \mathbb{S}^n_+)}(\mathscr{S})$, there exists $\mathcal{P}^{\text{BALL}}(\mu_\mathcal{I}, r_\mathcal{I}) \subseteq \mathscr{S}$ such that $\text{Proj}_{(\cdot, \hat{\Sigma})}(\mathcal{P}^{\text{ANY}}(\mu_\mathcal{I}, \cdot)) \subseteq \text{Proj}_{(\cdot, \hat{\Sigma})}(\mathcal{P}^{\text{BALL}}(\mu_\mathcal{I}, r_\mathcal{I}))$ for any $\hat{\Sigma} \in \mathbb{S}^n_+$, $\mathcal{P}^{\text{ANY}}(\mu_\mathcal{I}, \cdot) \in \mathscr{S}$.
\end{lemma}
\begin{lemma}\label{lemma: no_process_noise}
In the absence of process noise and state constraints, for any nodes $\mathcal{G} := \mathcal{P}^{\text{BALL}}(\mu, r)$ and $\mathcal{G}^- := \mathcal{P}^{\text{ANY}}$ such that $\text{Proj}_{(\cdot, \{\hat{\Sigma}\})}(\mathcal{G}^-) \subseteq \text{Proj}_{(\cdot, \{\hat{\Sigma}\})}(\mathcal{G})$ $\forall \hat{\Sigma}$ and such that $\text{Proj}_{(\cdot, \mathbb{S}^n_+)}(\mathcal{G}^-) \subset \text{Proj}_{(\cdot, \mathbb{S}^n_+)}(\mathcal{G})$, $h$-BRS($\mathcal{G}^-$) $\subset$ $h$-BRS($\mathcal{G}$).
\end{lemma}

In the presence of process noise and state constraints, MAX-COV-BALL achieves provably better coverage than MAXCOVAR under mild conditions, and achieves at least equal coverage to MAXCOVAR under any conditions. We state this property formally in Theorem \ref{thm: max_covar_dominance}.

\begin{theorem}\label{thm: max_covar_dominance}
(Dominance over MAXCOVAR) h-BRS($\mathcal{T}^{(r)}_{\text{OURS}}$) $\supseteq$ h-BRS($\mathcal{T}^{(r)}_{\text{MAXCOVAR}}$) for all planning scenes. Further, when conditions (a) and (b) required for Lemma \ref{lemma: mild_assumptions} are satisfied at the goal node $\mathcal{G}$ and when Condition \ref{condition: c} is satisfied for $\mathcal{T}^{(r)}_{\text{OURS}}$, there exist planning scenes such that h-BRS($\mathcal{T}^{(r)}_{\text{OURS}}$) $\supset$ h-BRS($\mathcal{T}^{(r)}_{\text{MAXCOVAR}}$) $\forall h \geq 1$, $\forall r \geq 1$.
\end{theorem}

In order for MAX-COV-BALL to provably achieve greater coverage than MAXCOVAR, it is sufficient for a distribution to exist that can reach the goal node in $h$ hops under MAX-COV-BALL but not under MAXCOVAR. Lemma \ref{lemma: mild_assumptions} and Condition \ref{condition: c} establish conditions under which such a distribution exists. Lemma \ref{lemma: mild_assumptions} states that the union of a ball-shaped ambiguity set $\mathcal{P}^{\text{BALL}}(\mu, r)$ and a MAXCOVAR ambiguity set $\mathcal{P}^{\text{MC}}(\mu, \Sigma)$ has a strictly larger $h$-BRS than $\mathcal{P}^{\text{MC}}(\mu, \Sigma)$ when (a) $r$ is sufficiently large and (b) the state constraints are sufficiently loose. Specifically, when conditions (a) and (b) hold, the distribution in the $h$-BRS of the union of the ambiguity sets that is furthest from the centroid $\mu$ in mean space is not in the $h-$BRS of $\mathcal{P}^{\text{MC}}(\mu, \Sigma)$.

\begin{lemma}\label{lemma: mild_assumptions}
Consider $\mathcal{G}$ to be the union of a set $\mathcal{P}^{\text{BALL}}(\mu, r)$ and a MAXCOVAR ambiguity set $\mathcal{P}^{\text{MC}}(\mu, \Sigma)$. Suppose that (a) $\forall \mathbf{x}_0 \in \text{Proj}_{(\cdot, \mathbb{S}^n_+)}(h-BRS(\mathcal{G}))$, $\Sigma_N \prec (r/\Phi^{-1}(1-\epsilon))^2$, and (b) $\forall \mathbf{x}_0 \in \text{Proj}_{(\cdot, \mathbb{S}^n_+)}(h-BRS(\mathcal{G}))$, the state constraints are strict inequalities (i.e. are not tight) at each timestep as $\mathbf{x}_0 \to \mathcal{G}$. 
Then, $h-\text{BRS}(\mathcal{G}) \supset h-\text{BRS}(\mathcal{P}^{\text{MC}}(\mu, \Sigma))$. Also, there exists $\mathbf{x}_0' \in \text{Proj}_{(\cdot, \mathbb{S}^n_+)}(h-\text{BRS}(\mathcal{G}))$ satisfying $\mathbf{x}_0' = \arg \max_{\mathbf{x}_0 \in \text{Proj}_{(\cdot, \mathbb{S}^n_+)}(h-\text{BRS}(\mathcal{G}))} ||\mathbf{x}_0 - \mu_\mathcal{G}||_2$ such that $\mathbf{x}_0' \notin \text{Proj}_{(\cdot, \mathbb{S}^n_+)}(h-\text{BRS}(\mathcal{P}^{\text{MC}}(\mu, \Sigma))$ for all $h \geq 1$.
\end{lemma}

In the case that the goal ambiguity set $\mathcal{G}$ is the union of a ball-shaped ambiguity set $\mathcal{P}^{\text{BALL}}(\mu, r)$ and a MAXCOVAR ambiguity set $\mathcal{P}^{\text{MC}}(\mu, \Sigma)$, Condition \ref{condition: c} states that the distribution in the $h$-BRS of the goal set $\mathcal{G}$ that is furthest from $\mu$ in mean space is not in the $h$-BRS of any other node in a backward reachable tree $\mathcal{T}$. In general, the sequence of node means used to construct the backward reachable tree $\mathcal{T}$ can be chosen such that this condition always holds.

\begin{condition}\label{condition: c}
Consider $\mathcal{G}$ to be the union of a set $\mathcal{P}^{\text{BALL}}(\mu, r)$ and a MAXCOVAR ambiguity set $\mathcal{P}^{\text{MC}}(\mu, \Sigma)$. There exists $\mathbf{x}_0' \in \text{Proj}_{(\cdot, \mathbb{S}^n_+)}(h-\text{BRS}(\mathcal{G}))$ satisfying $\mathbf{x}_0' = \arg \max_{\mathbf{x}_0 \in \text{Proj}_{(\cdot, \mathbb{S}^n_+)}(h-\text{BRS}(\mathcal{G}))} ||\mathbf{x}_0 - \mu_\mathcal{G}||_2$ such that $\mathbf{x}_0' \notin \text{Proj}_{(\cdot, \mathbb{S}^n_+)}(h-\text{BRS}(\nu))$ for any $\nu \in \nu(\mathcal{T}) \setminus \mathcal{G}$ and any $h \geq 1$.
\end{condition}

Recall from Section \ref{subsec: ambiguity_sets} that ball-shaped ambiguity sets of the form $\mathcal{P}^{\text{BALL}}$ and $\mathcal{P}^{\text{BALL}, \chi^2}$ underapproximate the set of Gaussian distributions that satisfy inclusion in a Euclidean ball with probability of at least $(1-\epsilon)$. However, 2-Wasserstein (or equivalently Gelbrich) sets also underapproximate this set. Suppose the goal set $\mathcal{G}$ is given by the set of Gaussian distributions that satisfy inclusion in a Euclidean ball with probability of at least $(1-\epsilon)$. Consider the maximal radius set of the form $\mathcal{P}^{\text{BALL}, \chi^2} \subseteq \mathcal{G}$ and the maximal radius set of the form $\mathcal{P}^{\mathbb{W}_2}(\mu, \Sigma, r) \subseteq \mathcal{G}$. The set of the form $\mathcal{P}^{\text{BALL}, \chi^2}$ has the same projections into mean space as $\mathcal{G}$ when the covariance is zero or when the covariance is maximal. For mild assumptions on $n$ and $\epsilon$ (which always hold when $n \geq 2, \epsilon \leq 0.2$), the set of the form $\mathcal{P}^{\mathbb{W}_2}(\mu, \Sigma, r)$ has a strictly smaller projection into mean space than the set of the form $\mathcal{P}^{\text{BALL}, \chi^2}$, and it excludes the maximal covariance distribution contained in $\mathcal{G}$.

\begin{remark}\label{remark: w2_projections}
Suppose the goal set $\mathcal{G}$ is defined by the set of Gaussian distributions $(\mu, \Sigma)$ such that a random variable $\mathbf{x} \sim \mathcal{N}(\mu, \Sigma)$ satisfies $\mathbb{P}(\mathbf{x} \in \mathbb{B}_n(\mu_\mathcal{G}, r_\mathcal{G})) \geq 1 -\epsilon$. There always exists $\mathcal{P}^{\text{BALL}, \chi^2}(\mu_{\chi^2}, r_{\chi^2}) \subseteq \mathcal{G}$ such that $\text{Proj}_{(\cdot, \mathbb{S}^n_+)}(\mathcal{P}^{\text{BALL}, \chi^2}(\mu_{\chi^2}, r_{\chi^2})) = \text{Proj}_{(\cdot, \mathbb{S}^n_+)}(\mathcal{G})$ and such that $\text{Proj}_{(\cdot, \{\Sigma_{\max}\})}(\mathcal{P}^{\text{BALL}, \chi^2}(\mu_{\chi^2}, r_{\chi^2})) = \text{Proj}_{(\cdot, \{\Sigma_{\max}\})}(\mathcal{G}) \neq \emptyset$, where $\Sigma_{\max} = \max_{(\mu, \Sigma) \in \mathcal{G}} \lambda_{\min}(\Sigma)$.
When $n > 1$ and $f^{-1}(1-\epsilon, n) \leq nf^{-1}(1-\epsilon, 1)$ (which holds $\forall n \geq 2$ when $\epsilon < 0.2$), for $\mathcal{P}^{\mathbb{W}_2, \max} = \arg \max_{r_W} \mathcal{P}^{\mathbb{W}_2}(\mu_W, \Sigma_W, r_W) \subseteq \mathcal{G}$, $\text{Proj}_{(\cdot, \mathbb{S}^n_+)}(\mathcal{P}^{\mathbb{W}_2, \max} ) \subset\text{Proj}_{(\cdot, \mathbb{S}^n_+)}(\mathcal{P}^{\text{BALL}, \chi^2}(\mu_{\chi^2}, r_{\chi^2}))$ and $\text{Proj}_{(\cdot, \{\Sigma_{\max}\})}(\mathcal{P}^{\mathbb{W}_2, \max}) = \emptyset$.
\end{remark}

Now, consider the maximal radius set of the form $\mathcal{P}^{\text{BALL}} \subseteq \mathcal{G}$. Although this set underapproximates $\mathcal{P}^{\text{BALL}, \chi^2}$ and so also underapproximates $\mathcal{G}$, it has the same projection into mean space as $\mathcal{G}$ when the covariance is maximal. Furthermore, for stronger but still realistic assumptions on $n$ and $\epsilon$ (which always hold when $2 \leq n \leq 6$, $\epsilon \leq 0.02$), the maximal radius set of the form $\mathcal{P}^{\mathbb{W}_2}(\mu, \Sigma, r)$ has a strictly smaller projection into mean space than the set of the form $\mathcal{P}^{\text{BALL}}$.

\begin{remark}\label{remark: w2_projections_phi}
Suppose the goal set $\mathcal{G}$ is defined as in Remark \ref{remark: w2_projections}. There always exists $\mathcal{P}^{\text{BALL}}(\mu, r) \subseteq \mathcal{G}$ such that $\text{Proj}_{(\cdot, \{\Sigma_{\max}\})}(\mathcal{P}^{\text{BALL}}(\mu, r)) = \text{Proj}_{(\cdot, \{\Sigma_{\max}\})}(\mathcal{G}) \neq \emptyset$, where $\Sigma_{\max} = \max_{(\mu, \Sigma) \in \mathcal{G}} \lambda_{\min}(\Sigma)$.
When $n > 1$ and $f^{-1}(1-\epsilon, n) \leq  \Phi^{-1}(1-\epsilon)^2f^{-1}(1-\epsilon, 1)$ (which holds e.g. when $2 \leq n \leq 6$ and $\epsilon < 0.02$), for $\mathcal{P}^{\mathbb{W}_2}(\mu_W, \Sigma_W, r_W) \subseteq \mathcal{G}$ maximizing $r_W$, $\text{Proj}_{(\cdot, \mathbb{S}^n_+)}(\mathcal{P}^{\mathbb{W}_2}(\mu_W, \Sigma_W, r_W)) \subset \text{Proj}_{(\cdot, \mathbb{S}^n_+)}(\mathcal{P}^{\text{BALL}}(\mu, r))$ and $\text{Proj}_{(\cdot, \{\Sigma_{\max}\})}(\mathcal{P}^{\mathbb{W}_2}(\mu_W, \Sigma_W, r_W)) = \emptyset$.
\end{remark}

%Also, the maximal radius 2-Wasserstein uncertainty set which is contained in $\mathcal{P}^{\text{BALL}, \chi^2}$ will have a smaller $h$-backward reachable set of distributions than $\mathcal{P}^{\text{BALL}, \chi^2}$ for any $h \geq 1$, and we define a condition under which the $h-$BRS of the 2-Wasserstein uncertainty set is strictly smaller. This condition always holds when state constraints are absent or are strict inequalities over the entire $h$-BRS of he 2-Wasserstein uncertainty set.

%\begin{remark}\label{remark: w2}
%Consider the set $\mathcal{P}^{\text{BALL}, \chi^2}(\mu_\mathcal{G}, r_\mathcal{G}))$. The 2-Wasserstein ball with maximal $r$ satisfying $\mathcal{P}^{\mathbb{W}_2}(\mu, \Sigma, r) \subseteq \mathcal{P}^{\text{BALL}, \chi^2}(\mu_\mathcal{G}, r_\mathcal{G}))$ is a strict subset of $\mathcal{P}^{\text{BALL}, \chi^2}$. Further, there exist planning scenes where $h-\text{BRS}(\mathcal{P}^{\text{BALL}, \chi^2}(\mu_\mathcal{G}, r_\mathcal{G})) \supset h-\text{BRS}(\mathcal{P}^{\mathbb{W}_2}(\mu, \Sigma, r))$ as long as $h-\text{BRS}(\mathcal{P}^{\text{BALL}, \chi^2}(\mu_\mathcal{G}, r_\mathcal{G}) \setminus \mathcal{P}^{\text{MC}}(\mu_\mathcal{G}, \frac{r_\mathcal{G}^2}{f^{-1}(1-\epsilon, n)}I)) \subset h-\text{BRS}(\mathcal{P}^{\text{BALL}, \chi^2}(\mu_\mathcal{G}, r_\mathcal{G}))$.
%\end{remark}

Formal proofs of all Lemmas, Theorems, and Remarks made in this section are provided in Appendix \ref{sec:appendix}.

\section{MAX-ELL-BALL: Constructing a BRT of Ellipsoidal Sets of Ball-shaped Ambiguity Sets of Distributions}\label{sec: max_ell_prob_reach}
In addition to planning over ball-shaped ambiguity sets of distributions as described in Section \ref{sec: max_cov_prob_reach}, we can plan over ellipsoidal sets of ball-shaped ambiguity sets of distributions. These ambiguity sets, as described in Section \ref{subsec: ambiguity_sets}, are parameterized by a central first-order moment $\hat{\mu}$, a shape matrix $\hat{\mathscr{P}}$, and a radius $\hat{r}$, such that:
\begin{align}
&\mathcal{P}^{\text{ELL-BALL}}(\hat{\mu}, \hat{\mathscr{P}}, \hat{r}) = \left\{\mathbb{P}: \begin{array}{l} (s -\hat{\mu})^T\mathscr{P}^{-1}(s -\hat{\mu}) \leq 1 \\ 
(\mu_\mathbb{P}, \Sigma_{\mathbb{P}}) \in \mathcal{P}^{\text{BALL}}(\hat{\mu}, \hat{r})  \end{array} \right\}. \notag
\end{align}
In order to maximize the size of these ambiguity sets, we alternate between maximizing $\det(\hat{\mathscr{P}})$ and $\hat{r}$. Maximizing $\det(\hat{\mathscr{P}})$ for a fixed $\hat{r}$ maximizes the set of centers of ball-shaped ambiguity sets of distributions that belong to $\mathcal{P}^{\text{ELL-BALL}}(\hat{\mu}, \hat{\mathscr{P}}, \hat{r})$, and maximizing $\hat{r}$ maximizes the size of the ball-shaped ambiguity sets of distributions that belong to $\mathcal{P}^{\text{ELL-BALL}}(\hat{\mu}, \hat{\mathscr{P}}, \hat{r})$.

This approach is similar to that taken in MAXELLIPSOID \cite{aggarwal2025tac}; in fact, we construct an algorithm that has volume projections into the state space provably equal to those of MAXELLIPSOID. We use the same approach as MAXELLIPSOID towards maximizing $\det(\hat{\mathscr{P}})$, so we focus here on maximizing $\hat{r}$ for a fixed $\hat{\mathscr{P}}$ by generalizing the approach described in Section \ref{sec: max_cov_prob_reach}. We first construct a program that generalizes Problem \ref{prob: ball_nlp} to our new ambiguity set formulation:
\begin{problem}\label{prob: ellipsoid_ambiguity_nlp} (MAX-ELL-BALL) Maximize the radius $r$ of the initial ambiguity set $\mathcal{P}^{\text{ELL-BALL}}(\mu_\mathcal{I}, \mathscr{P}_\mathcal{I}, r)$ for a fixed $\mathscr{P}_\mathcal{I}, \mu_\mathcal{I}$ by maximizing the radius of the ball $\mathbb{B}_n(\mu_\mathcal{I}, r)$ such that $\mathbb{B}_n(\mu_\mathcal{I}, r) \to \mathcal{P}^{\text{ELL-BALL}, \mathcal{G}}$. 
\begin{align}
\max_{\overline{\mathbf{u}}_k, K_k, \Sigma_k, \mathscr{P}_k, \mathscr{R}_k} r = \lambda_{\min}(\mathscr{R}_0)
\end{align}
such that:
\allowdisplaybreaks
\begin{align}
\mu_{k+1} &= A\mu_k + B\overline{\mathbf{u}}_k, \quad\mathscr{P}_{k+1} = A\mathscr{P}_kA^T, \label{eq: Pk_ell} \\
\mathscr{R}_{k+1} &= (A + BK)\mathscr{R}_k(A+BK)^T, \label{eq: Rk_ell} \\
\Sigma_{k+1} &= (A + BK)\Sigma_k(A + BK)^T + DD^T, \label{eq: Sigmak_ell} \\
\mu_0 &= \mu_\mathcal{I}, \label{eq: mean_init_ell} \\
\Sigma_0 &= 0, \label{eq: sigma_init_ell} \\
\mathscr{P}_0 &= \mathscr{P}_q, \label{eq: P_init_ell} \\
||s - \mu_N||_2 &+ \Phi^{-1}(1-\epsilon)\sqrt{\lambda_{\max}(\Sigma_N)} + \sqrt{\lambda_{\max}(\mathscr{R}_N)} \leq r_\mathcal{G}, \notag \\
\mathcal{E}(s, \mathscr{P}_N) & \subseteq \mathcal{E}(\mu_\mathcal{G}, \mathscr{P}_\mathcal{G}), \label{eq: ell_contain_ell} \\
{a_i^u}^T\overline{\mathbf{u}}_k &+ \Phi^{-1}(1-\epsilon)\sqrt{{a_i^u}^T K_k \Sigma_k K_k a_i^u} \notag \\ &+ \sqrt{{a_i^u}^T K_k \mathscr{R}_k K_k a_i^u} \leq b_i^u, \notag \\
{a_i^x}^T\mu_k &+ \Phi^{-1}(1-\epsilon)\sqrt{{a_i^x}^T \Sigma_k a_i^x} + \sqrt{{a_i^x}^T \mathscr{P}_k a_i^x} \notag \\ &+ \sqrt{{a_i^x}^T \mathscr{R}_k a_i^x} \leq b_i^x.
\end{align}
\end{problem}
Like Problem \ref{prob: ball_nlp}, this problem is challenging to solve because of the shape matrix and covariance dynamics. We construct another nonlinear optimization problem that maximizes an upper bound on $r$ for a fixed $\hat{\mathscr{P}}$:
\begin{problem}\label{prob: ellipsoid_nlp} (MAX-ELL-BALL-UB) Maximize an upper bound on the radius $r$ of the initial ambiguity set $\mathcal{P}^{\text{ELL-BALL}}(\mu_\mathcal{I}, \mathscr{P}_\mathcal{I}, r)$ for a fixed $\mathscr{P}_\mathcal{I}, \mu_\mathcal{I}$.
\begin{align}
\max_{\overline{\mathbf{u}}_k, K_k, \Sigma_k, \mathscr{P}_k} r = \lambda_{\min}(\Sigma_0)
\end{align}
such that:
\begin{align}
%\mu_{k+1} &= A\mu_k + B\overline{\mathbf{u}}_k, \notag \\
%\mathscr{P}_{k+1} &= A\mathscr{P}_kA^T, \notag \\
%\Sigma_{k+1} &= (A + BK)\Sigma_k(A + BK)^T + DD^T, \notag \\
%\mu_0 &= \mathbf{x}_0, \notag \\
%\mathscr{P}_0 &= \mathscr{P}_q, \notag \\
||s - \mu_N||_2 &+ \Phi^{-1}(1-\epsilon)\sqrt{\lambda_{\max}(\Sigma_N)} \leq r_\mathcal{G}, \notag \\
%\mathcal{E}(s, \mathscr{P}_N) & \subseteq \mathcal{E}(\mu_\mathcal{G}, \mathscr{P}_\mathcal{G}), \notag \\
{a_i^u}^T\overline{\mathbf{u}}_k &+ \Phi^{-1}(1-\epsilon)\sqrt{{a_i^u}^T K_k \Sigma_k K_k a_i^u} \leq b_i^u, \notag \\
{a_i^x}^T\mu_k &+ \Phi^{-1}(1-\epsilon)\sqrt{{a_i^x}^T \Sigma_k a_i^x} + \sqrt{{a_i^x}^T \mathscr{P}_k a_i^x} \leq b_i^x, \notag \\
\text{subject to: } &\text{(\ref{eq: Pk_ell}),  (\ref{eq: Sigmak_ell}), (\ref{eq: mean_init_ell}), (\ref{eq: P_init_ell}), (\ref{eq: ell_contain_ell}).} \notag
\end{align}
\end{problem}
This program is similar to Problem \ref{prob: ball_nlp_loose}, except for the addition of the shape matrix dynamics and the final ellipsoid containment constraint. We can relax the nonlinear chance constraints and terminal covariance constraint as before. We can relax the ellipsoid containment constraint as in \cite{aggarwal2025tac}. Recall that $\mathcal{E}(s, \mathscr{P}_N) \subseteq \mathcal{E}(s, \gamma \mathscr{P}_\mathcal{G}) \subseteq \mathcal{E}(\mu_\mathcal{G}, \mathscr{P}_\mathcal{G})$ is a sufficient condition for containment. The first part of this condition can simply be written as $\mathscr{P}_N \preceq \gamma \mathscr{P}_\mathcal{G}$. By the S-procedure \cite{polik2007survey}, the second part can be losslessly rewritten as
\begin{equation}
\frac{\lambda}{\gamma}
\begin{bmatrix}
\mathscr{P}_\mathcal{G}^{-1} & -s^T\mathscr{P}_\mathcal{G}^{-1} \\
-\mathscr{P}_\mathcal{G}^{{-1}^T}s & s^T \mathscr{P}_\mathcal{G}^{-1}s-1
\end{bmatrix} \succeq \hspace{-0.1cm}
\begin{bmatrix}
\mathscr{P}_\mathcal{G}^{-1} & -\mu_\mathcal{G}^T\mathscr{P}_\mathcal{G}^{-1} \\
-\mathscr{P}_\mathcal{G}^{{-1}^T}\mu_\mathcal{G} & \mu_\mathcal{G}^T \mathscr{P}_\mathcal{G}^{-1}\mu_\mathcal{G}-1
\end{bmatrix}\notag
\end{equation}
However, this condition is still non-convex due to the presence of $\frac{\lambda}{\gamma}$ and $s^T \mathscr{P}_\mathcal{G}^{-1}s$. Because $s^T \mathscr{P}_\mathcal{G}^{-1}s$ is a convex function, we can see that $s^T \mathscr{P}_\mathcal{G}^{-1}s \geq 2y^T\mathscr{P}_\mathcal{G}^{-1}s - y^T \mathscr{P}_\mathcal{G}^{-1}y$ $\forall y \in \mathbb{R}^n$ (e.g. the tangent line at any point provides a global lower bound). Then, if $y = \mu_\mathcal{G}$, we have $s^T \mathscr{P}_\mathcal{G}^{-1}s \geq 2\mu_\mathcal{G}^T\mathscr{P}_\mathcal{G}^{-1}s - \mu_\mathcal{G}^T \mathscr{P}_\mathcal{G}^{-1}\mu_\mathcal{G}$. We will define $\mathbb{M}_2(v_1, v_2, M) = 2v_2^TM^{-1}v_1 - v_2^T M^{-1}v_2 - 1$, with $\mathbb{M}_2^\mathcal{G} = \mathbb{M}_2(s, \mu_\mathcal{G}, \mathscr{P}_\mathcal{G})$. It follows (by the properties of the Schur complement) that
\begin{align}
&\begin{bmatrix}
\mathscr{P}_\mathcal{G}^{-1} & -s^T\mathscr{P}_\mathcal{G}^{-1} \\
-\mathscr{P}_\mathcal{G}^{{-1}^T}s & s^T \mathscr{P}_\mathcal{G}^{-1}s-1
\end{bmatrix} \succeq \begin{bmatrix}
\mathscr{P}_\mathcal{G}^{-1} & -s^T\mathscr{P}_\mathcal{G}^{-1} \\
-\mathscr{P}_\mathcal{G}^{{-1}^T}s & \mathbb{M}_2^\mathcal{G}
\end{bmatrix}
\end{align}
Substituting $\tau = \frac{\lambda}{\gamma}$ yields
\begin{align}
&\begin{bmatrix}
\mathscr{P}_\mathcal{G}^{-1} & -s^T\mathscr{P}_\mathcal{G}^{-1} \\
-\mathscr{P}_\mathcal{G}^{{-1}^T}s & \mathbb{M}_2^\mathcal{G}
\end{bmatrix} \succeq
\tau \begin{bmatrix}
\mathscr{P}_\mathcal{G}^{-1} & -\mu_\mathcal{G}^T\mathscr{P}_\mathcal{G}^{-1} \\
-\mathscr{P}_\mathcal{G}^{{-1}^T}\mu_\mathcal{G} & \mu_\mathcal{G}^T \mathscr{P}_\mathcal{G}^{-1}\mu_\mathcal{G}-1
\end{bmatrix} \notag
\end{align}
as a sufficient condition for ellipsoid containment. Rearranging yields the condition:
\begin{align}\label{eq: tau_condition}
\begin{bmatrix}
\mathscr{P}_\mathcal{G}^{-1}(1-\tau) & (\tau \mu_\mathcal{G}-s^T)\mathscr{P}_\mathcal{G}^{-1} \\
-\mathscr{P}_\mathcal{G}^{{-1}^T}(s-\tau \mu_\mathcal{G}) & \mathbb{M}_2^\mathcal{G} - \tau(\mu_\mathcal{G}^T \mathscr{P}_\mathcal{G}^{-1}\mu_\mathcal{G}-1)
%2\mu_\mathcal{G}^T\mathscr{P}_\mathcal{G}^{-1}s - (1 + \tau)\mu_\mathcal{G}^T \mathscr{P}_\mathcal{G}^{-1}\mu_\mathcal{G}-1 + \tau
\end{bmatrix} \succeq 0
\end{align}
Then, recalling the definition of $\mathbb{M}_1(a, M_r, M_k)$ from \eqref{eq: M1}, we can relax Problem \ref{prob: ellipsoid_nlp} to the following semidefinite program:
\begin{problem}\label{prob: ellipsoid_sdp_loose} Semidefinite relaxation of MAX-ELL-BALL-UB. %Find a control policy which maximizes an upper bound on the radius $r$ of the initial ambiguity set $\mathcal{P}^{\text{ELL-BALL}, \mathcal{I}}(\mu_0, \mathscr{P}_0, r_0)$ for a fixed $\mathscr{P}_0$.
\begin{align}
\max_{\overline{\mathbf{u}}_k, K_k, \Sigma_k, \mathscr{P}_k} r = \lambda_{\min}(\Sigma_0)
\end{align}
such that:
\begin{align}
%\mu_{k+1} &= A\mu_k + B\overline{\mathbf{u}}_k, \notag \\
&\Sigma_{k+1} = A\Sigma_kA^T + BU_kA^T + AU_kTB^T + BY_kB^T + DD^T, \notag \\
&\begin{bmatrix}
\Sigma_k & U_k^T \\
U_k & Y_k
\end{bmatrix}  \succeq 0, \notag \\
%\mu_0 &= \mathbf{x}_0, \notag \\
&||\mu_N - s||_2 + \left(\frac{(\Phi^{-1}(1-\epsilon))^2}{2r} \lambda_{\max}(\Sigma_N) + \frac{r}{2} \right) \leq r_\mathcal{G}, \notag \\
&{a_i^u}^T\overline{\mathbf{u}}_k + \Phi^{-1}(1-\epsilon)\mathbb{M}_1(a_i^u, Y_r, Y_k) \leq b_i^u, \notag \\
%&+ \frac{\Phi^{-1}(1-\epsilon)}{2 \sqrt{{a_i^u}^T Y_r a_i^u}}{a_i^u}^T Y_k a_i^u \notag \\ &+ \frac{\Phi^{-1}(1-\epsilon) \sqrt{{a_i^u}^T Y_r a_i^u}}{2}\leq b_i^u, \notag \\
&{a_i^x}^T\mu_k + \Phi^{-1}(1-\epsilon)\mathbb{M}_1(a_i^x, \Sigma_r, \Sigma_k) + \mathbb{M}_1(a_i^x, \mathscr{P}_r, \mathscr{P}_k) \leq \hspace{-0.1cm} b_i^x, \notag \\
%{a_i^x}^T\mu_k &+ \frac{\Phi^{-1}(1-\epsilon)}{2 \sqrt{{a_i^x}^T \Sigma_r a_i^x}}{a_i^x}^T \Sigma_k a_i^x \notag \\ &+ \frac{1}{2\sqrt{{a_i^x}^T \mathscr{P}_r a_i^x}}{a_i^x}^T  \mathscr{P}_k a_i^x \notag \\ &+ \frac{\Phi^{-1}(1-\epsilon)\sqrt{{a_i^x}^T \Sigma_r a_i^x}}{2}  + \frac{\sqrt{{a_i^x}^T \mathscr{P}_r a_i^x}}{2} \leq b_i^x, \notag \\
&\gamma, \tau  > 0; \tau < 1, \ \mathscr{P}_N \preceq \gamma \mathscr{P}_\mathcal{G}, \notag \\
%\mathscr{P}_{k+1} &= A \mathscr{P}_k A^T, \\
&\text{subject to: \eqref{eq: Pk_ell}, \eqref{eq: mean_init_ell}, \eqref{eq: tau_condition}.} \notag
\end{align}
\end{problem}
After finding a control policy that maximizes an upper bound on $r$, we need to find the true $r$ corresponding to the ambiguity set steered to the target set by the control policy. Accordingly, we derive a semidefinite relaxation of Problem \ref{prob: ellipsoid_ambiguity_nlp}. Under a fixed control policy, $K_k$, $\Sigma_k$, $\mu_k$, and $\mathscr{P}_k$ are not decision variables - the only decision variables are $\mathscr{R}_k$ and $s$. We can fix $s$ by first minimizing $||s-\mu_N||_2$ subject to $\mathscr{E}(s, \mathscr{P}_N) \subseteq \mathscr{E}(\mu_\mathcal{G}, \mathscr{P}_\mathcal{G})$. In this subproblem, with $\mathbb{M}_2^N = \mathbb{M}_2(s, \mu_\mathcal{G}, \mathscr{P}_\mathcal{N})$ the ellipsoid containment can be relaxed to
\begin{align}\label{eq: ellipsoid_ambiguity_s_procedure}
&\begin{bmatrix}
\mathscr{P}_\mathcal{N}^{-1} & -s^T\mathscr{P}_\mathcal{N}^{-1} \\
-\mathscr{P}_\mathcal{N}^{{-1}^T}s & \mathbb{M}_2^N
\end{bmatrix} \succeq \frac{1}{\lambda}\begin{bmatrix}
\mathscr{P}_\mathcal{G}^{-1} & -\mu_\mathcal{G}^T\mathscr{P}_\mathcal{G}^{-1} \\
-\mathscr{P}_\mathcal{G}^{{-1}^T}\mu_\mathcal{G} & \mu_\mathcal{G}^T \mathscr{P}_\mathcal{G}^{-1}\mu_\mathcal{G}-1
\end{bmatrix}
\end{align}
using the S-procedure and the global lower bound $s^T \mathscr{P}_\mathcal{N}^{-1}s\geq 2\mu_\mathcal{G}^T\mathscr{P}_\mathcal{N}^{-1}s - \mu_\mathcal{G}^T \mathscr{P}_\mathcal{N}^{-1}\mu_\mathcal{G}$.

Once $s$ is fixed, the terminal constraint can be convexified losslessly, and the chance constraints can be convexified by linearizing around reference values $\mathscr{R}_r$. So, defining $s^* = \arg \min_s ||s-\mu_N||_2$ subject to Eq. \eqref{eq: ellipsoid_ambiguity_s_procedure}, we can relax the nonlinear problem to a semidefinite prgram:
\begin{problem}\label{prob: ellipsoid_ambiguity_sdp} Semidefinite relaxation of MAX-ELL-BALL.
\begin{align}
\max_{{\mathscr{R}_k}} r = \lambda_{\min}(\mathscr{R}_0)
\end{align}
such that:
\begin{align}
%\mu_{k+1} &= A\mu_k + B\overline{\mathbf{u}}_k, \notag \\
%\mathscr{P}_{k+1} &= A\mathscr{P}_kA^T, \notag \\
%\mathscr{R}_{k+1} &= (A + BK)\mathscr{R}_k(A+BK)^T, \notag \\
%\Sigma_{k+1} &= (A + BK)\Sigma_k(A + BK)^T + DD^T, \notag \\
%\mu_0 &= \mathbf{x}_0, \notag \\
%\Sigma_0 &= 0, \notag \\
%\mathscr{P}_0 &= \mathscr{P}_q, \notag \\
\lambda_{\max}(\mathscr{R}_N)\hspace{-0.1cm} &\leq\hspace{-0.1cm} (r_\mathcal{G} - ||s^* - \mu_N||_2 - \Phi^{-1}(1-\epsilon)\sqrt{\lambda_{\max}(\Sigma_N)})^2, \notag \\
{a_i^u}^T\overline{\mathbf{u}}_k &+ \Phi^{-1}(1-\epsilon)\sqrt{{a_i^u}^T K_k \Sigma_k K_k a_i^u} \notag \\
&+ \mathbb{M}_1(a_i^u, K_k\mathscr{R}_rK_k, K_k\mathscr{R}_kK_k) \leq b_i^u, \notag \\
%&+ \frac{{a_i^u}^T K_k \mathscr{R}_k K_k a_i^u}{2\sqrt{{a_i^u}^T K_k \mathscr{R}_r K_k a_i^u}}   + \frac{\sqrt{{a_i^u}^T K_k \mathscr{R}_r K_k a_i^u}}{2} \leq b_i^u, \notag \\
{a_i^x}^T\mu_k &+ \Phi^{-1}(1-\epsilon)\sqrt{{a_i^x}^T \Sigma_k a_i^x} + \sqrt{{a_i^x}^T \mathscr{P}_k a_i^x} \notag \\
&+ \mathbb{M}_1(a_i^x, \mathscr{R}_r, \mathscr{R}_k \leq b_i^x, \notag \\
%&+ \frac{{a_i^x}^T \mathscr{R}_k a_i^x}{2\sqrt{{a_i^x}^T \mathscr{R}_r a_i^x}} + \frac{\sqrt{{a_i^x}^T \mathscr{R}_r a_i^x}}{2} \leq b_i^x, \notag \\
&\text{subject to: \eqref{eq: Pk_ell}, \eqref{eq: Rk_ell}, \eqref{eq: Sigmak_ell}, \eqref{eq: mean_init_ell}, \eqref{eq: sigma_init_ell}, \eqref{eq: P_init_ell}.} \notag
\end{align}
\end{problem}
If a feasible solution exists to Problem \ref{prob: ellipsoid_sdp_loose}, a solution also exists to Problem \ref{prob: ellipsoid_ambiguity_sdp}. However, in the presence of process noise, the solution may be overly conservative, returning an ambiguity set that excludes some distributions with large covariance that can actually reach the goal set. As in our construction of MAX-COV-BALL, we account for this conservatism by constructing a MIP which leverages multiple ambiguity set representations and has worst-case performance equivalent to the MAXELLIPSOID algorithm \cite{aggarwal2025tac}. This algorithm is fully described in Algorithm \ref{alg:construct_brt_max_ell_prob_reach}.

 The MAXELLIPSOID algorithm first maximizes the volume of the ellipsoid $\mathscr{P}_0$ for a fixed $\Sigma_0$ and then maximizes $\lambda_{\min}(\Sigma_0)$ for a fixed $\mathscr{P}_0$. In order to guarantee performance equal to or better than MAXELLIPSOID, Algorithm \ref{alg:construct_brt_max_ell_prob_reach} ensures that the ellipsoid $\mathcal{E}(\mathbf{x}_0, \mathscr{P}_0)$ found at each node is the same as the ellipsoid that would be found by the MAXELLIPSOID algorithm. Algorithm \ref{alg:construct_brt_max_ell_prob_reach} tracks seven elements at each node, which are parameters to the CREATENODE routine. The first five parameters, $\mu$, $\Sigma$,  $\Sigma^{\text{ELL}}$, $\mathscr{P}$, and $r$, describe the size and shape of the ambiguity set at each node. Each node is represented by an ambiguity set $\mathcal{P}^{\text{ELL}}(\mu, \Sigma, \mathscr{P}) \cup \mathcal{P}^{\text{ELL-BALL}}(\mu, r, \mathscr{P})$, where $\mathcal{E}(\mu, \mathscr{P})$ is an ellipsoid in Euclidean space, $\Sigma$ is the maximal backward reachable covariance under which any point in $\mathcal{E}(\mu, \mathscr{P})$ reaches the next node under a control policy $\mathscr{C}$, and $r$ is the maximal radius for the ball-shaped ambiguity set under which any point in $\mathcal{E}(\mu, \mathscr{P})$ reaches the next node under $\mathscr{C}$. The parameter $\Sigma^{\text{ELL}}$ describes the covariance that would be found by the MAXELLIPSOID algorithm at the node. The last three parameters, idx, $k$, and $\text{ch}_{\text{idx}}$, are pointers to the parent node, control policy $\mathscr{C}$, and the set of child nodes, respectively.

 At each iteration of Algorithm \ref{alg:construct_brt_max_ell_prob_reach}, the RAND routine is used to select a node for expansion and the RANDMEANAROUND routine is used to select a query node mean $\mu_q$. Then, MAXELLIPSOID (see \cite{aggarwal2025tac} for a full description) is run with the covariance at node $\nu_k$ equal to $\Sigma_k^{\text{ELL}}$, such that it returns the same ellipsoid $\mathscr{P}_0$ found by the MAXELLIPSOID algorithm. If this is infeasible, we rerun MAXELLIPSOID with the covariance at node $\nu_k$ equal to $\Sigma_k$ to check if we can still find a feasible node. Next, we run MAXCOVARELL (see \cite{aggarwal2025tac} for a full description) with the covariance at node $\nu_k$ equal to $\Sigma_k^{\text{ELL}}$ to find $\Sigma^{\text{ELL}}$ for the candidate node. MAX-ELL-BALL-UB is run in parallel with MAXCOVARELL to get two candidate initial covariances $\Sigma_{0, \text{MC}}$ and $\Sigma_{0, \text{ours}}$. Algorithm \ref{alg:construct_brt_max_ell_prob_reach} then selects the distribution (and corresponding control policy) that maximizes $\lambda_{\min}(\Sigma_0)$, before running MAX-ELL-BALL to find the maximal radius $r$ associated with the chosen distribution and control policy.

 We assign $\Sigma = \Sigma^{\text{ELL}}$ at the root node, and then at each node generated by Algorithm \ref{alg:construct_brt_max_ell_prob_reach}, $\Sigma^{\text{ELL}} \preceq \Sigma$ and $\mathscr{P}$ is the same as the ellipsoid shape matrix found by the MAXELLIPSOID algorithm. Therefore, Algorithm \ref{alg:construct_brt_max_ell_prob_reach} achieves coverage equal to or better than that of the MAXELLIPSOID algorithm. This property is formally stated in Theorem \ref{thm: max_ellipsoid_dominance} and proved in Appendix \ref{sec:appendix}. 

\begin{theorem}\label{thm: max_ellipsoid_dominance} (Dominance over MAXELLIPSOID) h-BRS($\mathcal{T}^{(r)}_\text{OURS}$) $\supseteq$ h-BRS($\mathcal{T}^{(r)}_\text{MAXELLIPSOID}$) for all planning scenes.
\end{theorem}
\begin{algorithm}[t]
\caption{Constructing a BRT with MAX-ELL-BALL}
\label{alg:construct_brt_max_ell_prob_reach}
\begin{algorithmic}[1]
    \Require $\mathcal{G}$, $N$, $n_\mathrm{iter}$ %Input
    \Ensure  $\mathcal{T}$ %Output
    \State $\nu \gets \phi$, $\varepsilon \gets \phi$
    \State $ \nu_{0} \gets \operatorname{CREATENODE}(\mu_{\mathcal{G},c}, \Sigma_{\mathcal{G}}, \Sigma_{\mathcal{G}}, \mathscr{P}_\mathcal{G}, r_{\mathcal{G}}, \emptyset, \emptyset, \{\})$
    \State $\nu \gets \nu \cup \{\nu_{0}\}$
    \For{$i \gets 1 \textrm{ to } n_\mathrm{iter}$}
        \State $\nu_{k} \gets \operatorname{RAND}(\nu)$ %{\tt rand}
        \State $ \mu_{q} \gets \operatorname{RANDMEANAROUND}(\nu_{k}, r_{\textrm{sample}}) $
        \State $\textrm{status}_\text{ELL}, \mathscr{P}_0, \mathscr{C} \gets$ \\
        \hspace*{2em} $\operatorname{MAXELLIPSOID} (\mu_q, \Sigma_q, \mu_k, \Sigma^{\text{ELL}}_k, \mathscr{P}_k, N)$
        \If{$\textrm{status}_\text{ELL} = \textrm{infeasible}$}
            \State $\textrm{status}_\text{ELL}, \mathscr{P}_0, \mathscr{C}  \gets$ \\
            \hspace*{4em} $\operatorname{MAXELLIPSOID} (\mu_q, \Sigma_q, \mu_k, \Sigma_k, \mathscr{P}_k, N)$
        \EndIf
        \If{$\textrm{status}_\text{ELL} \neq \textrm{infeasible}$}
            \State $\textrm{status}_\text{COV-ELL}, \Sigma^{\text{ELL}} \gets$ \\
            \hspace*{4em} $\operatorname{MAXCOVARELL}(\mu_q, \mathscr{P}_0, \mu_k, \Sigma^{\text{ELL}}_k, \mathscr{P}_k, N)$
            \State $\textrm{status}_\text{ours}, \Sigma_{0, \text{ours}}, \mathscr{C}_\text{ours} \gets$ \\
            \hspace*{4em} $\operatorname{MAX-ELL-BALL-UB}(\mu_q, \mathscr{P}_0, \mu_k, \Sigma_k, \mathscr{P}_k, N)$
            \State $\textrm{status}_\text{MC}, \Sigma_{0, \text{MC}}, \mathscr{C}_\text{MC} \gets$ \\
            \hspace*{4em} $\operatorname{MAXCOVARELL}(\mu_q, \mathscr{P}_0, \mu_k, \Sigma_k, \mathscr{P}_k, N)$
            \State $\textrm{status} \gets \textrm{status}_\text{ours} \vee \textrm{status}_\text{MC}$
            \State $\textrm{outputs} \gets \{(\Sigma_{0, \text{MC}}, \mathscr{C}_\text{MC}), (\Sigma_{0, \text{ours}}, \mathscr{C}_\text{ours})\}$
            \If{$\textrm{status} \neq \textrm{infeasible}$}
                \State $\Sigma_0, \mathscr{C} \gets \arg \max_{\text{outputs}} \lambda_{\min}(\Sigma_0)$
                \State $r \gets \operatorname{MAX-ELL-BALL}($\\
                \hspace*{10em}$\mu_q, \Sigma_0, \mathscr{P}_0, \mathscr{C}, \mu_k, \Sigma_k, \mathscr{P}_k, N)$
                \State $ \textrm{idx},  \mathrm{ch}_{\mathrm{idx}} \gets \operatorname{size}(\mathcal{\nu}) + 1, \{\} $
                %\State $ \mathrm{ch}_{\mathrm{idx}} \gets \{\}$
                \If{$\textrm{status}_\text{COV-ELL} = \textrm{infeasible}$}
                    \State $\Sigma^{\text{ELL}} \gets \Sigma_0$
                \EndIf
                \State $ \nu_{\mathrm{new}} \gets \operatorname{CREATENODE}($\\
                \hspace*{10em}$\mu_{q}, \Sigma_0, \Sigma^{\text{ELL}}, \mathscr{P}_0, r, \textrm{idx}, k, \mathrm{ch}_{\mathrm{idx}})$
                \State $ \nu \gets \nu \bigcup \{ \nu_{\mathrm{new}} \} $, $ \mathrm{ch}_{k} \gets \mathrm{ch}_{k} \bigcup \{ \textrm{idx}\} $
                \State $ \varepsilon_{\textrm{idx},k} \gets \mathscr{C}$, $ \varepsilon \gets \varepsilon \bigcup \{ \varepsilon_{\textrm{idx},k} \} $
            \EndIf
        \EndIf
    \EndFor
    \State $  \mathcal{T} \gets \{ \nu, \varepsilon \} $
    \State $ \textrm{\textbf{Return}} \ \  \mathcal{T}$
\end{algorithmic}
\end{algorithm}
\section{Experiments}\label{sec: experiments}
We demonstrate the efficacy of our method via simulation experiments. First, we evaluate our method on multi-query motion planning for a double integrator in a field of obstacles. We also evaluate our method on motion planning for a quadrotor in a 2D plane.
\subsection{Double Integrator with Obstacles}
The double integrator dynamics are modeled by
\begin{align}
A &= \begin{bmatrix}
I_{2 \times 2} & \Delta t I_{2 \times 2} \\
0_{2 \times 2} & I_{2 \times 2}
\end{bmatrix},
B = \begin{bmatrix}
\Delta t^2/2 I_{2 \times 2} \\
\Delta t I_{2 \times 2}
\end{bmatrix}, \notag \\
DD^T &= W_c\begin{bmatrix} \frac13 \Delta t^3 & \frac13 \Delta t^3 & \frac12 \Delta t^2 & \frac12 \Delta t^2 \\
\frac13 \Delta t^3 & \frac13 \Delta t^3 & \frac12 \Delta t^2 & \frac12 \Delta t^2 \\
\frac12 \Delta t^2 & \frac12 \Delta t^2 & \Delta t &\Delta t \\
\frac12 \Delta t^2 & \frac12 \Delta t^2 & \Delta t &\Delta t \\
\end{bmatrix}, \notag
\end{align}
such that the process noise matrix $D$ models the accumulation of the velocity noise over each timestep. 
We use $\Delta t = 0.1$s, $W_c = 0.1^3$ m$^2$/s$^3$, and a planning horizon of $N = 20$ timesteps. The goal ambiguity set is given by $\mathcal{P}^{\text{BALL}}(\mu_\mathcal{G}, \mathscr{P}_\mathcal{G}, r_\mathcal{G})$, with $\mu_\mathcal{G} = [9, 5, 0, 0]^T$, $\mathscr{P}_\mathcal{G} = 0.1 I_{4\times 4}$ and $r_\mathcal{G} = \sqrt{0.1 f^{-1}(0.99, 4)}$.%, where $f^{-1}(\cdot, n)$ is the inverse cumulative distribution function of the $\chi^2$ distribution with $n$ degrees of freedom.

We evaluate RANDCOVAR (as in \cite{aggarwal2025tac}), MAXCOVAR \cite{aggarwal2024sdp}, MAX-COV-BALL, MAXELLIPSOID with fixed $\Sigma$ (using ellipsoid maximization only), MAXELLIPSOID \cite{aggarwal2025tac}, and MAX-ELL-BALL on the task of building a reusable backward reachable tree from the goal ambiguity set $\mathcal{P}_\mathcal{G}$. We run each method for 2000 iterations with a sampling radius of $[\pm 1.25, \pm 1.25, \pm 1.25, \pm 1.25]^T$. We implement each method in Python using \texttt{cvxpy} \cite{cvxpy}. %Our code can be found at: \todo{TODO: make sure Draper is OK with publishing code, then add link.}
We generate 100 different random trees for each method, each with a different random seed. We sample 250 Gaussian distributions with $\Sigma_0 = 0.01 I_{4 \times 4}$ in the obstacle-free space, and try to connect each distribution to each tree. Table \ref{tab: di_results} shows the fraction of distributions that can be connected to trees generated by each method.

The control input space is bounded such that $a^u\!\!=\!\! \{[\pm 1, 0]^T,  [0, \pm 1]^T\}$, with $b^u \!\!=\!\! \{\pm 10, \pm 10 \}$ and $\epsilon^u \!\!=\!\! 0.01$ for each control constraint. Chance constraint linearization is performed with $Y_r = 15 I_{2 \times 2}$ and $\mathscr{P}_r = \text{diag}([0.05, 0.05, 0.5. 0.5])$. We set $\Sigma_r \!\!=\!\! \text{diag}([0.01, 0.01, 0.75, 0.75])$ for MAXCOVAR, RANDCOVAR and MAX-COV-BALL and $\Sigma_r \!\!=\!\! \text{diag}([0.05, 0.05, 0.5, 0.5])$ for MAXELLIPSOID and MAX-ELL-BALL. This choice of $\Sigma_r$ significantly improves performance for MAXELLIPSOID and MAX-ELL-BALL. We use $\Sigma_q \!\!=\!\! \text{diag}([0.05, 0.05, 0.025, 0.025])$ for  ellipsoid maximization in MAXELLIPSOID and MAX-ELL-BALL.

\begin{figure}[tbp]
\centering
\includegraphics[width=0.485\columnwidth]{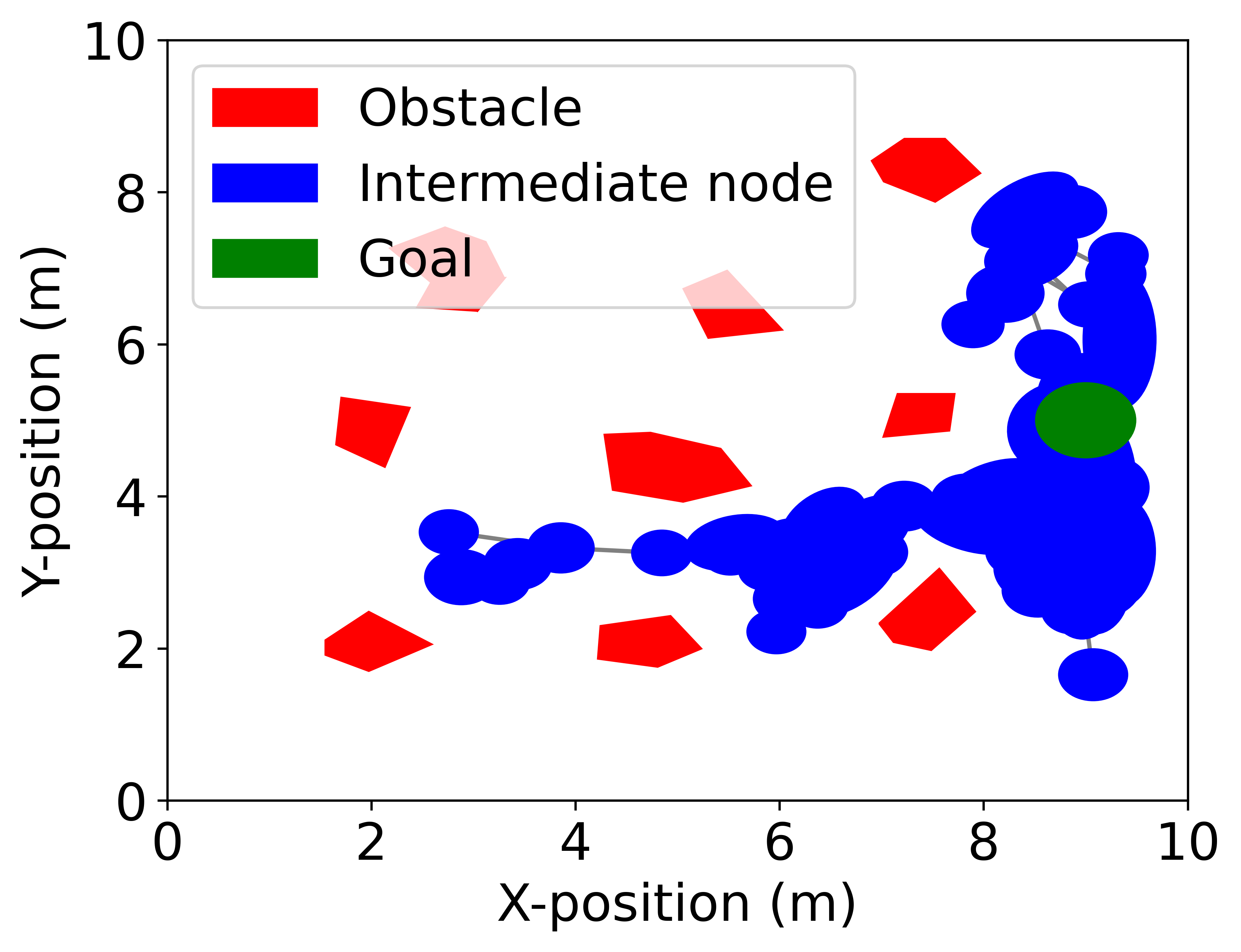}
\includegraphics[width=0.485\columnwidth]{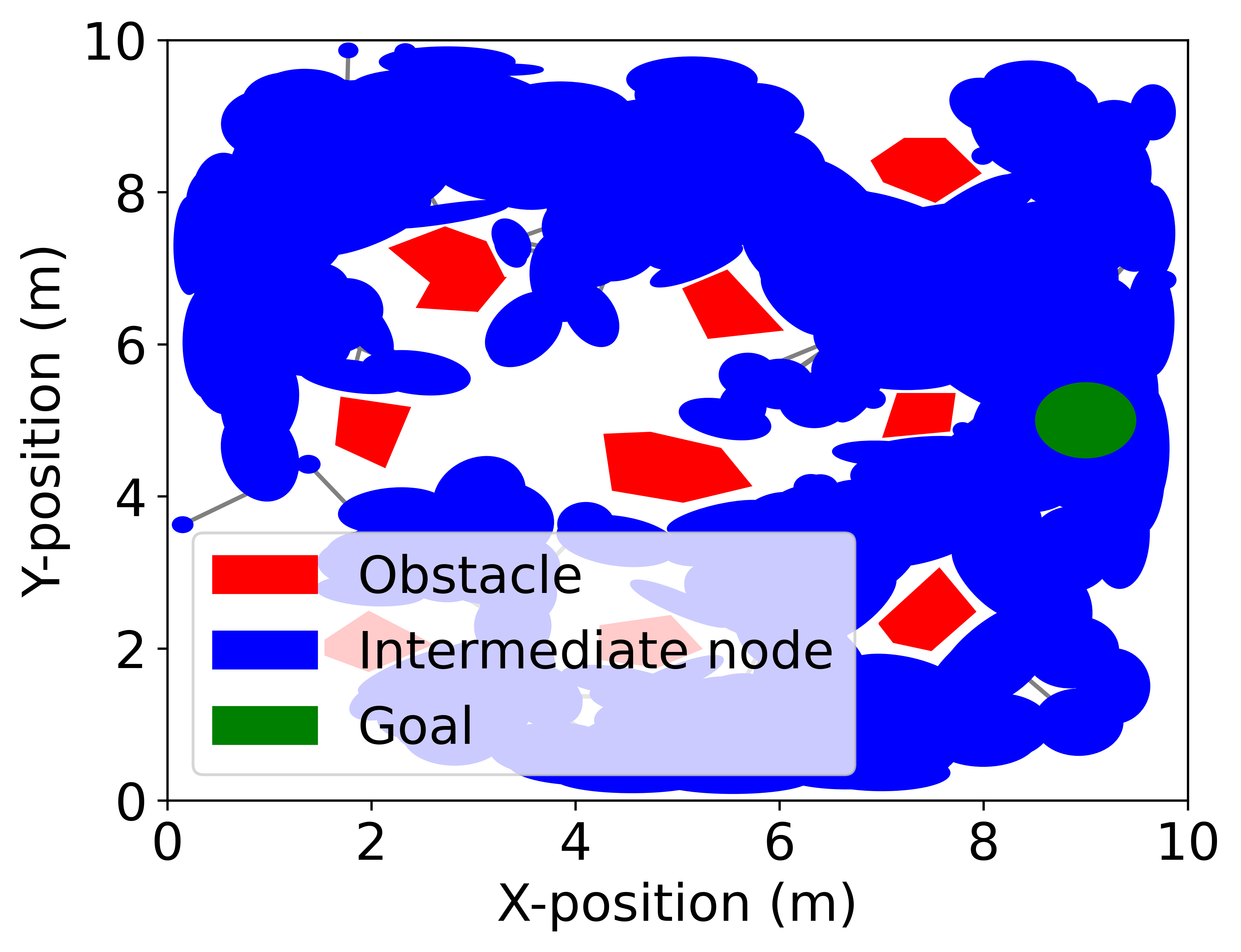}
\vspace*{-.15in}
\caption{Obstacles and goal region used in the double integrator planning experiment plotted against 99th percentile covariance contours for a backward reachable tree generated with MAX-ELL-BALL (left) and a backward reachable tree generated with MAX-COV-BALL (right).}
\label{fig: obstacle_config}
%\vspace{-3cm}
\end{figure}

The position space is bounded by $[0, 0]$, $[0, 10]$, $[10, 10]$, $[10, 0]$, and additional position-varying state chance constraints exist due to the presence of obstacles (see Figure \ref{fig: obstacle_config}). We decompose each obstacle into a union of convex sets. Then, when considering a candidate node center $\mu_q$, we project $\mu_q$ onto each convex obstacle and take the convex hull of the projected points as a safe set around $\mu_q$. All state chance constraints are enforced with $\epsilon^x = 0.01$. 

\begin{table}[t] 
\caption{Minimum, mean, median, 75th percentile, 90th percentile, and maximum for the fraction of initial distributions (out of 250) that can be connected to each tree.} \label{tab: di_results}
\begin{tabular}{|p{2.5cm}|c|c|c|c|c|c|}
\hline
\textbf{Method} & \textbf{Min} & \textbf{Mean} & \textbf{Med.} & \textbf{P75} & \textbf{P90} & \textbf{Max} \\
\hline
MAXELLIPSOID (fixed $\Sigma$) & 0.19 & 0.61 & 0.62 & 0.72 & 0.79 & 0.87 \\
MAXELLIPSOID & 0.24 & 0.60 & 0.60 & 0.68 & 0.80 & 0.90 \\
\textbf{MAX-ELL-BALL} & 0.26 & 0.67 & 0.66 & 0.79 & 0.85 & 0.92 \\
\hdashline
RANDCOVAR & 0.00 & 0.00 & 0.00 & 0.00 & 0.00 & 0.00 \\
MAXCOVAR & 0.42 & 0.89 & 0.90 & 0.94 & 0.96 & 0.99 \\
\textbf{MAX-COV-BALL} & \textbf{0.91} & \textbf{0.98} & \textbf{0.98} & \textbf{1.00} & \textbf{1.00} & \textbf{1.00} \\
\hline
\end{tabular}
%\vspace{-0.5cm}
\end{table}
We find that the methods that plan over ellipsoidal ambiguity sets of distributions (MAXELLIPSOID with fixed $\Sigma$, MAXELLIPSOID, and MAX-ELL-BALL) underperform MAX-COV-BALL and MAXCOVAR on this problem. This can be explained by the size of the second-order moments in each tree. When state constraints are tight, MAX-ELL-BALL and MAXELLIPSOID have a smaller maximal second-order moment associated with each node than MAX-COV-BALL and MAXCOVAR (see Figure \ref{fig: obstacle_config}), but have a larger set of first-order moments associated with each node. This means that MAX-ELL-BALL and MAXELLIPSOID sometimes fail to find edges, especially through narrow passages, because the second-order moment associated with the end node is too small for a feasible edge control to exist.

We see also that MAX-COV-BALL outperforms MAXCOVAR, and MAX-ELL-BALL outperforms MAXELLIPSOID. A Student's $t$-test indicates that these differences are statistically significant, with $p \ll 0.05$. Notably, this scenario includes process noise and tight state constraints, so this demonstrates that MAX-COV-BALL and MAX-ELL-BALL perform well empirically even in scenarios where the assumptions required for Lemmas \ref{lemma: no_process_noise} and \ref{lemma: mild_assumptions} do not hold. 

\subsection{Quadrotor with Varying Control Constraints and Process Noise}
The quadrotor state and control transition matrices are modeled by \cite{aggarwal2025tac}:
\begin{align}
A &= \begin{bmatrix}
I_{2 \times 2} & \Delta t I_{2 \times 2} & 0_{2 \times 2} \\
0_{2 \times 2} & I_{2 \times 2} & \Delta t I_{2 \times 2} \\
0_{2 \times 2} & 0_{2 \times 2} & I_{2 \times 2}
\end{bmatrix}, 
B = \begin{bmatrix} 
0_{2 \times 2} \\
0_{2 \times 2} \\
\Delta t I_{2 \times 2}
\end{bmatrix},
\end{align}
\begin{align}
DD^T &= W_c \begin{bmatrix}
        \frac{1}{20}\Delta t^5 & \frac{1}{20}\Delta t^5 &  \frac18 \Delta t^4 & \frac18 \Delta t^4 & \frac16 \Delta t^3 & \frac16 \Delta t^3 \\
        \frac{1}{20}\Delta t^5 & \frac{1}{20}\Delta t^5 &  \frac18 \Delta t^4 & \frac18 \Delta t^4 & \frac16 \Delta t^3 & \frac16 \Delta t^3 \\
        \frac18 \Delta t^4 & \frac18 \Delta t^4 & \frac13 \Delta t^3 & \frac13 \Delta t^3 & \frac12 \Delta t^2 & \frac12 \Delta t^2 \\
        \frac18 \Delta t^4 & \frac18 \Delta t^4 & \frac13 \Delta t^3 & \frac13 \Delta t^3 & \frac12 \Delta t^2 & \frac12 \Delta t^2 \\
        \frac16 \Delta t^3 & \frac16 \Delta t^3 & \frac12 \Delta t^2 & \frac12 \Delta t^2 & \Delta t &\Delta t \\
        \frac16 \Delta t^3 & \frac16 \Delta t^3 &  \frac12 \Delta t^2 & \frac12 \Delta t^2 & \Delta t &\Delta t \\
     \end{bmatrix}, \notag
\end{align}
such that the process noise matrix $D$ integrates the acceleration noise over each timestep. We use $\Delta t = 0.2$ sec and a planning horizon of $N = 20$ timesteps. We set $W_c = c^2 \Delta t$ m$^2$/s$^5$ and vary $c$ between 0.01 m/s$^3$ and 0.05 m$/s^3$. The goal ambiguity set is given by $\mathcal{P}^{\text{BALL}}(\mu_\mathcal{G}, \mathscr{P}_\mathcal{G}, r_\mathcal{G})$, with $\mu_\mathcal{G} = 0_{6 \times 1}$, $\mathscr{P}_\mathcal{G} = 0.5 I_{6 \times 6}$, and $r_\mathcal{G} = \sqrt{0.2 f^{-1}(0.99, 6)}$. The control input space is bounded such that $a^u = \{[\pm 1, 0]^T, [0, \pm 1]^T \}$ and $\epsilon^u = 0.01$ for each control chance constraint. We use $b^u = \{\pm u_{\max}, \pm u_{\max}\}$ for several values of $u_{\max}$. We use $\Sigma_r = 1.2 I_{6 \times 6}$, $\mathscr{P}_r = 1.2 I_{6 \times 6}$ and $Y_r = 0.5 I_{2 \times 2}$ for linearization. We use a sampling radius of $ \pm [5, 5, 2.5, 2.5, 0.675, 0.675]^T$.

We run MAXCOVAR and MAX-COV-BALL for 200 iterations, sampling the same node means for both methods. We then sample 250 distributions in a $20 \times 20$ bounding box and evaluate the number of distributions that can reach each tree. We average the number of distributions that can reach each tree over 100 random seeds.

\begin{figure}
\centering
\includegraphics[width=0.9\columnwidth]{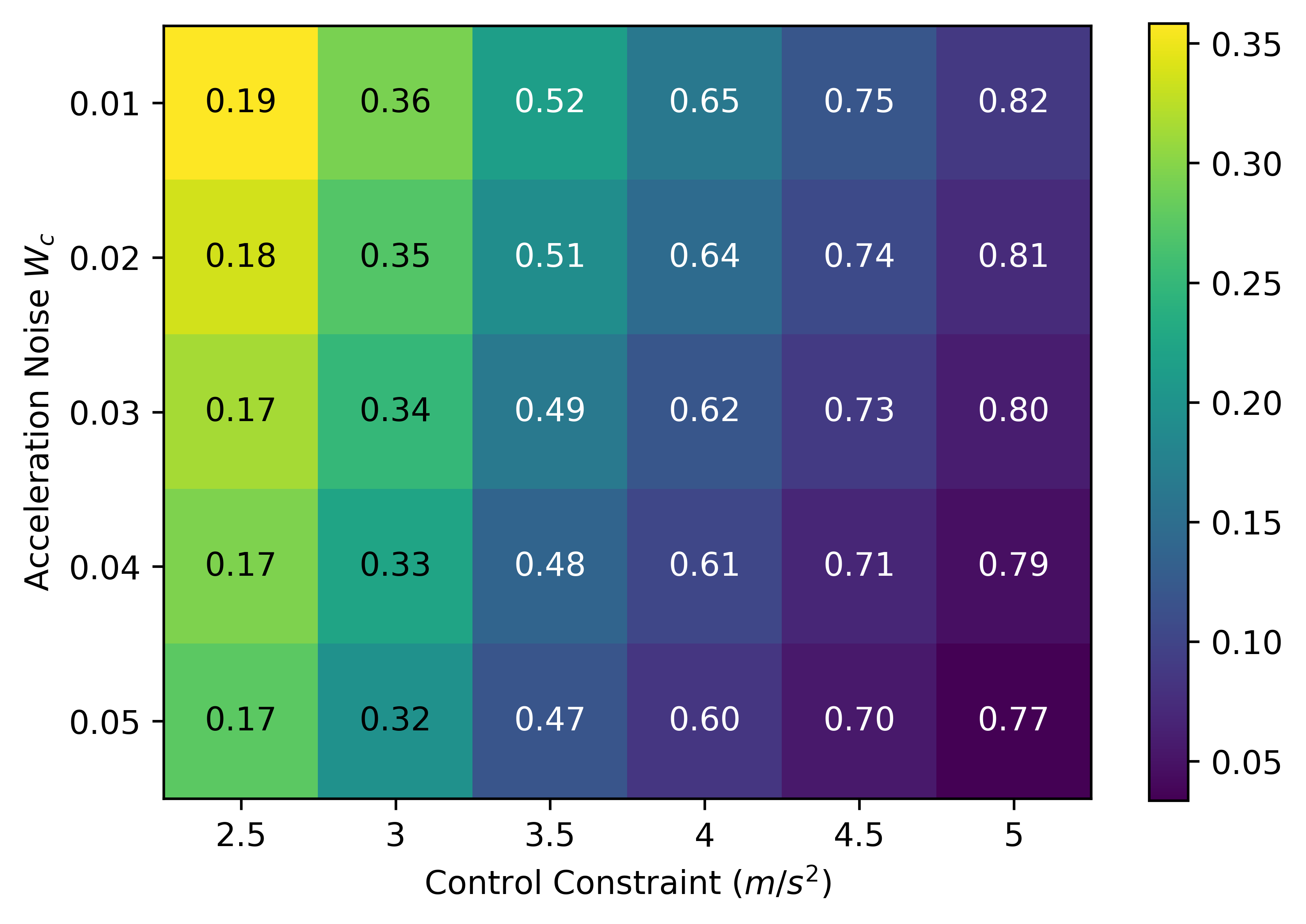}
\vspace{-0.3cm}
\caption{The color of each cell corresponds to the increase in coverage achieved by MAX-COV-BALL relative to MAXCOVAR, as a fraction of the coverage achieved by MAX-COV-BALL. The number in each cell corresponds to the fraction of initial distributions covered by MAX-COV-BALL (out of 250).}
\label{fig: heatmap}
\end{figure}

In Figure \ref{fig: heatmap}, we plot the gain in coverage achieved by MAX-COV-BALL with varying control constraints and varying levels of process noise. The number of each cell represents the mean fraction of randomly sampled initial distributions that can reach a tree generated by MAX-COV-BALL. The color of each cell is 1 minus the mean number of initial distributions that can reach a tree generated by MAXCOVAR divided by the number of such distributions that can reach a tree generated by MAX-COV-BALL. This gain in coverage is always greater than zero, indicating that MAX-COV-BALL achieves greater coverage than MAXCOVAR. The gain in coverage increases as the process noise decreases and as the control constraints tighten. The improvement in coverage achieved by MAX-COV-BALL as the process noise decreases is a result of the upper bound maximized by MAX-COV-BALL-UB tightening as the process noise decreases. Further, in the absence of state constraints, if the process noise is sufficiently small, the conditions required for Theorem \ref{thm: max_covar_dominance} to guarantee MAX-COV-BALL strictly achieving greater coverage than MAXCOVAR will be met. The relative improvement in coverage achieved by MAX-COV-BALL as the control constraints tighten can be explained by the fact that MAX-COV-BALL can often find nodes that MAXCOVAR cannot, but if the control constraints are very loose, both methods will find nodes at every iteration.

In Table \ref{tab: varying_control_results}, we fix $u_{\max} = 2.5 \text{ m}/\text{s}^2$ and $W_c = 0.03$ $s^{-3}$ and compare MAX-COV-BALL and MAXCOVAR to MAXELLIPSOID and MAX-ELL-BALL. We run MAXELLIPSOID and MAX-ELL-BALL for 50 iterations with the same parameters as MAXCOVAR and MAX-COV-BALL and with a query covariance $\Sigma_q = \text{diag}([0.1, 0.1, 0.05, 0.05, 0.0125, 0.0125])$. We also sample the same node means for MAXELLIPSOID and MAX-ELL-BALL for ease of comparison. We find that MAX-ELL-BALL performs the best, outperforming MAXELLIPSOID by about 3.5\% in the mean case and median case. We also find that MAX-ELL-BALL and MAXELLIPSOID outperform MAXCOVAR and MAX-COV-BALL in coverage, despite using many fewer iterations. This strong performance of MAXELLIPSOID and related methods in the absence of state constraints is consistent with our prior work \cite{aggarwal2025tac}.

\begin{table}[t]
\caption{Minimum, mean, median, 75th percentile, 90th percentile, and maximum for the fraction of initial distributions (out of 250) that can be connected to each tree.}  \label{tab: varying_control_results}
\begin{tabular}{|p{2.5cm}|c|c|c|c|c|c|}
\hline
\textbf{Method} & \textbf{Min} & \textbf{Mean} & \textbf{Med.} & \textbf{P75} & \textbf{P90} & \textbf{Max} \\
\hline
MAXCOVAR & 0.06 & 0.12 & 0.12 & 0.14 & 0.16 & 0.19 \\
\textbf{MAX-COV-BALL} & 0.10 & 0.17 & 0.17 & 0.20 & 0.22 & 0.28 \\
\hdashline
MAXELLIPSOID & 0.36 & 0.56 & 0.56 & 0.62 & 0.65 & 0.76 \\
\textbf{MAX-ELL-BALL} & \textbf{0.37} & \textbf{0.58} & \textbf{0.58} & \textbf{0.64} & \textbf{0.69} & \textbf{0.78} \\
\hline
\end{tabular}
\end{table}

\subsection{Results and Conclusion}
We demonstrate via simulation that MAX-COV-BALL outperforms MAXCOVAR and that MAX-ELL-BALL outperforms MAXELLIPSOID in a variety of scenarios, including a challenging scenario with obstacles and process noise. We find that MAX-COV-BALL outperforms all other methods in a scenario with obstacles, and that MAX-ELL-BALL outperforms all other methods when planning in an open space.

%\section{Conclusion}
%\label{sec:conclusion}
%\section{Future Work}

%\section*{Acknowledgment}
%This work was supported by the National Science
%Foundation Graduate Research Fellowship under grant no. 2141064. This work was also supported by the Draper Scholars program.
\appendix \label{sec:appendix}
\textbf{Proof of Lemma \ref{lemma: ball_invariance}}
First, suppose that $\text{Proj}_{(\cdot, \{\Sigma: \Sigma \succeq \lambda I \})}(\mathcal{P}^{\text{BALL}}(\mu_\mathcal{I}, r_\mathcal{I})) \xrightarrow[]{\mathscr{C}} \mathcal{P}^{\text{BALL}}(\mu_N, r_N)$, while satisfying all chance constraints with probability $\geq (1-\epsilon)$. 
Trivially, this also holds for $\text{Proj}_{(\cdot, \{\lambda I \})}(\mathcal{P}^{\text{BALL}}(\mu_\mathcal{I}, r_\mathcal{I}))$.

Now, suppose that $\text{Proj}_{(\cdot, \{\lambda I \})}(\mathcal{P}^{\text{BALL}}(\mu_\mathcal{I}, r_\mathcal{I})) \xrightarrow[]{\mathscr{C}} \mathcal{P}^{\text{BALL}}(\mu_N, r_N)$ (recalling that $\overline{\mathbf{x}}_N = \mu_N$) while satisfying all chance constraints with probability $\geq (1-\epsilon)$. Then, for any $\hat{\mu}_0 \in \text{Proj}_{(\cdot, \{\lambda I \})}(\mathcal{P}^{\text{BALL}}(\mu_\mathcal{I}, r_\mathcal{I}))$, applying the control policy $\mathscr{C}$ for $k$ timesteps to $\mathcal{N}(\hat{\mu}_0, \lambda I)$ results in  $\mathcal{N}(\hat{\mu}_k, \hat{\Sigma}_k)$ satisfying ${a_i^x}^T\hat{\mu}_k + \Phi^{-1}(1-\epsilon)\sqrt{{a_i^x}^T\hat{\Sigma}_ka_i^x} \leq b_i^x$ for any $a_i^x, b_i^x$ in the problem.

For any $\hat{\mu}_0$, the mean and covariance dynamics are
$
\hat{\mu}_{k+1} = A_k\hat{\mu}_k + B_k\overline{\mathbf{u}}_k + B_kK_k(\hat{\mu}_k - \overline{\mathbf{x}}_k)
$
and
$
\hat{\Sigma}_{k+1} = (A_k + B_kK_k)\hat{\Sigma}_k(A_k + B_kK_k)^T + DD^T,
$
with $
\overline{\mathbf{x}}_{k+1} = A_k \overline{\mathbf{x}}_k + B_k\overline{\mathbf{u}}_k.
$
and $\hat{\Sigma}_0 = \lambda I$.
We can define $\mathscr{A}^{\text{CL}}_{j,k} = \left(\prod_{t=j}^{k} (A_t + B_tK_t)\right)$. Without loss of generality, we can assume that $\overline{\mathbf{x}}_0 = \mu_\mathcal{I}$.
Then, based on the dynamics above, we have that
$
\hat{\mu}_k = \mathscr{A}^{\text{CL}}_{0, k-1}(\hat{\mu}_0 - \mu_\mathcal{I}) + \overline{\mathbf{x}}_k,$
and
$
\hat{\Sigma}_k = \mathscr{A}^{\text{CL}}_{0,k-1} \lambda I {\mathscr{A}^{\text{CL}}_{0,k-1}}^T + \sum_{r=1}^{k-1} \mathscr{A}^{\text{CL}}_{r,k-1} DD^T {\mathscr{A}^{\text{CL}}_{r,k-1}}^T.
$
We will define the contribution of the accumulated process noise to $\hat{\Sigma}_k$ as $
{\Sigma}_k^{(DD^T)} = \sum_{r=1}^{k} \mathscr{A}^{\text{CL}}_{r,k-1} DD^T {\mathscr{A}^{\text{CL}}_{r,k-1}}^T
$.
When calculating the maximum value of each chance constraint over all $\hat{\mu}_k$,
$
{a_i^x}^T \hat{\mu}_k = {a_i^x}^T \left(\overline{\mathbf{x}}_k + \mathscr{A}^{\text{CL}}_{0,k-1}(\hat{\mu}_0 - \mu_\mathcal{I}) \right),
$
so we can see that
$$
\max_{\mu_k} {a_i^x}^T \hat{\mu}_k = {a_i^x}^T\overline{\mathbf{x}}_k + \left|\left|{a_i^x}^T\mathscr{A}^{\text{CL}}_{0,k-1}\right|\right|_2r,
$$
where $r = \max_{\mu_0} ||\mu_0 - \mu_\mathcal{I}||_2$ for $(\mu_0, \lambda I) \in \mathcal{P}^{\text{BALL}}(\mu_\mathcal{I}, r_\mathcal{I})$, or equivalently $r = r_\mathcal{I} - \Phi^{-1}(1-\epsilon)\sqrt{\lambda}$.
It follows that
\begin{align}
{a_i^x}^T\overline{\mathbf{x}}_k + \left|\left|{a_i^x}^T\mathscr{A}^{\text{CL}}_{0,k-1}\right|\right|_2 r 
+ \Phi^{-1}(1-\epsilon)\sqrt{{a_i^x}^T\hat{\Sigma}_k {a_i^x}} \leq b_i^x. \notag
\end{align}
Now, suppose we choose an arbitrary $(\mu_0^*, \Sigma_0^*) \in \text{Proj}_{(\cdot, \{\Sigma: \Sigma \succeq \lambda I \})}(\mathcal{P}^{\text{BALL}}(\mu_\mathcal{I}, r_\mathcal{I}))$. Then,
$
\mu_k^* = \mathscr{A}^{\text{CL}}_{0,k-1}(\mu_0^* - \mu_\mathcal{I}) + \overline{\mathbf{x}}_k
$
and
$
\Sigma_k^* = \Sigma_k^{(DD^T)} + \mathscr{A}^{\text{CL}}_{0,k-1} \Sigma_0^* {\mathscr{A}^{\text{CL}}_{0,k-1}}^T.
$
Thus,
\begin{align}
\sqrt{{a_i^x}^T\mathscr{A}^{\text{CL}}_{0,k-1}\Sigma_0^* {\mathscr{A}^{\text{CL}}_{0,k-1}}^T{a_i^x} + {a_i^x}^T{\Sigma}_k^{(DD^T)}a_i^x}  = \sqrt{{a_i^x}^T \Sigma_k^* {a_i^x}} \notag
\end{align}
and similarly
\begin{align}
\sqrt{{a_i^x}^T\mathscr{A}^{\text{CL}}_{0,k-1}\lambda I{\mathscr{A}^{\text{CL}}_{0,k-1}}^T{a_i^x} + {a_i^x}^T{\Sigma}_k^{(DD^T)}a_i^x}  = \sqrt{{a_i^x}^T \hat{\Sigma}_k {a_i^x}}. \notag
\end{align}
It follows by properties of the square root that $\forall k, a_i^x$:
\begin{align}\label{eq: chance_constraint_diff}
&\sqrt{{a_i^x}^T \Sigma_k^* {a_i^x}} - \sqrt{{a_i^x}^T \hat{\Sigma}_k {a_i^x}} \notag \\
&\leq \sqrt{{a_i^x}^T\mathscr{A}^{\text{CL}}_{0,k-1}\Sigma_0^* {\mathscr{A}^{\text{CL}}_{0,k-1}}^T{a_i^x}} -\sqrt{{a_i^x}^T\mathscr{A}^{\text{CL}}_{0,k-1} \lambda I{\mathscr{A}^{\text{CL}}_{0,k-1}}^T{a_i^x}} \notag \\
&\leq \left| \left|{a_i^x}^T\mathscr{A}^{\text{CL}}_{0,k-1} \right| \right|_2 \left(\sqrt{\lambda_{\max}(\Sigma_0^*)} - \sqrt{\lambda}\right),
\end{align}
as long as $\Sigma_0^* \succeq \lambda I$, and similarly that $\forall k$ \begin{equation}\label{eq: eigval_diff}
\sqrt{\!\lambda_{\max}(\Sigma_k^*)} \!\!- \!\!\sqrt{\!\lambda_{\max}(\hat{\Sigma}_k)} \!\!\leq\!\! ||\mathscr{A}^{\text{CL}}_{0,k-1} ||_2 (\!\sqrt{\!\lambda_{\max}(\Sigma_0^*)} \!\!- \!\!\sqrt{\!\lambda}).
\end{equation}
Also,
$
{a_i^x}^T \mu_k^* \leq {a_i^x}^T\overline{\mathbf{x}}_k + \left|\left| {a_i^x}^T\mathscr{A}^{\text{CL}}_{0,k-1}\right|\right|_2 ||\mu_0^* - \mu_\mathcal{I}||_2,
$ so
\begin{align}
&{a_i^x}^T \mu_k^* + \Phi^{-1}\sqrt{{a_i^x}^T \Sigma_k^* {a_i^x}}  
\leq {a_i^x}^T\overline{\mathbf{x}}_k \notag \\
&+ \Phi^{-1}(1-\epsilon)\sqrt{{a_i^x}^T \hat{\Sigma}_k {a_i^x}} + \left|\left| {a_i^x}^T\mathscr{A}^{\text{CL}}_{0,k-1}\right|\right|_2 \Big(||\mu_0^* - \mu_\mathcal{I}||_2 \notag \\ &+ \Phi^{-1}(1-\epsilon)\sqrt{\lambda_{\max}(\Sigma_0^*)} - \Phi^{-1}(1-\epsilon)\sqrt{\lambda}\Big). \notag
\end{align}
Recalling that $\Phi^{-1}(1-\epsilon)\sqrt{\lambda_{\max}(\Sigma_0^*)} + ||\mu_0^* - \mu_\mathcal{I}||_2 \leq r_\mathcal{I}$ and that $r =  r_\mathcal{I} - \Phi^{-1}(1-\epsilon)\sqrt{\lambda}$, we can see that $\Phi^{-1}(1-\epsilon)\sqrt{\lambda_{\max}(\Sigma_0^*)} + ||\mu_0^* - \overline{\mathbf{x}}_0||_2 - \Phi^{-1}(1-\epsilon)\sqrt{\lambda} \leq r$. Then, all state chance constraints hold for any $(\mu_0^*, \Sigma_0^*) \in \text{Proj}_{(\cdot, \{\Sigma: \Sigma \succeq \lambda I \})}(\mathcal{P}^{\text{BALL}}(\mu_\mathcal{I}, r_\mathcal{I}))$.
By similar reasoning,
$$
{a_i^u}^T\overline{\mathbf{u}}_k + {a_i^u}^TK_k(\hat{\mu}_k - \overline{\mathbf{x}}_k) + \Phi^{-1}(1-\epsilon)\sqrt{{a_i^u}^TK_k\hat{\Sigma}_k K_k{a_i^u}} \leq b_i^u
$$
for all $\hat{\mu}_k, a_i^u, b_i^u$ implies control chance constraint satisfaction for any $(\mu_k^*, \Sigma_k^*)$ resulting from applying $\mathscr{C}$ to $(\mu_0^*, \Sigma_0^*) \in \text{Proj}_{(\cdot, \{\Sigma: \Sigma \succeq \lambda I \})}(\mathcal{P}^{\text{BALL}}(\mu_\mathcal{I}, r_\mathcal{I}))$ for $k$ timesteps. We can show this by simply substituting ${a_i^x}^T = {a_i^u}^TK_k$, $b_i^x = b_i^u - {a_i^u}^T\overline{\mathbf{u}}_k + {a_i^u}^TK_k\overline{\mathbf{x}}_k$ and then applying the proof above.

Finally, suppose that $(\hat{\mu}_N, \hat{\Sigma}_N) \in \mathcal{P}^\text{BALL}(\mu_N, r_N)$ for any $(\hat{\mu}_N, \hat{\Sigma}_N)$ resulting from applying $\mathscr{C}$ to $(\hat{\mu}_0, \hat{\Sigma}_0) \in \text{Proj}_{(\cdot, \{\lambda I \})}(\mathcal{P}^{\text{BALL}}(\mu_\mathcal{I}, r_\mathcal{I}))$ for $N$ timesteps. It follows that
$
\max_{\hat{\mu}_N} ||\hat{\mu}_N - \mu_N||_2 + \Phi^{-1}(1-\epsilon)\sqrt{\lambda_{\max}(\hat{\Sigma}_N)} \leq r_N.
$
Recalling that $\hat{\mu}_N = \mathcal{A}^\text{CL}_{0, N-1}(\hat{\mu}_0 - \mu_\mathcal{I}) + \overline{\mathbf{x}}_N$ and that $\mu_N = \overline{\mathbf{x}}_N$, we can rewrite this as
$
||\mathcal{A}^\text{CL}_{0, N-1}||_2 r + \Phi^{-1}(1-\epsilon)\sqrt{\lambda_{\max}(\hat{\Sigma}_N)} \leq r_N.
$
It follows from Eq. \eqref{eq: eigval_diff} that
$
||\mathcal{A}^\text{CL}_{0, N-1}||_2 r + \Phi^{-1}(1-\epsilon)\sqrt{\lambda_{\max}({\Sigma}_N^*)} - \Phi^{-1}(1-\epsilon)||\mathscr{A}^{\text{CL}}_{0,k-1} ||_2 (\sqrt{\lambda_{\max}(\Sigma_0^*)} - \sqrt{\lambda}) \leq r_N,
$ which simplifies to $||\mathcal{A}^\text{CL}_{0, N-1}||_2 (r_\mathcal{I} - \Phi^{-1}(1-\epsilon)\sqrt{\lambda_{\max}(\Sigma_0^*)}) + \Phi^{-1}(1-\epsilon)\sqrt{\lambda_{\max}({\Sigma}_N^*)} \leq r_N$. We can see that $\max_{\mu_N^*}||\mu_N^* - \mu_N||_2 = ||\mathcal{A}^\text{CL}_{0, N-1}||_2(r_\mathcal{I} - \Phi^{-1}(1-\epsilon)\sqrt{\lambda_{\max}(\Sigma_0^*)})$, and so it follows that $(\mu_N^*, \Sigma_N^*) \in \mathcal{P}^\text{BALL}(\mu_N, r_N)$.

\textbf{Proof of Lemma \ref{lemma: ball_invariance_no_process_noise}}
Suppose that $\text{Proj}_{(\cdot, \{\lambda I \})}(\mathcal{P}^{\text{BALL}}(\mu_\mathcal{I}, r_\mathcal{I})) \xrightarrow[]{\mathscr{C}} \mathcal{P}^{\text{BALL}}(\mu_N, r_N)$ (recalling that $\overline{\mathbf{x}}_N = \mu_N$) while satisfying all chance constraints with probability $\geq (1-\epsilon)$. In the absence of process noise, for any $\hat{\mu}_0$, the mean and covariance dynamics can be written as $\hat{\mu}_k = \mathcal{A}_{0, k-1}^\text{CL}(\hat{\mu}_0 - \mu_\mathcal{I}) + \overline{\mathbf{x}}_k$ and $\hat{\Sigma}_k = \mathcal{A}_{0, k-1}^\text{CL}\lambda I {\mathcal{A}_{0, k-1}^\text{CL}}^T$ (see the proof of Lemma \ref{lemma: ball_invariance}). Similarly, for any $(\mu_0^*, \Sigma_0^*) \in \mathcal{P}^{\text{BALL}}(\mu_\mathcal{I}, r_\mathcal{I})$, the mean and covariance dynamics can be written as $\mu_k^* = \mathcal{A}_{0, k-1}^\text{CL}(\mu_0^* - \mu_\mathcal{I}) + \overline{\mathbf{x}}_k$ and $\Sigma_k^* = \mathcal{A}_{0, k-1}^\text{CL}\Sigma_0^* {\mathcal{A}_{0, k-1}^\text{CL}}^T$. It follows that in the absence of proces noise, the inequalities in Eq. \ref{eq: chance_constraint_diff} and Eq. \ref{eq: eigval_diff} hold for any $\Sigma_0^*$, even if $\Sigma_0^* \not\succeq \lambda I$, and so it follows from the proof of Lemma \ref{lemma: ball_invariance} that $\mathcal{P}^{\text{BALL}}(\mu_\mathcal{I}, r_\mathcal{I}) \xrightarrow[]{\mathscr{C}} \mathcal{P}^{\text{BALL}}(\mu_N, r_N) \subseteq \mathcal{G}$.

Now, suppose that $\mathcal{P}^{\text{BALL}}(\mu_\mathcal{I}, r_\mathcal{I}) \xrightarrow[]{\mathscr{C}} \mathcal{G}$. It follows that $\mathcal{N}(\mu_\mathcal{I}, \frac{r_\mathcal{I}^2}{\Phi^{-1}(1-\epsilon)^2}) \xrightarrow{\mathscr{C}} \mathcal{N}(\mu_N, \Sigma_N)$ such that $(\mu_N, \Sigma_N) \in \mathcal{G}$. Recall that $\Sigma_N = \mathcal{A}_{0, N-1}^\text{CL}\frac{r_\mathcal{I}^2}{\Phi^{-1}(1-\epsilon)^2} {\mathcal{A}_{0, N-1}^\text{CL}}^T$, $\sqrt{\lambda_{\max}(\Sigma_N)} = ||\mathcal{A}_{0, N-1}^\text{CL}||_2\frac{r_\mathcal{I}}{\Phi^{-1}(1-\epsilon)}$ and therefore, $(\mu_N, \Sigma_N) \in \mathcal{P}^{\text{BALL}}(\mu_N, r_N)$ with $r_N = ||\mathcal{A}_{0, N-1}^\text{CL}||_2 r_\mathcal{I}$. Supposing that $\mathcal{G} = \mathcal{P}^{\text{BALL}, \chi^2}(\mu_\mathcal{G}, r_\mathcal{G})$, we must have $||\mu_N - \mu_\mathcal{G}||_2 + \sqrt{f^{-1}(1-\epsilon, n)}||\mathcal{A}_{0, N-1}^\text{CL}||_2\frac{r_\mathcal{I}}{\Phi^{-1}(1-\epsilon)} \leq r_\mathcal{G}$. Because $\sqrt{f^{-1}(1-\epsilon, n)} \geq \Phi^{-1}(1-\epsilon)$ for all $\epsilon, n$, it follows that $\mathcal{P}^{\text{BALL}}(\mu_N, r_N) \subseteq \mathcal{G}$. Similarly, if $\mathcal{G} = \mathcal{P}^{\text{BALL}}(\mu_\mathcal{G}, r_\mathcal{G})$, we must have $||\mu_N - \mu_\mathcal{G}||_2 + r_N \leq r_\mathcal{G}$ and $\mathcal{P}^{\text{BALL}}(\mu_N, r_N) \subseteq \mathcal{G}$. By the forward direction of this proof, it follows that $\mathcal{P}^{\text{BALL}}(\mu_\mathcal{I}, r_\mathcal{I}) \xrightarrow[]{\mathscr{C}} \mathcal{P}^{\text{BALL}}(\mu_N, r_N) \subseteq \mathcal{G}$ and then that $\text{Proj}_{(\cdot, \{\lambda I \})}(\mathcal{P}^{\text{BALL}}(\mu_\mathcal{I}, r_\mathcal{I})) \xrightarrow[]{\mathscr{C}} \mathcal{P}^{\text{BALL}}(\mu_N, r_N)$ for any $\lambda I$.

\textbf{Proof of Lemma \ref{lemma: maximal_brs_subset}}
Consider $\mathcal{P}^\text{BALL}(\mu_\mathcal{I}, r_\mathcal{I})$ such that $r_\mathcal{I} = \max_{r_0, \hat{\Sigma}_0} r_0 + \Phi^{-1}(1-\epsilon)\sqrt{\lambda_{\max}(\hat{\Sigma}_0)}$, where $\text{Proj}_{(\cdot, \{\hat{\Sigma}_0\}}(\mathcal{P}^\text{ANY}(\mu_\mathcal{I}, \cdot)) = \mathbb{B}_n(\mu_\mathcal{I}, r_0)$ for some $\mathscr{C}', \mathcal{P}^\text{ANY}(\mu_\mathcal{I}, \cdot)$  such that $\mathcal{P}^\text{ANY}(\mu_\mathcal{I}, \cdot) \xrightarrow{\mathscr{C}'} \mathcal{P}^\text{ANY}(\mu_N, \cdot) \subseteq \mathcal{G}$ (with $\overline{\mathbf{x}}_N = \mu_N$). It follows that $\text{Proj}_{(\cdot, \{\Sigma_0\}}(\mathcal{P}^\text{BALL}(\mu_\mathcal{I}, r_\mathcal{I})) \supseteq \text{Proj}_{(\cdot, \{\Sigma_0\}}(\mathcal{P}^\text{ANY}(\mu_\mathcal{I}, \cdot))$ for any $\Sigma_0 \in \mathbb{S}^n_+$, $\mathcal{P}^\text{ANY}(\mu_\mathcal{I}, \cdot) \xrightarrow{\mathscr{C}'} \mathcal{P}^\text{ANY}(\mu_N, \cdot) \subseteq \mathcal{G}$.
Recall from the definition of $\mathcal{P}^\text{ANY}$ that for any $\hat{\Sigma} \in \mathbb{S}^n_+$, $\hat{\Sigma} = QDQ^T$ for some $Q \in \mathrm{SO}(n), D \in \mathbf{D}^n$, $\text{Proj}_{(\cdot, \{VDV^T: V \in \textrm{SO}(n)\}}(\mathcal{P}^\text{ANY}(\mu_\mathcal{I}, \cdot)) = \text{Proj}_{(\cdot, \{\hat{\Sigma}\}}(\mathcal{P}^\text{ANY}(\mu_\mathcal{I}, \cdot)) = \mathbb{B}_n(\mu_\mathcal{I}, \cdot)$. In the absence of process noise, for any $(\hat{\mu}_0, \hat{\Sigma}_0) \in \text{Proj}_{(\cdot, \{VDV^T: V \in \textrm{SO}(n)\}}(\mathcal{P}^\text{ANY}(\mu_\mathcal{I}, \cdot)) = \mathbb{B}_n(\mu_\mathcal{I}, r_0)$, the mean and covariance dynamics are given by $\hat{\mu}_k = \mathcal{A}_{0, k-1}^\text{CL}(\hat{\mu}_0 - \mu_\mathcal{I}) + \overline{\mathbf{x}}_k$ and $\hat{\Sigma}_k = \mathcal{A}_{0, k-1}^\text{CL}\hat{\Sigma}_0{\mathcal{A}_{0, k-1}^\text{CL}}^T$ (see the proof of Lemma \ref{lemma: ball_invariance}). Similarly, $(\mu_\mathcal{I}, \frac{r_\mathcal{I}^2}{\Phi^{-1}(1-\epsilon)^2}I) \to (\mu_k, \Sigma_k) = (\overline{\mathbf{x}}_k,  \mathcal{A}_{0, k-1}^\text{CL}\frac{r_\mathcal{I}^2}{\Phi^{-1}(1-\epsilon)^2}I{\mathcal{A}_{0, k-1}^\text{CL}}^T)$.

Without loss of generality, we can represent each state or control chance constraint as $a^T\mu_k + \Phi^{-1}(1-\epsilon)\sqrt{a^T\Sigma_k a} \leq b$. The maximum value of the constraint over $\hat{\mu}_k, \hat{\Sigma}_k$ must be less than $b$. We see then that
$
\max_{\hat{\mu}_k, \hat{\Sigma}_k} a^T\hat{\mu}_k + \Phi^{-1}(1-\epsilon)\sqrt{{a}^T\hat{\Sigma}_k a}
$ is the sum of $\max_{\hat{\mu}_k} {a}^T\hat{\mu}_k = a^T\overline{\mathbf{x}}_k + ||a^T\mathcal{A}_{0, k-1}^\text{CL}||_2 r_0$ and $\max_{\hat{\Sigma}_k} \Phi^{-1}(1-\epsilon)\sqrt{a^T\hat{\Sigma}_k a} = \max_{V \in \mathrm{SO}(n)} \Phi^{-1}(1-\epsilon)\sqrt{{a}^T\mathcal{A}_{0, k-1}^\text{CL}VDV^T{\mathcal{A}_{0, k-1}^\text{CL}}^T a} = \Phi^{-1}(1-\epsilon)||{a}^T\mathcal{A}_{0, k-1}^\text{CL}||_2\sqrt{\hat{\Sigma}_0}$. It follows that $a^T\overline{\mathbf{x}}_k + ||a^T\mathcal{A}_{0, k-1}^\text{CL}||_2r_\mathcal{I} \leq b$, so the constraint holds for $\mu_k, \Sigma_k$.

Recalling that $\forall \hat{\mu}_N, \hat{\Sigma}_N$, $(\hat{\mu}_N, \hat{\Sigma}_N) \in \mathcal{P}^\text{ANY}(\mu_N, \cdot)$, we must have $\text{Proj}_{(\cdot, \hat{\Sigma}_N)}(\mathcal{P}^\text{ANY}(\mu_N, \cdot)) = \mathbb{B}_n(\mu_N, r_N)$ $\forall \hat{\Sigma}_N$. We have also that $\max_{\hat{\mu}_N}||\hat{\mu}_N - \mu_N||_2 = \max_{\hat{\mu}_N}||\mathcal{A}_{0, N-1}^\text{CL}(\hat{\mu}_0 - \mu_\mathcal{I})||_2 = ||\mathcal{A}_{0, N-1}^\text{CL}||_2r_0$, so $\forall \hat{\Sigma}_N$, $r_N \geq ||\mathcal{A}_{0, N-1}^\text{CL}||_2r_0$. Also, $\max_{\hat{\Sigma}_N} \lambda_{\max}(\hat{\Sigma}_N) = ||\mathcal{A}_{0, k-1}^\text{CL}||_2\sqrt{\lambda_{\max}(\hat{\Sigma}_0)}$. Supposing $\mathcal{P}^\text{ANY}(\mu_N, \cdot) \subseteq \mathcal{G}$, $\mathcal{G} := \mathcal{P}^{\text{BALL}, \chi^2}(\mu_\mathcal{G}, r_\mathcal{G})$, we must have $||\mu_N - \mu_\mathcal{G}||_2 + ||\mathcal{A}_{0, k-1}^\text{CL}||_2r_0 + \sqrt{f^{-1}(1-\epsilon, n)}||\mathcal{A}_{0, k-1}^\text{CL}||_2\sqrt{\lambda_{\max}(\hat{\Sigma}_0)} \leq r_\mathcal{G}$. It follows that $\mathcal{P}^{\text{BALL}}(\mu_N, ||\mathcal{A}_{0, k-1}^\text{CL}||_2r_\mathcal{I}) \subseteq \mathcal{G}$. This similarly holds when $\mathcal{G} := \mathcal{P}^{\text{BALL}}(\mu_\mathcal{G}, r_\mathcal{G})$. So, $\text{Proj}_{(\cdot, \{\frac{r_\mathcal{I}^2}{\Phi^{-1}(1-\epsilon)^2}I\})}(\mathcal{P}^\text{BALL}(\mu_\mathcal{I}, r_\mathcal{I})) \xrightarrow{\mathscr{C}'} \mathcal{P}^{\text{BALL}}(\mu_N, ||\mathcal{A}_{0, k-1}^\text{CL}||_2r_\mathcal{I}) \subseteq \mathcal{G}$. It follows from Lemma \ref{lemma: ball_invariance_no_process_noise} that $\mathcal{P}^\text{BALL}(\mu_\mathcal{I}, r_\mathcal{I}) \xrightarrow{\mathscr{C}'} \subseteq \mathcal{G}$. 

\textbf{Proof of Lemma \ref{lemma: no_process_noise}}
Because $\mathcal{G}^- \subset \mathcal{G}$, $(\mu, \Sigma) \in h-\text{BRS}(\mathcal{G}^-) \implies (\mu, \Sigma) \in h-\text{BRS}(\mathcal{G})$. Equivalently, $h$-BRS$(\mathcal{G}^-) \subseteq h-\text{BRS}(\mathcal{G})$. Recall that there are no state constraints, and that there is no process noise. Recall also that $\text{Proj}_{(\cdot, \mathbb{S}^n_+)}(\mathcal{G}^-) \subset \text{Proj}_{(\cdot, \mathbb{S}^n_+)}(\mathcal{G})$ and that $\mathcal{G}, \mathcal{G}^-$ have ball-shaped projections. Then, $\exists c, r$ such that $\text{Proj}_{(\cdot, \mathbb{S}^n_+)}(\mathcal{G}^-) \subseteq \mathbb{B}_n(c, r)$ and that $\mathbb{B}_n(c, r+\epsilon) \subseteq \text{Proj}_{(\cdot, \mathbb{S}^n_+)}(\mathcal{G})$, such that $\epsilon > 0$.

Consider the $h-$BRS, or backward reachable set via $hN$ steps, of a ball of point distributions $\mathbb{B}_n(c, r)$, such that $(\mu_\mathcal{G}, 0) \in \mathbb{B}_n(c, r)$ iff $||\mu_\mathcal{G} - c||_2 \leq r$. This $h-$BRS can be defined as the set of distributions $(\mu_0, \Sigma_0)$ such that $\mu_{k+1} = A\mu_k + B\overline{\mathbf{u}}_k$, $\Sigma_{k+1} = (A + BK)\Sigma_k(A + BK)^T$, $||\mu_{hN} - c||_2 \leq r$, $\Sigma_{hN} = 0$, and such that the chance constraints are satisfied at each timestep. Supposing that $\Sigma_0 = 0$ and that $K_k = 0 \forall k$ and that there are no state constraints, this reduces to the set of distributions $(\mu_0, 0)$ such that $\mu_{k+1} = A\mu_k + B\overline{\mathbf{u}}_k$, ${a_i^u}^T\overline{\mathbf{u}}_k \leq b_i^u$ for all $i, k$, and $||\mu_{hN} - c||_2 \leq r$. This is a convex set in $\mathbb{R}^n$.

Because $h-$BRS($\mathbb{B}_n(c, r)$) is convex and nonempty and $||\mu_0 - c||_2$ is continuous, by the Weierstrass extreme value theorem, there must exist $(\mu_0', 0) \in h-\text{BRS}(\mathbb{B}_n(c, r))$ which maximizes $||\mu_0 - c||_2$. It follows that $\nexists (\mu_0, 0) \in h-\text{BRS}(\mathbb{B}_n(c, r))$ such that $||\mu_0 - c||_2 > ||\mu_0' - c||_2$ and that there exists a control policy $\mathscr{C}$ which drives $(\mu_0', 0)$ to $(\mu_{hN}', 0)$ for some $\mu_{hN}' \in \mathbb{B}_n(c, r)$. Without loss of generality, we will assume that $\overline{\mathbf{x}}_0 = \mu_0'$ in $\mathscr{C}$, which implies also that $\overline{\mathbf{x}}_{k} = \mu_{k}'$ for all $k$.

Recall from the proof of Lemma \ref{lemma: ball_invariance} that when $\mathscr{C}$ is applied to any $(\mu_0, 0)$, $\mu_k = \mathscr{A}^{\text{CL}}_{0, k-1}(\mu_0 - \mu_0') + \mu'_{k}$ for all $k$. It follows that under $\mathscr{C}$, $(\mu_0, 0) \to (\mu_{hN}, 0)$, with $||\mu_{hN} - \mu_{hN}'||_2 \leq ||\mathscr{A}^{\text{CL}}_{0, hN-1}||_2||\mu_0 - \mu_0'||_2$, and also with $||\mu_k - \mu_k'||_2 \leq ||\mathscr{A}^{\text{CL}}_{0, k-1}||_2||\mu_0 - \mu_0'||_2$ $\forall k=0,\ldots,hN$.

Consider the Euclidean ball $\mathbb{B}_n(\mu_0', \frac{\epsilon}{||\mathscr{A}^{\text{CL}}_{0, hN-1}||_2})$. For any $\mu_0 \in \mathbb{B}_n(\mu_0', \frac{\epsilon}{||\mathscr{A}^{\text{CL}}_{0, hN-1}||_2})$, $(\mu_0, 0) \to (\mu_{hN}, 0)$ under the control policy above, with $||\mu_{hN} - \mu_{hN}'||_2 \leq \epsilon$. Recall that $\mu_{hN}' \in \mathbb{B}(c, r)$, so $||\mu_{hN}' - c||_2 \leq r$. It follows from the triangle inequality that $||\mu_{hN} - c||_2 \leq r + \epsilon$. Recall also that $\mathbb{B}_n(c, r + \epsilon) \subseteq \text{Proj}_{(\cdot, \mathbb{S}^n_+)}(\mathcal{G})$. It follows that $(\mu_{hN}, 0) \in \mathcal{G}$.

Because $\mathbb{B}_n(\mu_0', \frac{\epsilon}{||\mathscr{A}^{\text{CL}}_{0, hN-1}||_2}) \subset h-\text{BRS}(\mathcal{G})$ and $\exists \mu_0 \in \mathbb{B}_n(\mu_0', \frac{\epsilon}{||\mathscr{A}^{\text{CL}}_{0, hN-1}||_2})$ such that $||\mu_0 - c||_2 > ||\mu_0' - c||_2$, there must exist $\mu_0$ such that $(\mu_0, 0) \in h-\text{BRS}(\mathcal{G})$ and $(\mu_0, 0) \notin h-\text{BRS}(\mathbb{B}_n(c, r))$. For this initial distribution $(\mu_0, 0)$, no control policy exists such that $\mu_{hN} \in \mathbb{B}_n(c, r)$. Because $\mathbb{B}_n(c, r) \supseteq \text{Proj}_{(\cdot, \mathbb{S}^n_+)}(\mathcal{G}^-)$, this implies that $\nexists$ a control policy which drives $(\mu_0, 0)$ to $\mathcal{G}^-$. So, $(\mu_0, 0) \notin h-\text{BRS}(\mathcal{G}^-)$.

Then, because $\exists (\mu_0, 0) \in h-\text{BRS}(\mathcal{G})$ such that $(\mu_0, 0) \notin h-\text{BRS}(\mathcal{G}^-)$ and because $h-\text{BRS}(\mathcal{G}^-) \subseteq h-\text{BRS}(\mathcal{G})$, we can say strictly that $h-\text{BRS}(\mathcal{G}^-) \subset h-\text{BRS}(\mathcal{G})$.

\textbf{Proof of Lemma \ref{lemma: mild_assumptions}}
Because $\mathcal{P}^{\text{MC}}(\mu, \Sigma) \subseteq \mathcal{G}$, $(\mu, \Sigma) \in h-\text{BRS}(\mathcal{P}^{\text{MC}}(\mu, \Sigma)) \implies (\mu, \Sigma) \in h-\text{BRS}(\mathcal{G})$. Equivalently, $h-\text{BRS}(\mathcal{P}^{\text{MC}}(\mu, \Sigma)) \subseteq h-\text{BRS}(\mathcal{G})$. Now, we will show that $\exists (\mu, \Sigma) \in h-\text{BRS}(\mathcal{G})$ such that $(\mu, \Sigma) \notin h-\text{BRS}(\mathcal{P}^{\text{MC}}(\mu, \Sigma))$. Let us consider $\text{Proj}_{(\cdot, \mathbb{S}^n_+)}(h-\text{BRS}(\mathcal{P}^{\text{MC}}(\mu, \Sigma)))$,
which is a set in $\mathbb{R}^n$ bounded by equalities and non-strict inequalities, so it is closed and bounded and is therefore compact. It follows by the Weierstrass extreme value theorem that $||\mu_0 - \mu_\mathcal{G}||_2$ for $\mu_0 \in \text{Proj}_{(\cdot, \mathbb{S}^n_+)}(h-\text{BRS}(\mathcal{P}^{\text{MC}}(\mu, \Sigma)))$ is bounded from above and below, and also that $||\mu_0 - \mu_\mathcal{G}||_2$ attains its supremum and infimum (not necessarily uniquely) on $\text{Proj}_{(\cdot, \mathbb{S}^n_+)}(h-\text{BRS}(\mathcal{P}^{\text{MC}}(\mu, \Sigma)))$. Then, we can choose $\mu_0'$ such that $\mu_0' = \arg \sup_{\mu_0 \in \text{Proj}_{(\cdot, \mathbb{S}^n_+)}(h-\text{BRS}(\mathcal{P}^{\text{MC}}(\mu, \Sigma)))} ||\mu_0 - \mu_\mathcal{G}||_2.$

It follows that $\nexists \mu_0 \in \text{Proj}_{(\cdot, \mathbb{S}^n_+)}(h-\text{BRS}(\mathcal{P}^{\text{MC}}(\mu, \Sigma)))$ such that $||\mu_0 - \mu_\mathcal{G}||_2 > ||\mu_0' - \mu_\mathcal{G}||_2$ and that there exists a control policy $\mathscr{C}$ and some $\Sigma_0'$ under which $(\mu_0', \Sigma_0') \to (\mu_{hN}', \Sigma_{hN}')$, with $\mu_{hN}' = \mu$. Without loss of generality, we can define $\overline{\mathbf{x}}_0 = \mu_0'$ in $\mathscr{C}$. Recall from the proof of Lemma \ref{lemma: ball_invariance} that when $\mathscr{C}$ is applied to $(\mu_0, \Sigma_0')$ for any $\mu_0$, $\mu_k = \mathscr{A}_{0, k-1}^\text{CL}(\mu_0 - \mu_0') + \mu_k'$. Then, we can see that $||\mu_k - \mu_k'||_2 \leq ||\mathscr{A}_{0, k-1}^\text{CL}||_2||\mu_0 - \mu_0'||_2$ $\forall k=0,\ldots,hN$.
Recall that the state contraints are loose at each timestep, such that $\exists \delta_c$ such that ${a_i^x}^T\mu_k' + \Phi^{-1}(1-\epsilon)\sqrt{{a_i^x}^T\Sigma_k {a_i^x}} + \delta_c \leq b_i^x $ for all $k=0,\ldots,hN$. Recall also that $\Sigma_{hN}' \prec (r_\mathcal{G}/\Phi^{-1}(1-\epsilon)^2)$. It follows then that $\exists \delta_f$ such that $\Phi^{-1}(1-\epsilon)\sqrt{\lambda_{\max}(\Sigma_{hN}')} + \delta_f \leq r_\mathcal{G}$. Then, $\exists \delta_0$ such that $||{a_i^x}^T||_2||\mathscr{A}_{0, k-1}^\text{CL}||_2\delta_0 \leq \delta_c$ $\forall i, k$, and such that $||\mathscr{A}_{0, hN-1}^\text{CL}||_2\delta_0 \leq \delta_f$. Then, for any $\mu_0$ s.t. $||\mu_0 - \mu_0'||_2 \leq \delta_0$, the state constraints hold for any $\mu_k$ and $\Phi^{-1}(1-\epsilon)\sqrt{\lambda_{\max}(\Sigma_{hN}')} + ||\mu_{hN} - \mu_{hN}'||_2 \leq r_\mathcal{G}$. Because $\mu_{hN}' = \mu_\mathcal{G}$, equivalently $(\mu_{hN}, \Sigma_{hN}') \in \mathcal{G}$. It follows then that $(\mu_0, \Sigma_0') \in h-\text{BRS}(\mathcal{G})$.

Then, $\exists (\mu_0, \Sigma_0') \in h-\text{BRS}(\mathcal{G})$ such that $||\mu_0 - \mu_\mathcal{G}||_2 > ||\mu_0' - \mu_\mathcal{G}||_2$, so $\mu_0 \notin \text{Proj}_{(\cdot, \mathbb{S}^n_+)}(h-\text{BRS}(\mathcal{P}^{\text{MC}}(\mu, \Sigma)))$ and therefore $(\mu_0, \Sigma_0') \notin h-\text{BRS}(\mathcal{P}^{\text{MC}}(\mu, \Sigma))$. Then, we must have $h-\text{BRS}(\mathcal{G}) \subset h-\text{BRS}(\mathcal{P}^{\text{MC}}(\mu, \Sigma)))$. It follows that for $\hat{\mu_0} = \arg \max_{\text{Proj}_{(\cdot, \mathbb{S}^n_+)}(h-\text{BRS}(\mathcal{G}))} ||\mu_0 - \mu_\mathcal{G}||_2$, $\hat{\mu_0} \notin \text{Proj}_{(\cdot, \mathbb{S}^n_+)}(h-\text{BRS}(\mathcal{P}^{\text{MC}}(\mu, \Sigma)))$.

\textbf{Proof of Remark \ref{remark: w2_projections}}
Suppose $\mathcal{G}$ is the goal ambiguity set, with $\mathcal{G} := \{(\mu, \Sigma): \mathbb{P}(\mathbf{x} \in \mathbb{B}(\mu_\mathcal{G}, r_\mathcal{G})) \geq 1-\epsilon \text{ for } \mathbf{x} \sim \mathcal{N}(\mu, \Sigma)\}$. To maximize $r_{\chi^2}$ such that $\mathcal{P}^{\text{BALL}, \chi^2}(\mu_{\chi^2}, r_{\chi^2}) \subseteq \mathcal{G}$, $\mu_{\chi^2}, r_{\chi^2} = \mu_\mathcal{G}, r_\mathcal{G}$. Further, $\text{Proj}_{(\cdot, \mathbb{S}^n_+)}(\mathcal{P}^{\text{BALL}, \chi^2}(\mu_{\chi^2}, r_{\chi^2})) = \mathbb{B}_n(\mu_\mathcal{G}, r_\mathcal{G}) = \text{Proj}_{(\cdot, \mathbb{S}^n_+)}(\mathcal{G})$. By the properties of Mahalanobis distance, $\Sigma_{\max} = \frac{r_\mathcal{G}^2}{f^{-1}(1-\epsilon, n)}I$, and $\text{Proj}_{(\cdot, \{\Sigma_{\max}\})}(\mathcal{P}^{\text{BALL}, \chi^2}(\mu_{\chi^2}, r_{\chi^2})) = \{\mu_\mathcal{G}\} = \text{Proj}_{(\cdot, \{\Sigma_{\max}\})}(\mathcal{G})$.
Now suppose $\mathcal{P}^{\mathbb{W}_2}(\mu_W, \Sigma_W, r_W) \subseteq \mathcal{G}$. In order to maximize $r_W$, we must have $\mu_W = \mu_\mathcal{G}$, $\Sigma_W = 0$. Consider the distribution $(\mu_\mathcal{G}, \Sigma_c)$ with $\Sigma_c = \text{diag}([\frac{r_\mathcal{G}^2}{f^{-1}(1-\epsilon, 1)} + \delta, 0, \ldots, 0]))$. It follows that $\mathbb{W}_2((\mu_\mathcal{G}, 0); (\mu_\mathcal{G}, \Sigma_c)) = \text{tr}(\Sigma_c) = \frac{r_\mathcal{G}^2}{f^{-1}(1-\epsilon, 1)} + \delta$. By the properties of Mahalanobis distance, $(\mu_\mathcal{G}, \Sigma_c) \notin \mathcal{G}$ for any $\delta > 0$. Under the conditions stated in Remark \ref{remark: w2_projections}, $r_W \leq \frac{r_\mathcal{G}}{\sqrt{f^{-1}(1-\epsilon, 1)}} < \frac{r_\mathcal{G} \sqrt{n}}{\sqrt{f^{-1}(1-\epsilon, n)}} < r_\mathcal{G}$. It follows then that $\text{Proj}_{(\cdot, \mathbb{S}^n_+)}(\mathcal{P}^{\mathbb{W}_2}(\mu_W, \Sigma_W, r_W)) = \mathbb{B}_n(\mu_\mathcal{G}, r_W) \subset \text{Proj}_{(\cdot, \mathbb{S}^n_+)}(\mathcal{P}^{\text{BALL}, \chi^2}(\mu_{\chi^2}, r_{\chi^2}))$ and $\text{Proj}_{(\cdot, \{\Sigma_{\max}\})}(\mathcal{P}^{\mathbb{W}_2}(\mu_W, \Sigma_W, r_W)) = \emptyset$.

\textbf{Proof of Remark \ref{remark: w2_projections_phi}}
Suppose $\mathcal{G}$ is the goal ambiguity set, with $\mathcal{G} := \{(\mu, \Sigma): \mathbb{P}(\mathbf{x} \in \mathbb{B}(\mu_\mathcal{G}, r_\mathcal{G})) \geq 1-\epsilon \text{ for } \mathbf{x} \sim \mathcal{N}(\mu, \Sigma)\}$. To maximize $r$ such that $\mathcal{P}^{\text{BALL}}(\mu, r) \subseteq \mathcal{G}$, $\mu, r = \mu_\mathcal{G}, \frac{\Phi^{-1}(1-\epsilon)}{\sqrt{f^{-1}(1-\epsilon, n)}}r_\mathcal{G}$. Further, $\text{Proj}_{(\cdot, \mathbb{S}^n_+)}(\mathcal{P}^{\text{BALL}}(\mu, r)) = \mathbb{B}_n(\mu_\mathcal{G}, \frac{\Phi^{-1}(1-\epsilon)}{\sqrt{f^{-1}(1-\epsilon, n)}}r_\mathcal{G})$. By the properties of Mahalanobis distance, $\Sigma_{\max} = \frac{r_\mathcal{G}^2}{f^{-1}(1-\epsilon, n)}I$, and $\text{Proj}_{(\cdot, \{\Sigma_{\max}\})}(\mathcal{P}^{\text{BALL}}(\mu, r)) = \{\mu_\mathcal{G}\} = \text{Proj}_{(\cdot, \{\Sigma_{\max}\})}(\mathcal{G})$.

Now suppose $\mathcal{P}^{\mathbb{W}_2}(\mu_W, \Sigma_W, r_W) \subseteq \mathcal{G}$. In order to maximize $r_W$, we must have $\mu_W = \mu_\mathcal{G}$, $\Sigma_W = 0$. Recall also from the proof of Remark \ref{remark: w2_projections} that $r_W \leq \frac{r_\mathcal{G}}{\sqrt{f^{-1}(1-\epsilon, 1)}}$. Recall from the statement of Remark \ref{remark: w2_projections_phi} that $n \geq 2$, and $f^{-1}(1-\epsilon, n) \leq \Phi^{-1}(1-\epsilon)^2f^{-1}(1-\epsilon, 1)$. This implies also that $\epsilon \leq 0.2$. Therefore, we can see that $r_W \leq \frac{r_\mathcal{G}}{\sqrt{f^{-1}(1-\epsilon, 1)}} \leq \frac{r_\mathcal{G} \Phi^{-1}(1-\epsilon)}{\sqrt{f^{-1}(1-\epsilon, n)}}$ and also that $r_W \leq \frac{r_\mathcal{G}}{\sqrt{f^{-1}(1-\epsilon, 1)}} \leq \frac{r_\mathcal{G} \sqrt{n}}{\sqrt{f^{-1}(1-\epsilon, n)}}$. It follows then that $\text{Proj}_{(\cdot, \mathbb{S}^n_+)}(\mathcal{P}^{\mathbb{W}_2}(\mu_W, \Sigma_W, r_W)) \subset \text{Proj}_{(\cdot, \mathbb{S}^n_+)}(\mathcal{P}^{\text{BALL}}(\mu, r))$ and $\text{Proj}_{(\cdot, \{\Sigma_{\max}\})}(\mathcal{P}^{\mathbb{W}_2}(\mu_W, \Sigma_W, r_W)) = \emptyset$.

\textbf{Proof of Theorem \ref{thm: maximal_coverage}}
Recall the definition of an $h$-BRS of a tree of ambiguity sets of distributions from Definition \ref{def: h_BRS_tree}. Let $\mathcal{T}^{(n)}$ be a tree of ambiguity sets of distributions after $n$ iterations, such that each node $\nu$ in $\mathcal{T}^{(n)}$ is represented by an ambiguity set $\mathcal{P}^{\text{BALL}}(\mu, r)$. We will show that for all planning scenes, $h-\text{BRS}(\mathcal{T}^{(n+1)}_\text{OURS}) \supset h-\text{BRS}(\mathcal{T}^{(n+1)}_\text{ANY})$, where $\mathcal{T}^{(n+1)}_\text{OURS}$ and $\mathcal{T}^{(n+1)}_\text{ANY}$ are trees obtained from $\mathcal{T}^{(n)}$ by adding a node using Algorithm 1 and the ANY procedure, respectively.

At iteration $n+1$, an edge is constructed from an existing node $\nu$ in the tree $\mathcal{T}^{(n)}$ to a new node with central query mean $\mu_q$. By Lemma \ref{lemma: ball_invariance_no_process_noise}, in the absence of process noise, maximizing the upper bound on the radius $r$ of an initial ambiguity set (see Problem \ref{prob: ball_nlp_loose}) and maximizing the actual radius $r$ of the initial ambiguity set (see Problem \ref{prob: ball_nlp}) are equivalent, as the upper bound becomes tight. It follows that MAX-COV-BALL-UB reduces exactly to MAX-COV-BALL. Recall also that $\nu$ is represented by an ambiguity set $\mathcal{P}^{\text{BALL}}(\mu_\mathcal{G}, r_\mathcal{G})$. Then, MAX-COV-BALL-UB (and therefore Algorithm 1) yields an ambiguity set $\mathcal{P}^{\text{BALL}}(\mu_q, r_q)$ such that $r_q$ is maximal. 

It follows by Lemma \ref{lemma: maximal_brs_subset} that for $\mathcal{P}^{\text{ANY}}$ found by the ANY procedure by connecting $\nu$ to $\mu_q$, $\text{Proj}_{(\cdot, \{\hat{\Sigma}\})}(\mathcal{P}^{\text{ANY}}) \subseteq \text{Proj}_{(\cdot, \{\hat{\Sigma}\})}(\mathcal{P}^{\text{BALL}}(\mu_q, r_q))$ and $\text{Proj}_{(\cdot, \mathbb{S}^n_+)}(\mathcal{P}^{\text{ANY}}) \subseteq \text{Proj}_{(\cdot, \mathbb{S}^n_+)}(\mathcal{P}^{\text{BALL}}(\mu_q, r_q))$. Recalling that $\text{Proj}_{(\cdot, \mathbb{S}^n_+)}(\mathcal{P}^{\text{ANY}}) \neq \text{Proj}_{(\cdot, \mathbb{S}^n_+)}(\mathcal{P}^{\text{BALL}}(\mu_q, r_q))$ (by the definition of ANY), we can say then strictly that $\text{Proj}_{(\cdot, \mathbb{S}^n_+)}(\mathcal{P}^{\text{ANY}}) \subset \text{Proj}_{(\cdot, \mathbb{S}^n_+)}(\mathcal{P}^{\text{BALL}}(\mu_q, r_q))$. It follows then by Lemma \ref{lemma: no_process_noise} that $t-\text{BRS}(\mathcal{P}^{\text{ANY}}) \subset t-\text{BRS}(\mathcal{P}^{\text{BALL}}(\mu_q, r_q))$ $\forall t \geq 1$. Therefore, $(h-d^{(n+1)})-BRS(\mathcal{P}^{\text{BALL}}(\mu_q, r_q)) \supset (h-d^{(n+1)})-BRS(\mathcal{P}^\text{ANY})$ $\forall h > d^{(n+1})$, where $d^{(n+1)}$ is the number of hops from the root node to $\mathcal{\nu}$.

From Definition \ref{def: h_BRS_tree}, 
$
h-\text{BRS}(\mathcal{T}^{(n+1)}_\text{OURS}) = h-\text{BRS}(\mathcal{T}^{(n)}) \bigcup (h-d^{(n+1)})-BRS(\mathcal{P}^{\text{BALL}}(\mu_q, r_q))
$
and
$
h-\text{BRS}(\mathcal{T}^{(n+1)}_\text{ANY}) = h-\text{BRS}(\mathcal{T}^{(n)}) \bigcup (h-d^{(n+1)})-BRS(\mathcal{P}^\text{ANY}).
$
Recalling that $(h-d^{(n+1)})-BRS(\mathcal{P}^{\text{BALL}}(\mu_q, r_q)) \supset (h-d^{(n+1)})-BRS(\mathcal{P}^\text{ANY})$ $\forall h > d^{(n+1})$, we will assume that $\exists (\mu, \Sigma) \in (h-d^{(n+1)})-BRS(\mathcal{P}^{\text{BALL}}(\mu_q, r_q))$ such that $(\mu, \Sigma) \notin (h-d^{(n+1)})-BRS(\mathcal{P}^\text{ANY})$ and $(\mu, \Sigma) \notin h-\text{BRS}(\mathcal{T}^{(n)})$. In general, it is possible to sample $\mu_q$ such that this statement is guaranteed to hold. It follows then that when starting from a common tree state $\mathcal{T}^{(n)}$, $h-\text{BRS}(\mathcal{T}^{(n+1)}_\text{OURS}) \supset h-\text{BRS}(\mathcal{T}^{(n+1)}_\text{ANY})$.

Suppose then that $n = 0$, such that $\mathcal{T}^{(0)} = \{\mathcal{P}^{\text{BALL}}(\mu_\mathcal{G}, r_\mathcal{G})\}$. It follows from the argument above that $h-\text{BRS}(\mathcal{T}^{(1)}_\text{OURS}) \supset h-\text{BRS}(\mathcal{T}^{(1)}_\text{ANY})$. It follows then by induction that for any $n$, $h-\text{BRS}(\mathcal{T}^{(n)}_\text{OURS}) \supset h-\text{BRS}(\mathcal{T}^{(n)}_\text{ANY})$.

\textbf{Proof of Theorem \ref{thm: max_covar_dominance}}
%First, we will show that $h-\text{BRS}(\mathcal{T}_{OURS}^{(r)}) \supseteq h-\text{BRS}(\mathcal{T}_\text{MAXCOVAR}^{(r)})$ for all planning scenes. 
Let $\mathcal{T}^{(n)}$ be a tree of ambiguity sets of distributions after $n$ iterations, such that each node $\nu$ in $\mathcal{T}^{(n)}$ is represented by an ambiguity set $\mathcal{P}^{\text{BALL} \cup \text{MC}}(\mu, \Sigma, r) := \mathcal{P}^{\text{BALL}}(\mu, r) \cup \mathcal{P}^{\text{MC}}(\mu, \Sigma)$. We will show that for all planning scenes, $h-\text{BRS}(\mathcal{T}^{(n+1)}_\text{OURS}) \supseteq h-\text{BRS}(\mathcal{T}^{(n+1)}_\text{MAXCOVAR})$, where $\mathcal{T}^{(n+1)}_\text{OURS}$ and $\mathcal{T}^{(n+1)}_\text{MAXCOVAR}$ are trees obtained from $\mathcal{T}^{(n)}$ by adding a node using Algorithm \ref{alg:construct_brt_max_prob_reach} and the MAXCOVAR procedure \cite{aggarwal2024sdp}, respectively.

At iteration $n+1$, an edge is constructed from the existing node $\nu$ in $\mathcal{T}^{(n)}$ to a new node with central query mean $\mu_q$. Using Algorithm \ref{alg:construct_brt_max_prob_reach}, the new node is given by an ambiguity set $\mathcal{P}^{\text{BALL} \cup \text{MC}}(\mu_q, \Sigma, r) = \mathcal{P}^{\text{BALL}}(\mu_q, r) \cup \mathcal{P}^{\text{MC}}(\mu_q, \Sigma)$ with $\lambda_{\min}(\Sigma) \geq \lambda_{\min}(\Sigma_{\text{MAXCOVAR}})$, where the MAXCOVAR procedure returns an ambiguity set $\mathcal{P}^{\text{MC}}(\mu_q, \Sigma_{\text{MAXCOVAR}})$. It follows that $\mathcal{P}^{\text{MC}}(\mu_q, \Sigma_{\text{MAXCOVAR}}) \subseteq \mathcal{P}^{\text{MC}}(\mu_q, \Sigma) \subseteq \mathcal{P}^{\text{BALL} \cup \text{MC}}(\mu_q, \Sigma, r)$. It follows then that $t-\text{BRS}(\mathcal{P}^{\text{MC}}(\mu_q, \Sigma_{\text{MAXCOVAR}})) \subseteq t-\text{BRS}(\mathcal{P}^{\text{BALL} \cup \text{MC}}(\mu_q, \Sigma, r))$ $\forall t \geq 1$. Therefore, $(h - d^{(n+1)})-\text{BRS}(\mathcal{P}^{\text{MC}}(\mu_q, \Sigma_{\text{MAXCOVAR}})) \subseteq (h - d^{(n+1)})-\text{BRS}(\mathcal{P}^{\text{BALL} \cup \text{MC}}(\mu_q, \Sigma, r))$, where $d^{(n+1)}$ is the number of hops from the root node to $\nu$.
%From Definition \ref{def: h_BRS_tree}, 
%\begin{align}
%h-\text{BRS}&(\mathcal{T}^{(n+1)}_{OURS}) = h-\text{BRS}(\mathcal{T}^{(n)})\notag \\
%&\bigcup (h-d^{(n+1)})-BRS(\mathcal{P}^{\text{BALL} \cup \text{MC}}(\mu_q, \Sigma, r))
%\end{align}
%and
%\begin{align}
%h-\text{BRS}&(\mathcal{T}^{(n+1)}_\text{MAXCOVAR}) = h-\text{BRS}(\mathcal{T}^{(n)})\notag \\
%&\bigcup (h-d^{(n+1)})-BRS(\mathcal{P}^{\text{MC}}(\mu_q, ^{\text{BALL}\cup \text{MC}})).
%\end{align}
It follows that $h-\text{BRS}(\mathcal{T}^{(n+1)}_\text{OURS}) \supseteq h-\text{BRS}(\mathcal{T}^{(n+1)}_\text{MAXCOVAR})$. Suppose then that $n = 0$, such that $\mathcal{T}^{(0)} = \{\mathcal{P}^{\text{BALL}}(\mu_\mathcal{G}, r_\mathcal{G}) \}$. It follows from the argument above by induction that for any $n$, $h-\text{BRS}(\mathcal{T}^{(n)}_\text{OURS}) \supseteq h-\text{BRS}(\mathcal{T}^{(n)}_\text{MAXCOVAR})$ $\forall h \geq 1$.

Next, we will show that if conditions (a) and (b) required for Lemma \ref{lemma: mild_assumptions} hold for the root node $\mathcal{G} := \mathcal{P}^{\text{BALL}}(\mu_\mathcal{G}, r_\mathcal{G})$, and if Condition \ref{condition: c} holds,
%$\exists (\mathbf{x}_0', \Sigma_0 = 0) \in h-\text{BRS}(\mathcal{G})$ satisfying $\mathbf{x}_0' = \arg \max_{\mathbf{x}_0 \in \text{Proj}_\mu(h-\text{BRS}(\mathcal{G}))} ||\mathbf{x}_0 - \mu_\mathcal{G}||_2$ such that $(\mathbf{x}_0', \Sigma_0 = 0) \notin h-\text{BRS}(\nu_j)$ for any $\nu_j \in \nu(\mathcal{T}^{(n)}_\text{OURS})\setminus \mathcal{G}$ and any $h \geq 1$,
then $h-\text{BRS}(\mathcal{T}^{(n)}_\text{OURS}) \supset h-\text{BRS}(\mathcal{T}^{(n)}_{\text{MAXCOVAR}})$ $\forall n \geq 1, h \geq 1$. Consider $\mathcal{G}_\text{MC} := \mathcal{P}^{\text{MC}}(\mu_\mathcal{G}, \Sigma_\mathcal{G})$, with $\Sigma_\mathcal{G} = \frac{r_\mathcal{G}^2}{\Phi^{-1}(1-\epsilon, n)^2}I$. Then, $\mathcal{G}_\text{MC} \subseteq \mathcal{G}$, where $\mathcal{G}_\text{MC}$ represents the set of candidate terminal distributions when planning with the MAXCOVAR algorithm.
%From Definition \ref{def: h_BRS_tree},
%\begin{align}
%h-\text{BRS}(\mathcal{T}^{(n+1)}_{OURS}) &:= h-\text{BRS}(\mathcal{G}) \notag \\
%&\cup \left(\bigcup_{i \in \nu(\mathcal{T}^{(n+1)}_{OURS} \setminus \mathcal{G})} (h - d_i) - BRS(\nu_i)\right)
%\end{align}
%\begin{align}
%h-\text{BRS}&(\mathcal{T}^{(n+1)}_{\text{MAXCOVAR}}) := h-\text{BRS}(\mathcal{G}_\text{MAXCOVAR}) \notag \\
%&\cup \left(\bigcup_{i \in \nu(\mathcal{T}^{(n+1)}_{\text{MAXCOVAR}} \setminus \mathcal{G})} (h - d_i) - BRS(\nu_i)\right)
%\end{align}
By Lemma \ref{lemma: mild_assumptions}, $h-\text{BRS}(\mathcal{G}) \supset h-\text{BRS}(\mathcal{G}_\text{MC})$, and $\forall h \geq 1$, $\exists \mathbf{x}_0' = \arg \max_{\mathbf{x}_0 \in \text{Proj}_{(\cdot, \mathbb{S}^n_+)}(h-\text{BRS}(\mathcal{G}))} ||\mathbf{x}_0 - \mu_\mathcal{G}||_2$ such that $\mathbf{x}_0' \notin 
\text{Proj}_{(\cdot, \mathbb{S}^n_+)}(h-\text{BRS}(\mathcal{G}_{\text{MC}}))$. Recall that $(\mathbf{x}_0', \Sigma_0 = 0) \notin h-\text{BRS}(\nu_j)$ for any $\nu_j \in \nu(\mathcal{T}^{(n)}_\text{OURS})\setminus \mathcal{G}$ and any $h \geq 1$. Because $h-\text{BRS}(\mathcal{T}^{(n)}_\text{OURS}) \supseteq h-\text{BRS}(\mathcal{T}^{(n)}_\text{MAXCOVAR})$ for all planning scenes and for all $h, n \geq 1$, it follows that $(\mathbf{x}_0', \Sigma_0 = 0) \notin h-\text{BRS}(\nu_j)$ for any $\nu_j \in \nu(\mathcal{T}^{(n)}_\text{MAXCOVAR})\setminus \mathcal{G}$ and any $h \geq 1$. Therefore, $\forall h \geq 1$, $\exists (\mathbf{x}_0', \Sigma_0=0) \in h-\text{BRS}(\mathcal{T}^{(n)}_\text{OURS})$ such that $(\mathbf{x}_0', \Sigma_0=0) \notin h-\text{BRS}(\mathcal{T}^{(n)}_\text{MAXCOVAR})$ and therefore $h-\text{BRS}(\mathcal{T}^{(n)}_\text{OURS}) \supset h-\text{BRS}(\mathcal{T}^{(n)}_\text{MAXCOVAR})$ $\forall h \geq 1$.

\textbf{Proof of Theorem \ref{thm: max_ellipsoid_dominance}}
We will show that $h-\text{BRS}(\mathcal{T}_{\text{OURS}}^{(r)}) \supseteq h-\text{BRS}(\mathcal{T}_{\text{MAXELLIPSOID}}^{(r)})$ for all planning scenes. %This follows easily from the proof of Theorem \ref{thm: max_covar_dominance} and the property that $\mathcal{T}_{\text{OURS}}^{(r)}$ preserves the same ellipsoids at each node as $\mathcal{T}_{\text{MAXELLIPSOID}}^{(r)}$. 
Let $\mathcal{T}^{(n)}$ be a tree of ambiguity sets of distributions after $n$ iterations, such that each node $\nu$ in $\mathcal{T}^{(n)}$ is represented by an ambiguity set $\mathcal{P}^{\text{ELL-BALL} \cup \text{ELL}}(\mu, \Sigma, \mathcal{P}, r) = \mathcal{P}^{\text{ELL-BALL}}(\mu, \mathcal{P}, r) \cup \mathcal{P}^{\text{ELL}}(\mu, \mathcal{P}, \Sigma)$. We will show that for all planning scenes, $h-\text{BRS}(\mathcal{T}^{(n+1)}_\text{OURS}) \supseteq h-\text{BRS}(\mathcal{T}^{(n+1)}_\text{MAXELLIPSOID})$, where $\mathcal{T}^{(n+1)}_\text{OURS}$ and $\mathcal{T}^{(n+1)}_\text{MAXELLIPSOID}$ are trees obtained from $\mathcal{T}^{(n)}$ by adding a node using Algorithm \ref{alg:construct_brt_max_ell_prob_reach} and the MAXELLIPSOID procedure, respectively.

At iteration $n+1$, an edge is constructed from the existing node $\nu$ in $\mathcal{T}^{(n)}$ to a new node with central query mean $\mu_q$. Using Algorithm \ref{alg:construct_brt_max_ell_prob_reach}, the new node is given by an ambiguity set $\mathcal{P}^{\text{ELL-BALL} \cup \text{ELL}}(\mu_q, \Sigma, \mathcal{P}, r)$ with $\lambda_{\min}(\Sigma) \geq \lambda_{\min}(\Sigma_{\text{MAX-ELL}})$ and $\mathcal{P} = \mathcal{P}_\text{MAX-ELL}$, where $\mathcal{P}^{\text{ELL}}(\mu_q, \Sigma_{\text{MAX-ELL}}, \mathcal{P}_{\text{MAX-ELL}})$ is the ambiguity set returned by the MAXELLIPSOID procedure. It follows that $\mathcal{P}^{\text{ELL}}(\mu_q, \Sigma_{\text{MAX-ELL}}, \mathcal{P}_{\text{MAX-ELL}}) \subseteq \mathcal{P}^{\text{ELL-BALL} \cup \text{ELL}}(\mu_q, \Sigma, \mathcal{P}, r)$ and then that $t-\text{BRS}(\mathcal{P}^{\text{ELL}}(\mu_q, \Sigma_{\text{MAX-ELL}}, \mathcal{P}_{\text{MAX-ELL}})) \subseteq t-\text{BRS}(\mathcal{P}^{\text{ELL-BALL} \cup \text{ELL}}(\mu_q, \Sigma, \hat{\mathcal{P}}, r))$ $\forall t \geq 1$. Therefore, $(h - d^{(n+1)})-\text{BRS}(\mathcal{P}^{\text{ELL}}(\mu_q, \Sigma_{\text{MAX-ELL}}, \mathcal{P}_{\text{MAX-ELL}})) \subseteq (h - d^{(n+1)})-\text{BRS}(\mathcal{P}^{\text{ELL-BALL} \cup \text{ELL}}(\mu_q, \Sigma, \hat{\mathcal{P}}, r))$, where $d^{(n+1)}$ is the number of hops from the root node to $\nu$.
\begin{comment}
From Definition \ref{def: h_BRS_tree}, 
$h-\text{BRS}(\mathcal{T}^{(n+1)}_{\text{OURS}}) = h-\text{BRS}(\mathcal{T}^{(n)})\bigcup (h-d^{(n+1)})-BRS(\mathcal{P}^{\text{ELL-BALL} \cup \text{ELL}}(\mu_q, \Sigma, \mathcal{P}, r))$
and
$h-\text{BRS}(\mathcal{T}^{(n+1)}_{\text{MAXELLIPSOID}}) = h-\text{BRS}(\mathcal{T}^{(n)})\bigcup (h-d^{(n+1)})-BRS(\mathcal{P}^{\text{ELL}}(\mu_q, \Sigma_{\text{MAX-ELL}}, \mathcal{P}_{\text{MAX-ELL}})).
$
\end{comment}
It follows by Definition \ref{def: h_BRS_tree} that $h-\text{BRS}(\mathcal{T}^{(n+1)}_{\text{OURS}}) \supseteq h-\text{BRS}(\mathcal{T}^{(n+1)}_{\text{MAXELLIPSOID}})$. Suppose then that $n = 0$, such that $\mathcal{T}^{(0)} = \{\mathcal{P}^{\text{ELL-BALL}}(\mu_\mathcal{G}, \mathcal{P}_\mathcal{G}, r_\mathcal{G}) \}$. It follows from the argument above by induction that for any $n$, $h-\text{BRS}(\mathcal{T}^{(n)}_{\text{OURS}}) \supseteq h-\text{BRS}(\mathcal{T}^{(n)}_{\text{MAXELLIPSOID}})$ $\forall h \geq 1$.
%\section*{Conclusion}
%\input{conclusion}
%% Use plainnat to work nicely with natbib.
\small
%\baselinestretch{0.9}
\renewcommand{\baselinestretch}{0.9}
\bibliographystyle{plain}
\bibliography{references}
\begin{IEEEbiography}[{\includegraphics[width=1in,height=1.25in,clip,keepaspectratio]{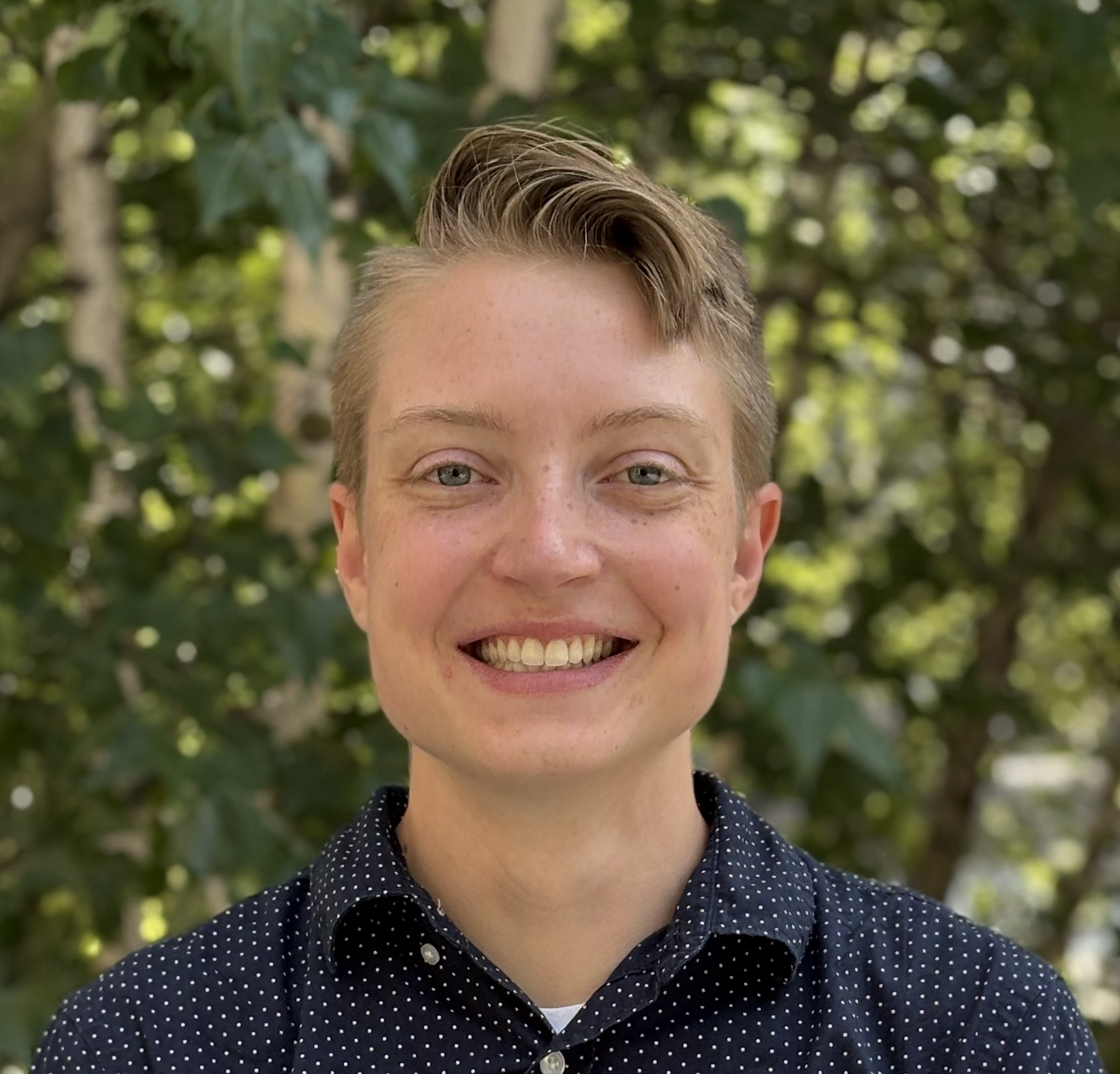}}]%
{Alex Rose} received the S.B. degree and S.M. degrees in aeronautics and astronautics from the Massachusetts Institute of Technology (MIT), Cambridge, MA, USA, in 2021 and 2023, respectively, and is currently working towards the Ph.D. degree in aeronautics and astronautics at MIT.

Alex's research interests include control theory and optimization with applications to space systems and robotics.
\end{IEEEbiography}
\begin{IEEEbiography}[{\includegraphics[width=1in,height=1.25in,clip,keepaspectratio]{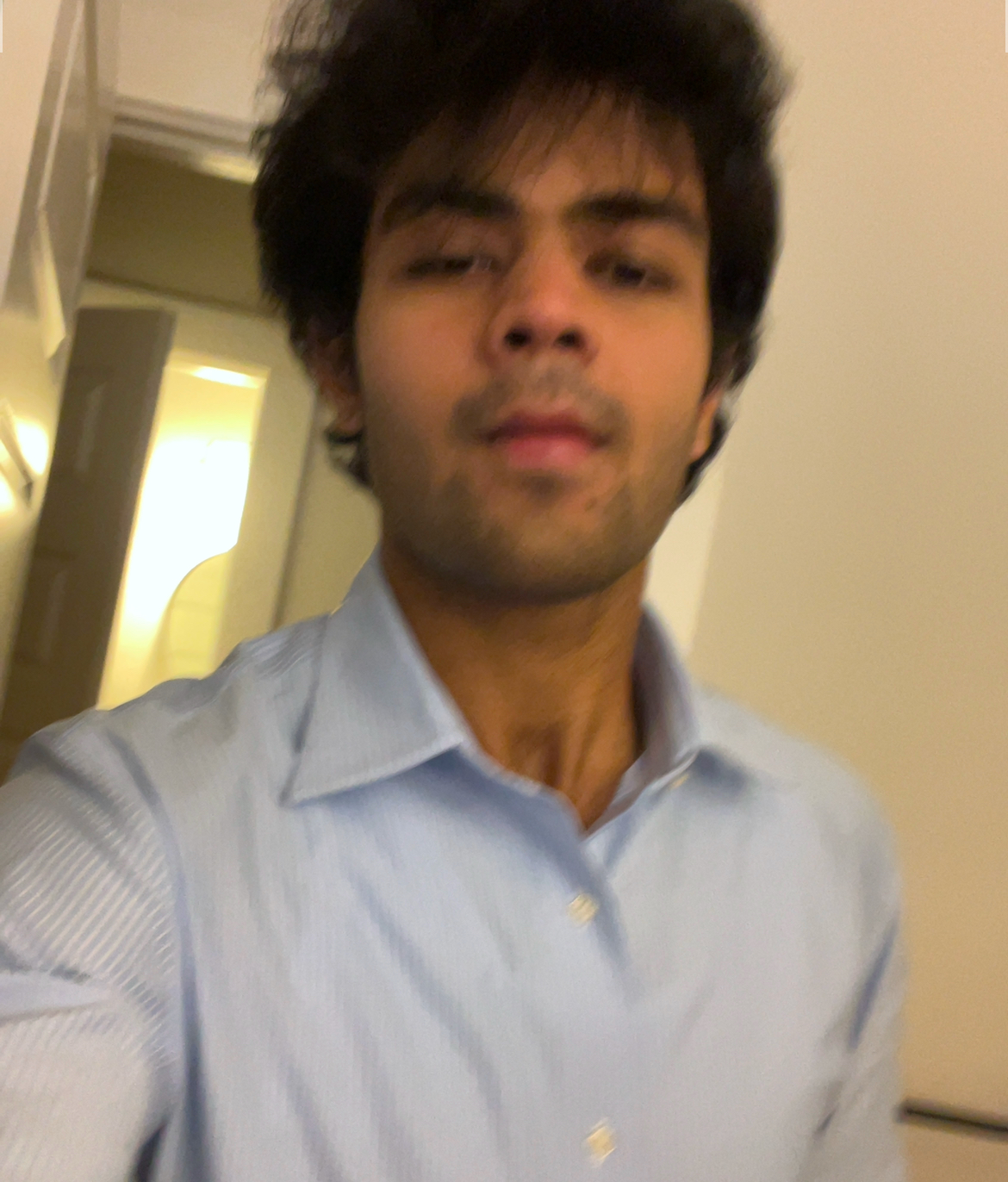}}]%
{Naman Aggarwal} received the Inter-Disciplinary Dual Degree (B.Tech + M.Tech) in Aerospace Engineering and Systems and Control Engineering from the Indian Insitute of Technology Bombay, Mumbai, India in 2021. He is currently  working towards the Ph.D. degree in the Department of Aeronautics and Astronautics and the Laboratory of Information and Decision Systems (LIDS) from Massachusetts Institute of Technology (MIT), Cambridge, MA, USA. He is a member of the Aerospace Controls Laboratory, led by Prof. Jonathan How. His current research interests include control theory and optimization with applications at the intersection of learning, games, and multi-agent control.
\end{IEEEbiography}
\begin{IEEEbiography}[{\includegraphics[width=1in,height=1.25in,clip,keepaspectratio]{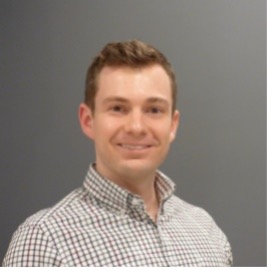}}]%
{Christopher Jewison} is a GN\&C Engineer at Draper. He received his bachelor’s degree (2012) from Cornell University and his master’s degree (2014) and PhD (2017) from MIT.
\end{IEEEbiography}
\begin{IEEEbiography}[{\includegraphics[width=1in,height=1.25in,clip,keepaspectratio]{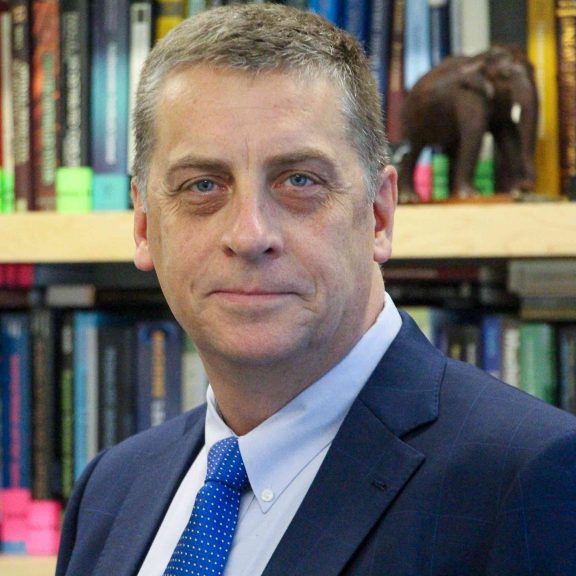}}]%
{Jonathan P. How}
(Fellow, IEEE) received the B.A.Sc. degree in Engineering Science (Aerospace) from the University of Toronto, Toronto, ON, Canada, in 1987, and the S.M. and Ph.D. degrees in aeronautics and astronautics from the Massachusetts Institute of Technology (MIT), Cambridge, MA, USA, in 1990 and 1993, respectively.
In 2000, he joined MIT, where he is currently the Ford Professor of Engineering. Prior to this, he was an Assistant
Professor with Stanford University, Stanford, CA, USA.
Dr. How was the Recipient of the 
IEEE Transactions on Robotics King-Sun Fu Memorial Best Paper Award for 2022 and 2024, the IEEE Control Systems Society Distinguished Member Award in 2020, and the AIAA Intelligent Systems Award in 2020.  He was the Editor-in-Chief for IEEE Control Systems Magazine from 2015 to 2019. He is a Fellow of the AIAA. He was elected to the National Academy of Engineering in 2021.
\end{IEEEbiography}
\end{document}